\documentclass{aa}  
\usepackage{graphicx}
\usepackage{txfonts}
\usepackage{threeparttable}
\usepackage{CJK}

\begin{document}

   \title{Cloud formation in the atomic and molecular phase: \ion{H}{I} self absorption (HISA) towards a Giant Molecular Filament}
	\titlerunning{HISA study towards a GMF}
\author{
	Y. Wang\inst{1}
	\and
	S. Bihr\inst{1}
	\and
	H. Beuther\inst{1}       
	\and
	M. R. Rugel\inst{1,2}
	\and
	J. D. Soler\inst{1}
	\and
          J. Ott\inst{3}
          \and
          J. Kainulainen\inst{4}
          \and
          N. Schneider\inst{5}
          \and
          R. S. Klessen\inst{6,7}
           \and
          S. C. O. Glover\inst{6}
            \and
          N. M. McClure-Griffiths\inst{8}
          \and
           P. F. Goldsmith\inst{9}
           \and
           K. G. Johnston\inst{10}
          \and
         K. M. Menten\inst{2}
          \and
          S. Ragan \inst{11}
          \and
           L. D. Anderson\inst{12,13,14}
          \and
          J. S. Urquhart\inst{15}
          \and
           H. Linz\inst{1}
	 \and
          N. Roy\inst{16}
          \and
          R. J. Smith\inst{17}
          \and
          F. Bigiel\inst{18}
          \and
          T. Henning\inst{1}
           \and
          S. N. Longmore\inst{19}
}
\institute{
Max Planck Institute for Astronomy, K\"onigstuhl 17, 69117 Heidelberg, Germany\\
\email{wang@mpia.de}
\and
Max-Planck-Institut f\"ur Radioastronomie, Auf dem H\"ugel 69, 53121 Bonn, Germany
\and
National Radio Astronomy Observatory, P.O. Box O, 1003 Lopezville Road, Socorro, NM 87801, USA
\and
Chalmers University of Technology, Department of Space, Earth and Environment, SE-412 93 Gothenburg, Sweden
\and
I. Physik. Institut, University of Cologne, Z\"ulpicher Str.77, 50937 Cologne, Germany
\and
Universit\"at Heidelberg, Zentrum f\"ur Astronomie, Institut f\"ur Theoretische Astrophysik, Albert-Ueberle-Str. 2, 69120 Heidelberg, Germany
 \and
Universit\"at Heidelberg, Interdisziplin\"ares Zentrum f\"ur Wissenschaftliches Rechnen, INF 205, 69120, Heidelberg, Germany
 \and
Research School of Astronomy and Astrophysics, The Australian National University, Canberra, ACT, Australia
\and
Jet Propulsion Laboratory, California Institute of Technology, 4800 Oak Grove Drive, Pasadena, CA 91109, USA
\and
School of Physics and Astronomy, University of Leeds, Leeds LS2 9JT, UK
\and
School of Physics and Astronomy, Cardiff University, Queen’s Buildings, The Parade, Cardiff, CF24 3AA, UK
\and
Department of Physics and Astronomy, West Virginia University, Morgantown, WV 26506, USA
\and
Adjunct Astronomer at the Green Bank Observatory, P.O. Box 2, Green Bank WV 24944, USA
\and
Center for Gravitational Waves and Cosmology, West Virginia University, Chestnut Ridge Research Building, Morgantown, WV 26505, USA
  \and
 School of Physical Sciences, University of Kent, Ingram Building, Canterbury, Kent CT2 7NH, UK
\and 
Department of Physics, Indian Institute of Science, Bengaluru 560012, India
\and
Jodrell Bank Centre for Astrophysics, School of Physics and Astronomy, The University of Manchester, Oxford Road, Manchester, M13 9PL, UK 
\and
Argelander Institut f\"ur Astronomie, Auf dem H\"ugel 71, 53121 Bonn, Germany
\and
 Astrophysics Research Institute, Liverpool John Moores University, IC2, Liverpool Science Park, 146 Brownlow Hill, Liverpool L3 5RF, UK
}

\date{Received dd, mm, yyyy; accepted dd, mm, yyyy}

 
  \abstract
 {Molecular clouds form from the atomic phase of the interstellar medium. However, characterizing the transition between the atomic and the molecular interstellar medium (ISM) is a difficult observational task. Here we address cloud formation processes by combining \ion{H}{I} self absorption (HISA) with molecular line data. Column density probability density functions (N-PDFs) are a common tool to examine molecular clouds. One scenario proposed by numerical simulations is that the N-PDF evolves from a log-normal shape at early times to a power-law-like shape at later times. To date, investigations of N-PDFs are mostly limited to the molecular component of the cloud. In this paper, we study the cold atomic component of the giant molecular filament GMF38.1-32.4a (GMF38a, distance=3.4~kpc, length$\sim230$~pc), calculate its N-PDFs and study its kinematics. We identify an extended HISA feature, which is partly correlated with the $^{13}$CO emission. The peak velocities of the HISA and $^{13}$CO observations agree well on the eastern side of the filament, whereas a velocity offset of approximately 4~km~s$^{-1}$ is found on the western side. The sonic Mach number we derive from the linewidth measurements shows that a large fraction of the HISA, which is ascribed to the cold neutral medium (CNM), is at subsonic and transonic velocities. The column density of the CNM part is on the order of 10$^{20}$ to 10$^{21}$\,cm$^{-2}$. The column density of molecular hydrogen, traced by $^{13}$CO, is an order of magnitude higher. The N-PDFs from HISA (CNM), \ion{H}{i} emission (the warm and cold neutral medium), and $^{13}$CO (molecular component) are well described by log-normal functions, which is in agreement with turbulent motions being the main driver of cloud dynamics. The N-PDF of the molecular component also shows a power law in the high column-density region, indicating self-gravity. We suggest that we are witnessing two different evolutionary stages within the filament. The eastern subregion seems to be forming a molecular cloud out of the atomic gas, whereas the western subregion already shows high column density peaks, active star formation and evidence of related feedback processes. }


   \maketitle
%

\section{Introduction}
\label{sect_intro}

Stars, one of the key components of our universe, form in molecular clouds which are composed mainly of molecular hydrogen \citep[e.g.,][]{Larson2003, Stahler2005, McKee2007, Dobbs2014, Tan2014}, yet the formation process of molecular clouds is still under debate. Various studies show molecular clouds form out of relatively diffuse atomic hydrogen gas \citep[e.g., ][]{Larson1981, Blitz2007, Clark2012, Hennebelle2012, Dobbs2014, Sternberg2014, Klessen2016}. Different processes have been proposed \citep[e.g.,][]{Hennebelle2012, Dobbs2014, Klessen2016}. Basically, the atomic gas contracts and the increased column density shield the cloud from the interstellar UV radiation and cool down to form molecular gas. The cold neutral medium (CNM) with a typical temperature in the range of 40 to 100~K and volume density $>10\sim100$~cm$^{-3}$ \citep{McKee1977, Wolfire1995, wilson2010} is the key component connecting the diffuse atomic gas with the molecular gas. Thus, observational constraints on the physical properties of the CNM, such as density distribution and kinematics, are crucial to understand the formation of the molecular cloud.

Although 21cm \ion{H}{i} line emission offers a straightforward tool to study atomic hydrogen, it is difficult to determine the properties of the gas from which it arises. The main challenge is the coexistence of the warm neutral medium (WNM) and the CNM assumed to be in pressure equilibrium \citep{McKee1977, Wolfire1995, Wolfire2003}. Studies of \ion{H}{i} self absorption (HISA) overcome this problem, as they only trace the CNM. HISA was first detected by \citet{Heeschen1954, Heeschen1955} towards the Galactic center. HISA features occur if cold, dense atomic hydrogen is in front of a warmer emission background \citep[e.g.,][]{Knapp1974}. Since then a number of observations that have been carried out with single dish telescopes and interferometers, and HISA features were found to be widespread in the Milky Way \citep[e.g.,][]{Riegel1972, Knapp1974, Heiles1975, McCutcheon1978, Levinson1980, Minn1981, Shuter1987, vanderWerf1988, vanderWerf1989, Montgomery1995, Gibson2000, Kavars2003, Gibson2005, Kavars2005, Denes2018}. The spin temperature of the cold \ion{H}{i} responsible for HISA ranges from $\sim$10-60\,K, \citep[e.g.][]{Gibson2000, Kavars2005, McClure-Griffiths2006}. A special case of the HISA features are so called \ion{H}{i} narrow self absorption (HINSA) features were studied towards nearby molecular clouds, revealing small linewidths on the order of $\lesssim$1\,km\,s$^{-1}$ \citep{Li2003, Goldsmith2005, Krco2008, Krco2010, Zuo2018}. However, studies characterizing the column density and the kinematic distribution of the CNM in large maps are still rare. 

Recent observations have revealed a group of large ($\geq$ 100~pc) and massive ($\geq$ 10$^5$~M$_\odot$) filaments, known as giant molecular filaments (GMFs), which may be linked to Galactic dynamics and trace the gravitational mid-plane in the Milky Way (MW) \citep{Jackson2010, Goodman2014, Wang2015, Zucker2015, Abreu-Vicente2016, Li2016, Wang2016, Zucker2018, Zhang2019}. These observations show that GMFs are the largest coherent gas structures in our Milky Way, and often contain different evolutionary stages of the star formation regions simultaneously in the same filament \citep{Goodman2014, Zucker2015}, which makes them ideal targets to study the CNM properties in different environments that lead to molecular cloud formation. 

A common tool to study molecular clouds are the probability density functions of the column density (N-PDFs) \citep[see e.g.,][]{Ostriker2001, Lombardi2008, Kainulainen2009, AlvesdeOliveira2014, Sadavoy2014, Abreu-Vicente2015, Stutz2015, Schneider2015, Lin2017,Chen2018}. The shape of N-PDFs is predicted to depend on the physical processes acting within the cloud. In the early evolution of a molecular cloud, turbulent motions within the cloud dominate and the N-PDF reveals a log-normal shape. The width of the log-normal N-PDF is also determined by the turbulent motions \citep[see e.g.,][]{Federrath2010, Ballesteros-Paredes2011, Kritsuk2011, Federrath2013, Burkhart2015B, Bialy2017}. In this scenario, more evolved clouds develop a high-density power-law tail, indicating that the cloud structure has evolved and gravity dominates. Observations indicate that star-forming clouds show such tails, lending support to this scenario \citep[e.g.,][]{Kainulainen2009, Schneider2013}. The slope of the power-law N-PDF can be related to evolutionary stages of the clouds with steeper slopes possibly indicating with earlier evolutionary stages \citep[e.g.,][]{Kritsuk2011, Federrath2013, Ward2014}. 



High-mass star-forming regions reveal multiple power-laws, having a shallower slope for the highest density regions. This indicates a slower collapse for such regions \citep{Schneider2015}. \citet{Lombardi2015} and \citet{Alves2017} present a contrasting argument, reporting that all N-PDFs have a power-law shape and the log-normal shape could be an observational bias, a view point that has triggered considerable controversy \citep{Ossenkopf2016, Chen2018, Koertgen2019}. Theoretical work and simulations of molecular clouds also reproduce N-PDFs in different forms \citep[e.g.,][]{Vazquez-Semadeni1994, Federrath2010, Federrath2012, Burkhart2015B}. \citet{Burkhart2015} and \citet{Imara2016} studied nearby molecular clouds and report \ion{H}{i} N-PDFs with a log-normal shape, without any power-law tail. \citet{Rebolledo2017} studied the Carina and Gum~31 molecular complex, where the \ion{H}{i} N-PDF also shows a log-normal shape.

To investigate the transition of atomic to molecular hydrogen in more detail, we examine the hydrogen content with HISA measurements in detail for GMF38.1-32.4a (GMF38a, \citealt{Ragan2014}). With a velocity range between 50 and 60~km~s$^{-1}$ \citep{Ragan2014}, GMF38a is at a median distance of 3.4~kpc from the Sun (Galactocentric distance $\sim$5.9~kpc) estimated from the Bayesian Distance Estimator tool \citep{Reid2016}. The top panel of Fig.\,\ref{fig_filament_overview} shows the integrated $^{13}$CO emission from the GRS survey \citep{Jackson2006}. Our goal is to study the kinematics of this GMF in the molecular and atomic hydrogen traced by $^{13}$CO emission and HISA, respectively. Furthermore, we analyze N-PDFs for the atomic and molecular hydrogen and compare their properties. 

\section{Observations and Methods}
\label{sect_obs}
\subsection{The \ion{H}{i} 21~cm line and continuum}
The \ion{H}{i} 21~cm line observations of GMF38.1-32.4 are part of the \ion{H}{i}, OH, Recombination line survey of the Milky Way (THOR; \citealt{beuther2016}). The survey observed a part of the first quadrant of the Galactic plane ($l=14.0-67.4^\circ$ and $\lvert b \rvert \leq 1.25^\circ$) with the {\it Karl G. Jansky} Very Large Array (VLA) in C-configuration at L band from 1 to 2~GHz, covering the \ion{H}{i} 21~cm line, 4 OH lines, 19 H$\alpha$ recombination lines, and eight continuum bands between 1 and 2~GHz \citep{beuther2016}. Each pointing was observed for $4\times2$~minutes to ensure a uniform $uv$-coverage. The spectral window for the \ion{H}{i} 21~cm line was set to have a bandwidth of 2~MHz ($\sim400$~km~s$^{-1}$) and a spectral resolution of 3.91~kHz ($\sim0.82$~km~s$^{-1}$). The data calibration was done with CASA\footnote{\url{http://casa.nrao.edu}; version 4.1.0} \citep{McMullin2007}. The flux and bandpass were calibrated with the quasar 3C~286. J1822-0938 was used for the phase and gain calibration \citep[see also][]{beuther2016}.  

To recover the large scale structure we combined the C-configuration data with the \ion{H}{i} Very Large Array Galactic Plane Survey (VGPS, \citealt{stil2006}), which consists of VLA D-configuration data combined with single-dish observations from the Green Bank Telescope (GBT). We subtracted the continuum in the visibility datasets (with the CASA command uvcontsub), and used the multiscale CLEAN in CASA\footnote{version 5.1.1} to image the three adjacent tiles of continuum-subtracted C-configuration data together with D-configuration data. A pixel size of 4$\arcsec$, a spectral resolution of 1.5~km~s$^{-1}$, and a robust weighting value of 0.45 were used. The resulting images, which have a synthesized beam between 20$\arcsec$ to 40$\arcsec$ over the entire coverage of the THOR survey, were all smoothed to a common resolution of 40$\arcsec$. The images were further combined with the VGPS images (D+GBT) using the task ``feather''  in CASA to recover the large-scale structure. We compared the flux of the combined \ion{H}{i} data with the single dish GBT data from VGPS. The flux agrees with each other within 5.7\%. Considering that the typical absolute flux calibration uncertainty for the VLA at 1.4~GHz is $\sim$5\% \citep{beuther2016}, it is reasonable to conclude that our combined \ion{H}{I} data fully recover the extended emission. The noise level in the line-free channel is about 4~K per 1.5~km~s$^{-1}$.

Additionally, the THOR C-configuration only \ion{H}{i} line with the continuum data \citep{beuther2016} are used to measure the \ion{H}{i} optical depth towards bright continuum sources in the background. The THOR+VGPS 1.4~GHz continuum data \citep[VLA C+D+GBT,][]{Wang2018} are employed to estimate the diffuse continuum emission in the background. By comparing the flux density of the known SNRs \citep{green2014}, \citet{anderson2017} showed that the flux retrieved from the combined continuum data is consistent with the literature. Thus the continuum data also recover the extended emission.

\subsection{\ion{H}{i} self absorption}
\label{sec_HI_self_absorption}
\begin{figure*}
\centering
 \includegraphics[width=1\textwidth]{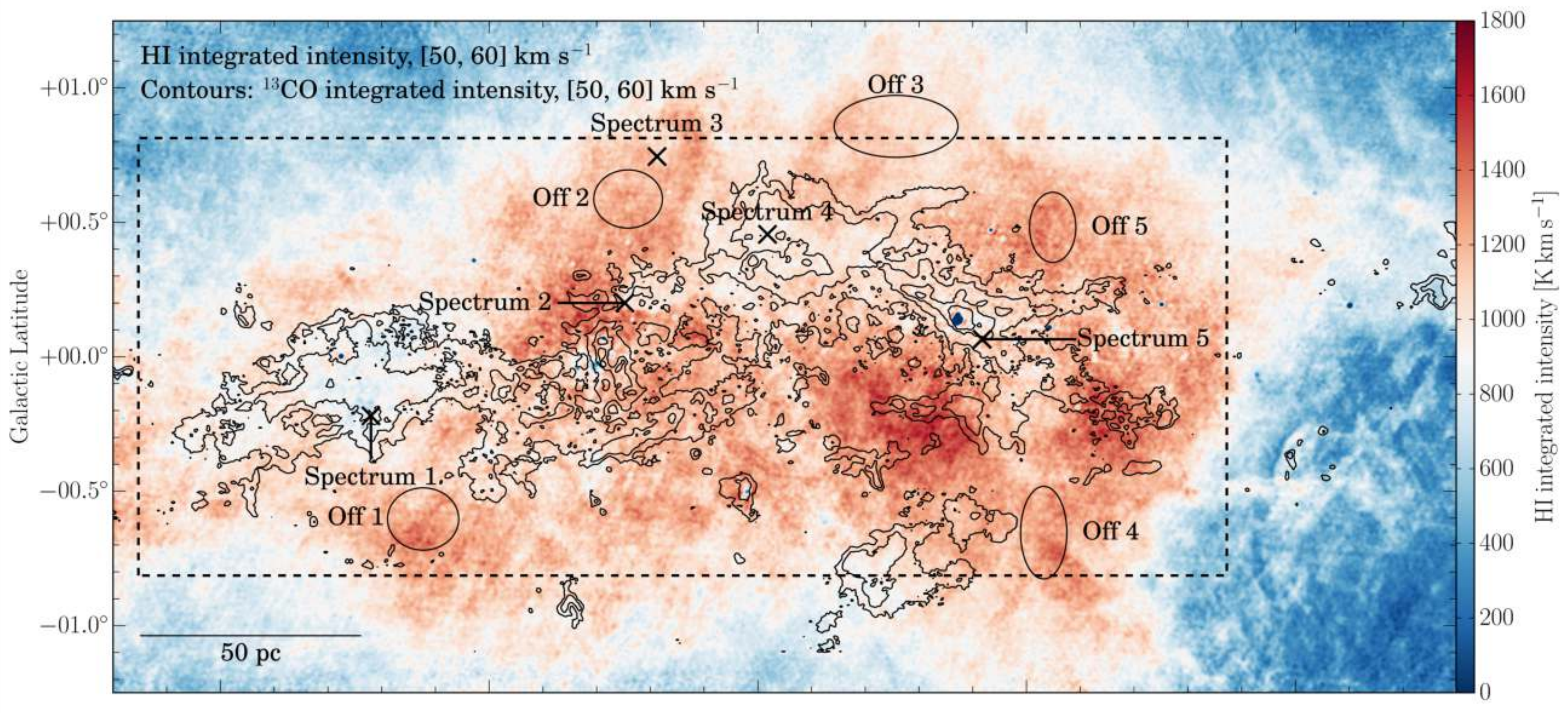}\\
 \includegraphics[width=1\textwidth]{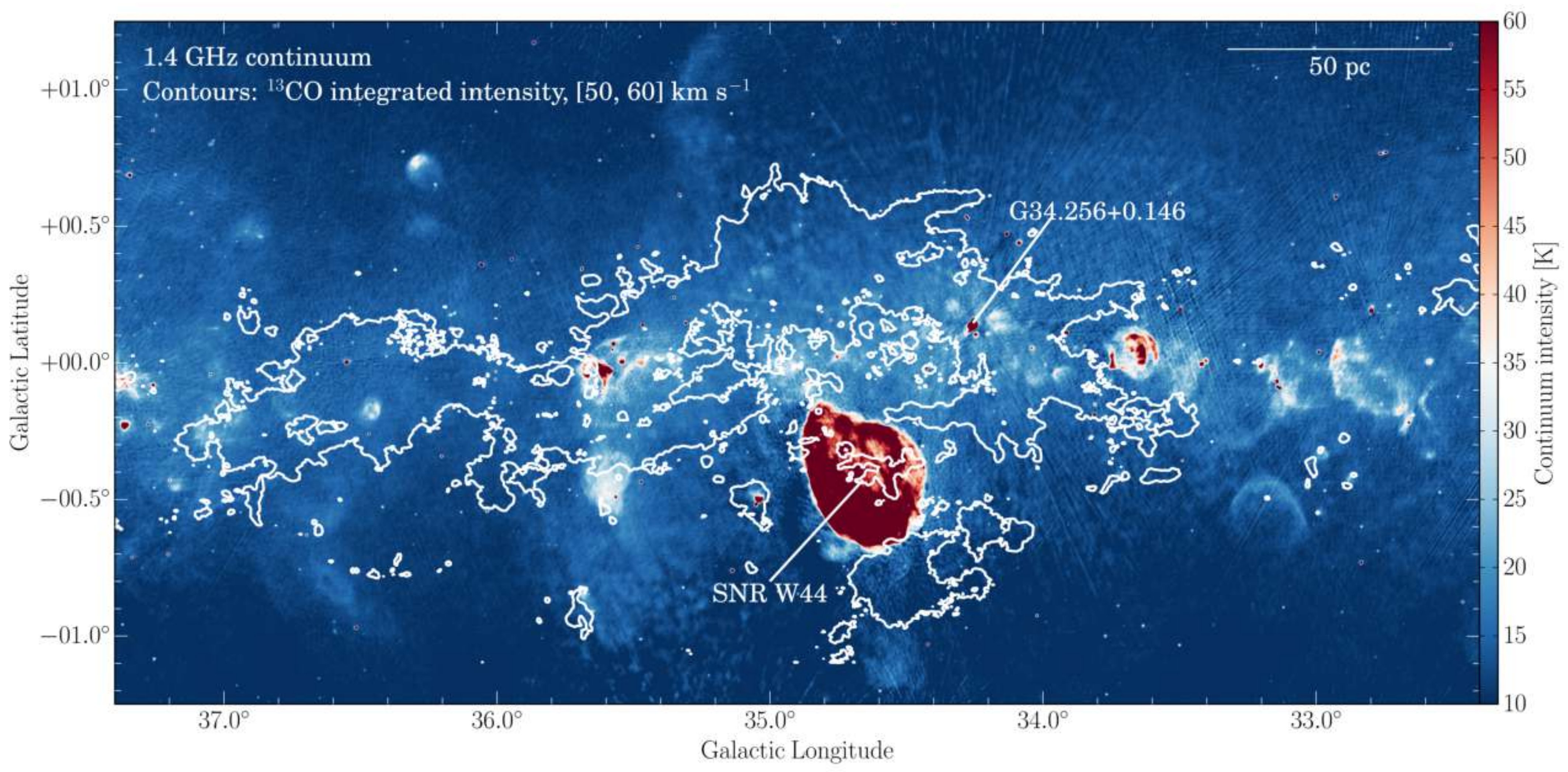}\\
\caption{Overview of the giant molecular filament GMF38a. The top panel shows the $^{13}$CO \citep[GRS,][]{Jackson2006} integrated intensity contours in the range $v_{\rm{LSR}} = 50-60$\,km\,s$^{-1}$ overlaid on integrated \ion{H}{i} emission in the same velocity range. The black ellipses show the “off-positions” whose spectra are shown in Fig.\,\ref{fig_different_off_positions} and the black ``$\times$'' signs mark positions whose spectra are show in Fig.\,\ref{fig_different_HISA_spectra}. The bottom panel shows the same $^{13}$CO integrated intensity contours overlaid on 1.4\,GHz continuum emission from the THOR survey \citep{Wang2018}. The contours in the top panel indicate integrated $^{13}$CO emission levels of 5, 10, 20 and 30\,K\,km\,s$^{-1}$. The contours in the bottom panel indicate integrated $^{13}$CO emission levels of 5~K~km\,s$^{-1}$ for reference. The dashed box in the top panel outlines the region that is discussed in the following sections and shown in Fig.\ref{fig_HISA_peak_emission_overlay}, \ref{fig_overview_velocity_map}, \ref{fig_overview_FWHM_map}, \ref{fig_column_density}, and \ref{fig_maximum_spin_temperature}.}
  \label{fig_filament_overview}
\end{figure*}

The integrated \ion{H}{i} emission over the velocity range 50--60~km~s$^{-1}$, shown in the top panel of Fig.\,\ref{fig_filament_overview}, reveals diffuse emission covering a larger area than the $^{13}$CO emission. The strongest \ion{H}{i} emission does not coincide with $^{13}$CO emission, but an anti-correlation between the \ion{H}{i} and $^{13}$CO emission is suggested. Our analysis in the following section shows that this anti-correlation is due to the HISA: the cold atomic hydrogen absorbs the emission from an emitting atomic hydrogen cloud in the background, i.e., \ion{H}{i} self absorption. The terminology ``\ion{H}{i} self absorption'' can be misleading. The emission and absorption processes can occur in the same cloud, but it is possible that the \ion{H}{i} emission originates from a distant background cloud, which covers a similar or larger range of LSR velocities as the absorbing foreground cloud as illustrated in Fig.\,\ref{fig_HISA_position_sketch}. A comprehensive discussion about the radiative transfer of HISA features can be found in \citet{Gibson2000}, \citet{Kavars2003}, \citet{Li2003}, and \citet{Goldsmith2005}. In general, we observe an emitting foreground and background \ion{H}{i} cloud, which have spin temperatures, $T_{\rm{fg}}$, and $T_{\rm{bg}}$, respectively. The cold, absorbing \ion{H}{i} cloud can be located between these two emitting clouds, having the spin temperature, $T_{\rm{HISA}}$. Furthermore, we observe 1.4\,GHz continuum emission, which can be a diffuse Galactic component or arise from discrete strong sources. For simplicity, we assume that the continuum emission is situated in the background. In this, we will exclude the possibility of strong, discrete continuum sources and consider only the weak diffuse continuum background when estimating the HISA properties. In Sect.\,\ref{sect_HI_optical_depth_continuum_source}, we will utilize strong continuum sources to determine the optical depth of the atomic hydrogen, which can help us to constrain the spin temperature of HISA. Following the equation of radiative transfer in \citet{Rybicki1979}, the measured on and off position brightness temperatures of the line above the continuum at a certain velocity are:
\begin{figure}
\centering
 \includegraphics[width=0.5\textwidth]{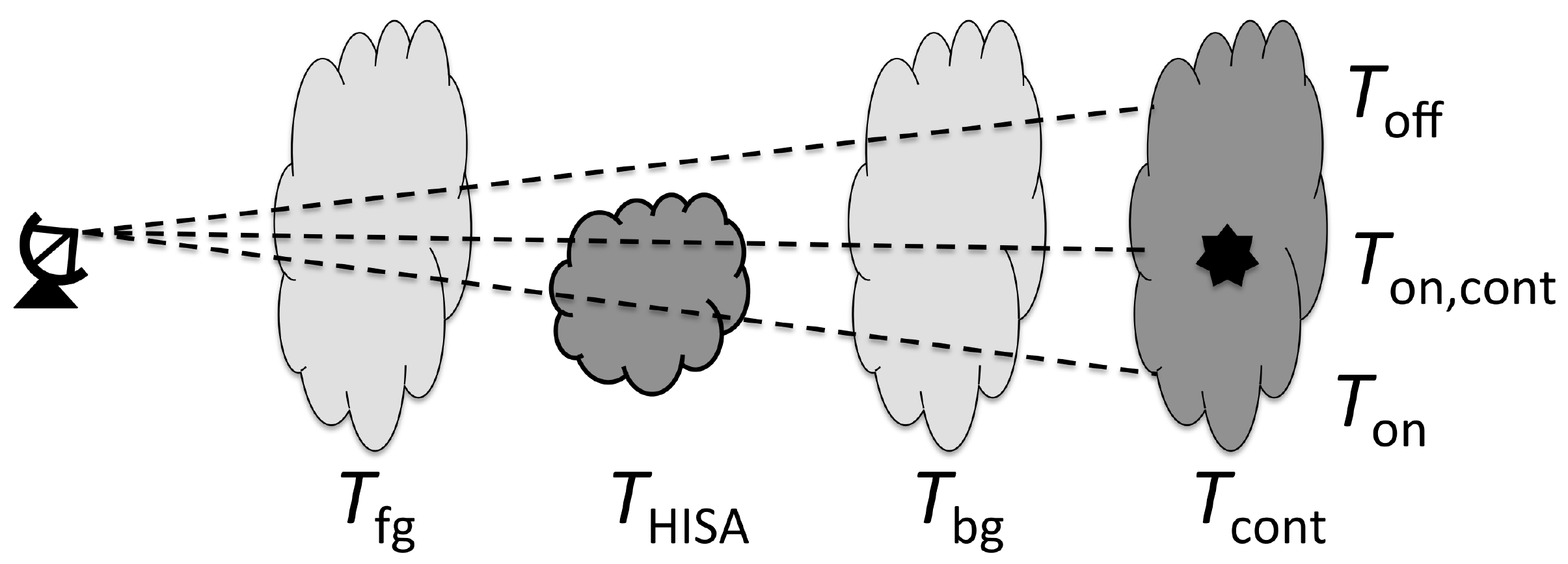}
\caption{Schematic view of the observed \ion{H}{i} components. The cold absorbing cloud (HISA) with temperature $T_{\rm HISA}$ is surrounded by emitting clouds with temperature $T_{\rm{fg}}$ and $T_{\rm{bg}}$. Behind the \ion{H}{i} clouds, several continuum sources can be situated, either diffuse or discrete (marked with a star).}
  \label{fig_HISA_position_sketch}
\end{figure}
\begin{figure}
\centering
 \includegraphics{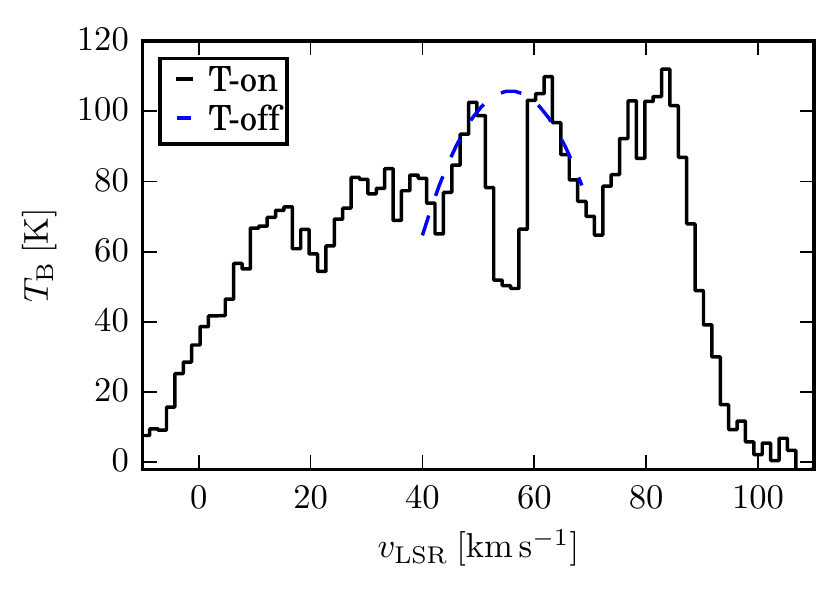}
\caption{Example spectrum showing a prominent HISA feature around $v_{_{\rm{LSR}}}\sim 55$\,km\,s$^{-1}$. The actual \ion{H}{i} spectra is shown in black ($T_{\rm{on}}$) and the estimated background emission using a second order polynomial fit (see Sect.\,\ref{sect_background_estimate_T_off}) is shown in blue ($T_{\rm{off}}$).}
  \label{fig_HISA_spectrum_intro}
\end{figure}
\begin{equation}
\begin{split}
T_{\rm{off}} = & \: T_{\rm{fg}} (1-{\rm e}^{-\tau_{\rm{fg}}}) + T_{\rm{bg}} (1-{\rm e}^{-\tau_{\rm{bg}}}){\rm e}^{-\tau_{\rm{fg}}}+ T_{\rm{cont}} {\rm e}^{-(\tau_{\rm{fg}} + \tau_{\rm{bg}})} -T_{\rm{cont}}\\
T_{\rm{on}} \, = & \: T_{\rm{fg}} (1-{\rm e}^{-\tau_{\rm{fg}}}) + T_{\rm{HISA}} (1-{\rm e}^{-\tau_{\rm{HISA}}}){\rm e}^{-\tau_{\rm{fg}}} + \\ 
& \: T_{\rm{bg}} (1-{\rm e}^{-\tau_{\rm{bg}}}){\rm e}^{-(\tau_{\rm{fg}}+\tau_{\rm{HISA}})} + T_{\rm{cont}} {\rm e}^{-(\tau_{\rm{fg}}+\tau_{\rm{HISA}} + \tau_{\rm{bg}})}-T_{\rm{cont}},
\end{split}
\label{eq_on_off_start}
\end{equation}
where $\tau_{\rm{fg}}$, $\tau_{\rm{bg}}$, $\tau_{\rm{HISA}}$ are the corresponding optical depths of each component shown in Fig.\,\ref{fig_HISA_position_sketch} and $T_{\rm{cont}}$ is the continuum brightness temperature. During the data reduction, we subtract the continuum emission from the \ion{H}{i} visibility data (see also Sect.~\ref{sect_obs}), which is indicated by the last term ($-T_{\rm{cont}}$, see Sect.~\ref{sect_obs}). An example spectrum illustrating $T_{\rm{on}}$ and fitted $T_{\rm{off}}$ is shown in Fig.~\ref{fig_HISA_spectrum_intro}. Assuming on and off spectra share the same $T_{\rm cont}$ and calculating the difference, we get:
\begin{equation}
\begin{split}
T_{\rm{on}-\rm{off}}  =\: & T_{\rm{HISA}} (1-{\rm e}^{-\tau_{\rm{HISA}}}){\rm e}^{-\tau_{\rm{fg}}} - T_{\rm{bg}} (1-{\rm e}^{-\tau_{\rm{bg}}}){\rm e}^{-\tau_{\rm{fg}}}(1-{\rm e}^{-\tau_{\rm{HISA}}})\\
& - T_{\rm{cont}} (1-{\rm e}^{-\tau_{\rm{HISA}}}){\rm e}^{-(\tau_{\rm{fg}}+\tau_{\rm{bg}})}\\
	=\: & \left(T_{\rm{HISA}} - T_{\rm{bg}} (1-{\rm e}^{-\tau_{\rm{bg}}}) - T_{\rm{cont}} {\rm e}^{-\tau_{\rm{bg}}}\right) \times (1-{\rm e}^{-\tau_{\rm{HISA}}}){\rm e}^{-\tau_{\rm{fg}}}
\end{split}
\label{eq_HISA_on_minus_off}
\end{equation}
This equation can be further simplified by introducing the dimensionless parameter $p$ \citep[e.g., ][]{Feldt1993, Gibson2000}:
\begin{equation}
p\equiv \frac{T_{\rm{bg}} (1-{\rm e}^{-\tau_{\rm{bg}}})}{T_{\rm{off}}}.
\label{eq_definition_of_p}
\end{equation}
That means for $p=1$, there is no foreground emission and for $p=0.5$, the foreground and background emission are equal. Measuring $p$ is difficult and it usually has to be assumed. As a last simplification, we assume that the foreground and background clouds are optically thin and therefore $\tau_{\rm{fg}}$ and $\tau_{\rm{bg}}$ are small  \citep{Gibson2000}. This results in:
\begin{equation}
T_{\rm{on-off}}  = \left( T_{\rm{HISA}} - p \: T_{\rm{off}} - T_{\rm{cont}} \right) \times (1-{\rm e}^{-\tau_{\rm{HISA}}}).
\label{eq_HISA_on_minus_off_solution}
\end{equation}
$T_{\rm{on}}$ and $T_{\rm{off}}$ can be derived from our \ion{H}{i} emission line observations, $T_{\rm{cont}}$ is from our THOR+VGPS 1.4~GHz continuum data (see Sect.~\ref{sect_obs}). With these observable quantities we can estimate the properties of the HISA using Eq.~\ref{eq_HISA_on_minus_off_solution}. Specifically we derive the cloud spin temperature $T_{\rm{HISA}}$ and the optical depth $\tau_{\rm{HISA}}$. We cannot disentangle the spin temperature and the optical depth. 
Fig~\ref{fig_Ts_vs_tau} shows an example of the relation between $\tau_{\rm{HISA}}$ and $T_{\rm{HISA}}$ for the region close to the strong continuum source G34.256+0.146 of $T_{\rm{off}} = 103$\,K, $T_{\rm{on}} = 50$\,K, and $T_{\rm{cont}} = 17$\,K. The different colors represent different values of $p$ from 0.4 to 1. The black vertical line indicates the temperature of the cosmic microwave background radiation \citep{Fixsen1996, fixsen2009, Planck2016} $T=2.7$\,K. Since the L band continuum background emission in the Galactic plane is larger than 0 (bottom panel Fig.~\ref{fig_filament_overview}), the spin temperature must be larger than 2.7~K. The general interpretation of the curves is that a higher optical depth is necessary to produce the assumed absorption feature for higher spin temperatures. This dependency becomes very steep at a certain point, depending on $p$. In Sect.~\ref{sect_discuss} we shall discuss the relations among $\tau$, spin temperature and $p$ in detail. 

\begin{figure}
\centering
 \includegraphics[width=0.5\textwidth]{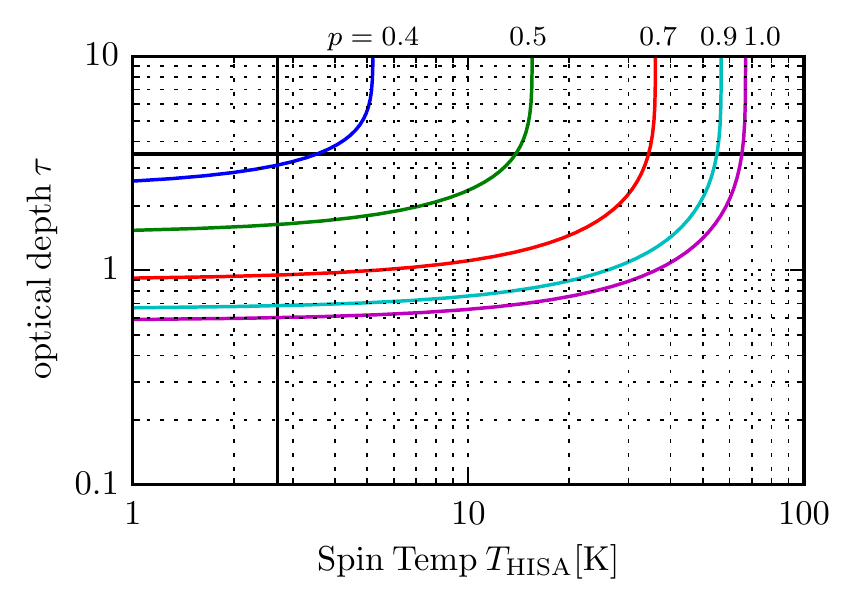}
 \caption{Optical depth as a function of the spin temperature (as introduced in Eq.\,\ref{eq_HISA_on_minus_off_solution}) for the region close to the strong continuum source G34.256+0.146 with $T_{\rm{off}} = 103$\,K, $T_{\rm{on}} = 50$\,K and $T_{\rm{cont}} = 17$\,K. The optical depth of $\tau=3.5$ is indicated with a black horizontal line. The temperature of the CMB is indicated at $T=2.7$\,K with a black vertical line.}
  \label{fig_Ts_vs_tau}
\end{figure}

\subsection{Background estimate to measure $T_{\rm{off}}$}
\label{sect_background_estimate_T_off}
\begin{figure}
\centering
 \includegraphics[width=0.5\textwidth]{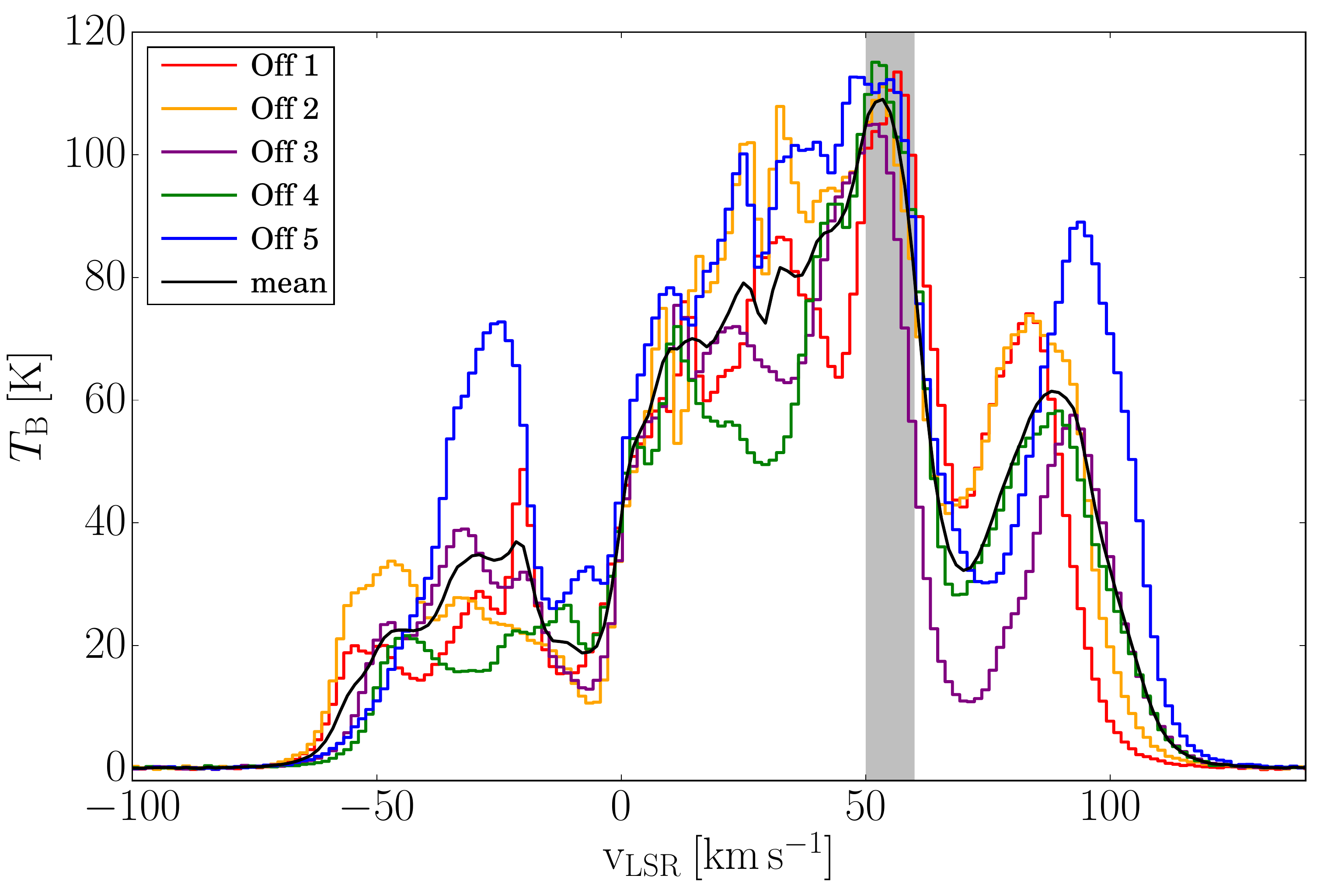}\\
\caption{Selected \ion{H}{i} emission spectra around the GMF38a, which can be used as ``off-positions''. The regions we used for the extraction are shown in Fig.\,\ref{fig_filament_overview}. The black line shows the mean spectrum of all five off-spectra and the gray shaded area indicates the velocity range of the HISA feature ($v_{\rm{LSR}} = 50$ to 60~km~s$^{-1}$).}
  \label{fig_different_off_positions}
\end{figure}
To extract a reliable HISA feature, it is crucial to know the background \ion{H}{i} emission. Different methods can be found in the literature to perform this task. The first one is to use absorption-free \ion{H}{i} emission spectra, located close to the absorption feature \citep[e.g., ][]{Gibson2000}, referred to as ``off-positions''. This method assumes that the \ion{H}{i} background emission stays spatially constant over the absorption feature, which might be true for spatially small HISA features. We tested this method by extracting \ion{H}{i} spectra from five different regions, which are labeled as ``Off 1'' to ``Off 5'' in Fig.\,\ref{fig_filament_overview}. These off-positions were chosen to be regions without significant 1.4\,GHz continuum emission and without $^{13}$CO emission. Furthermore, these regions did not show significant self absorption features at the velocity range of $v_{\rm{LSR}} = 50-60$\,km\,s$^{-1}$. The corresponding spectra are presented in Fig.\,\ref{fig_different_off_positions}. These spectra reveal large variations, which makes it difficult to use them as a common off-position.

\begin{figure}
\centering
 \includegraphics[width=0.24\textwidth]{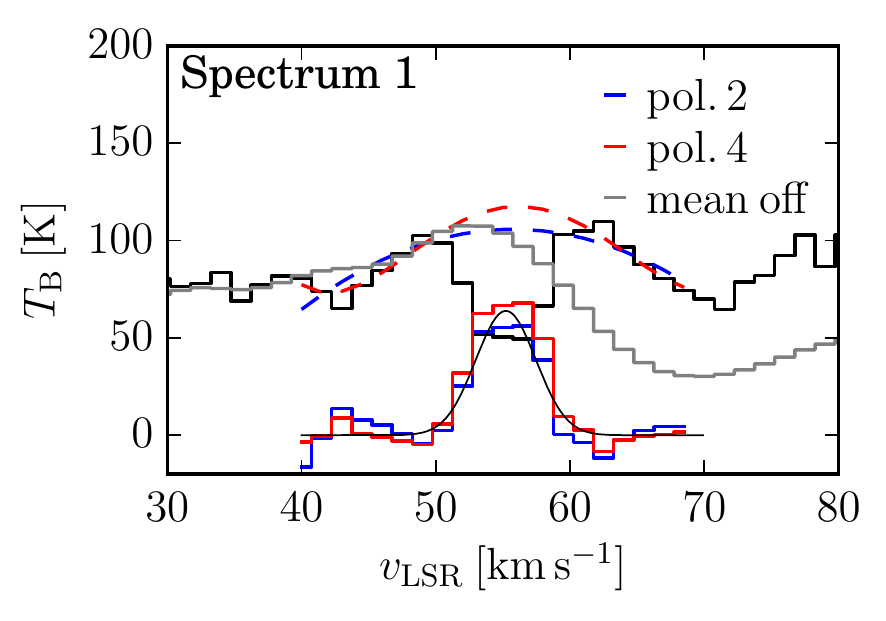}    
 \includegraphics[width=0.24\textwidth]{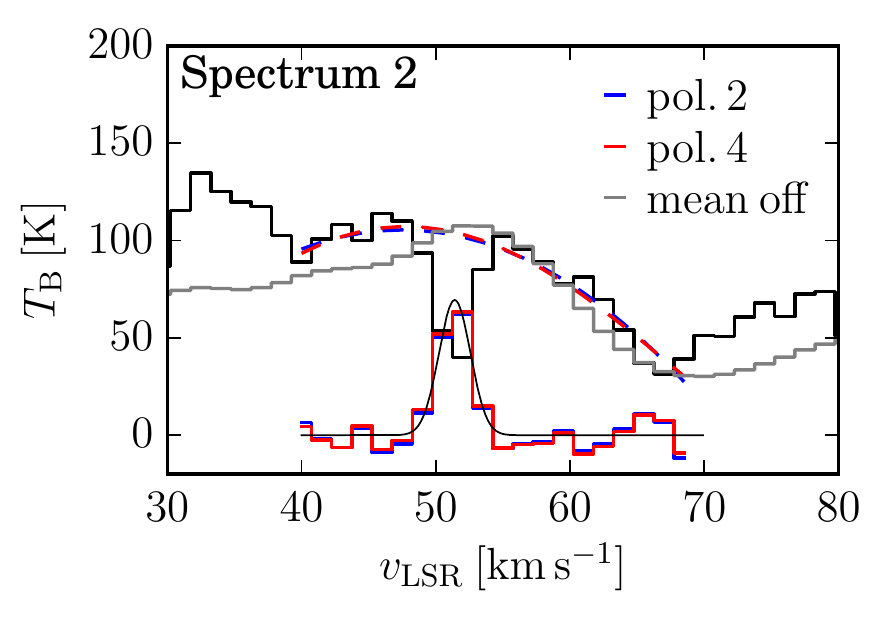}
 \includegraphics[width=0.24\textwidth]{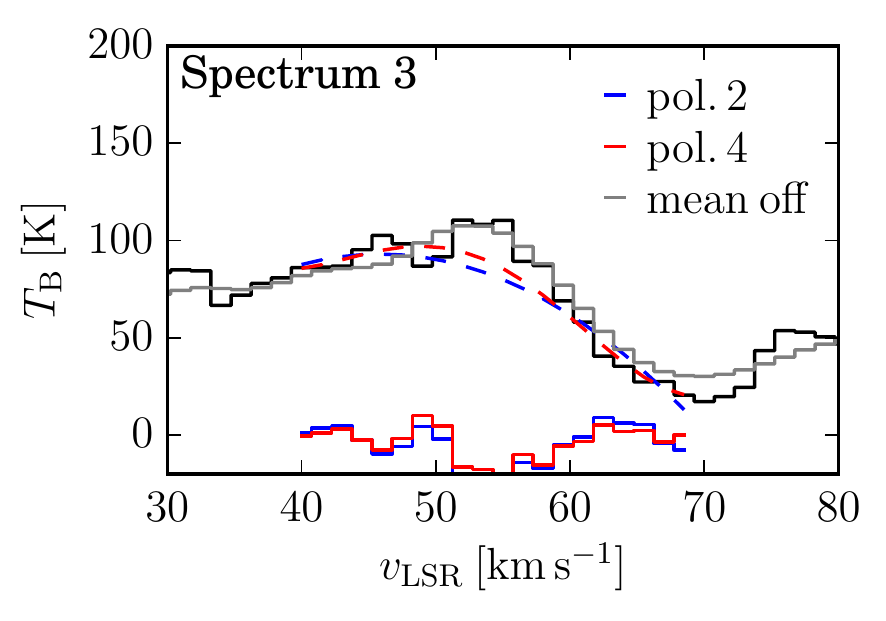}
 \includegraphics[width=0.24\textwidth]{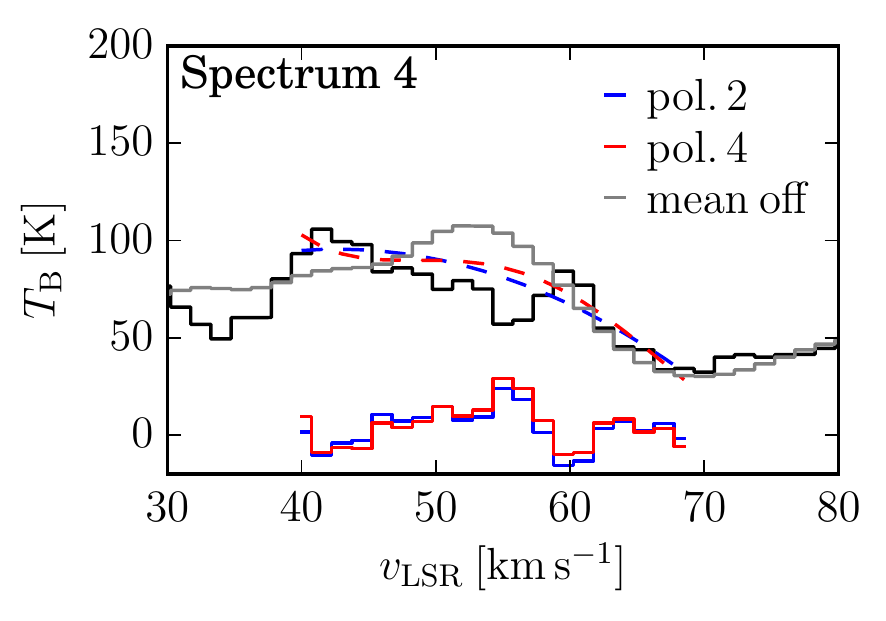}
 \includegraphics[width=0.24\textwidth]{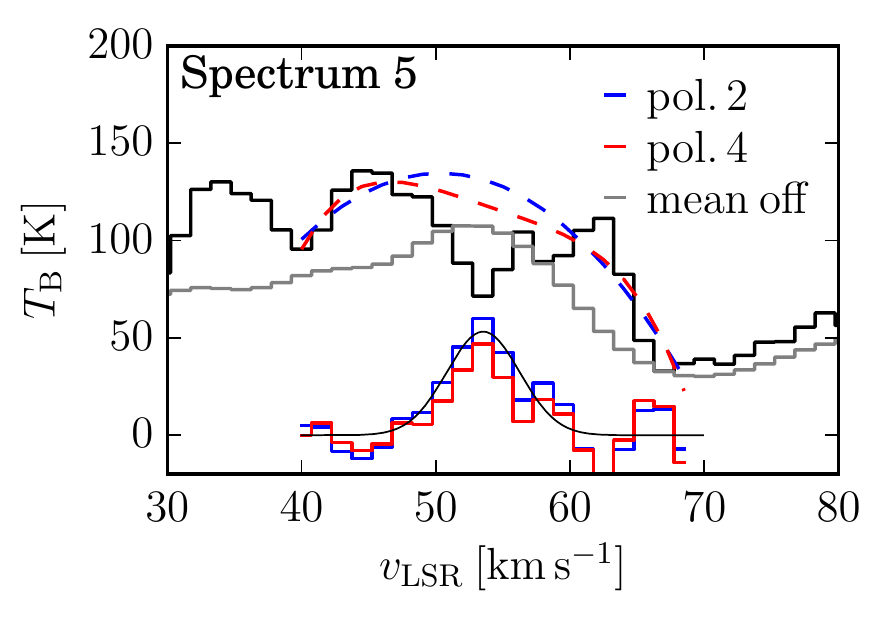}
\caption{Selected \ion{H}{i} emission spectra corresponding the positions indicated in Fig.~\ref{fig_filament_overview}. The black lines represent the \ion{H}{i} spectra, the colors correspond to the HISA. The gray spectra indicate the mean of the five off-positions presented in Fig.\,\ref{fig_different_off_positions}. The blue and red dashed lines represent the fits of the background to the HISA spectrum for a polynomial of second and fourth order, respectively, using the velocity range of $v_{\rm{LSR}} = 40-50$ and $60-70$\,km\,s$^{-1}$ for the the baseline of the fit. The blue and red solid lines show the difference between fitted \ion{H}{i} spectra and the measured \ion{H}{i} spectra ($T_{\rm{on-off}}$ in Eq.\,\ref{eq_HISA_on_minus_off_solution}) for a polynomial of second and fourth order, respectively. We fitted Spectra 1, 2, and 5 using a Gaussian function, shown by the black solid curve on top of the HISA spectra.}
  \label{fig_different_HISA_spectra}
\end{figure}

The second method utilizes a fit to the absorption free channels of the \ion{H}{i} spectra to get $T_{\rm{off}}$. This method is applied frequently, using different functions to fit the \ion{H}{i} emission, for example, linear fits \citep[e.g., ][]{Minn1981, Montgomery1995, McClure-Griffiths2006} or polynomials with different order \citep[e.g.][]{Myers1978, Bowers1980, Shuter1987, Kavars2003, Li2003}. Fig.\,\ref{fig_different_HISA_spectra} presents five different spectra from different positions indicated in Fig.\,\ref{fig_filament_overview}. We used second and fourth order polynomials to fit the spectra for the velocity range around the HISA ($v_{\rm{LSR}} = 40-50$ and $60-70$\,km\,s$^{-1}$). A polynomial function of the third order gave very similar results as the polynomial of the second order, thus for clarity we do not show it here.

It is difficult to estimate which function is more suitable to fit the \ion{H}{i} spectra. For regions without absorption, we expect that the fitted spectra represents the actual spectra. Spectrum 3 in Fig.\,\ref{fig_different_HISA_spectra} shows such a region and both functions represent the \ion{H}{i} spectra well. Spectra 2 and 4 in Fig.\,\ref{fig_different_HISA_spectra} represents \ion{H}{i} absorption features and the difference between the second and forth order polynomial is small. In contrast to this, Spectra 1 and 5 in Fig.\,\ref{fig_different_HISA_spectra} reveal a large difference between the fit functions. The fourth order polynomial fit is much higher ($\sim$50\,K) than the second order polynomial for Spectrum 1, but much lower than the second order polynomial for Spectrum 5. It is not obvious which function describes the \ion{H}{i} spectra more accurately. However, the fourth order polynomial might overestimate the actual spectra as steep slopes within the fitted velocity range would result in high values for the fitted spectra. In contrast to this, the second order polynomial might underestimate the \ion{H}{i} emission for this spectra. Hence, the fourth order polynomial might be an upper limit and the second order polynomial might be a lower limit. We will use both functions in the following analysis to estimate the uncertainty of $T_{\rm{off}}$ and to extract HISA.

Another method was to utilize the second derivative representation of the spectrum as described in \citet{Krco2008}. \citet{Krco2008} demonstrated that HINSA feature would become dominant in the second derivative representation. We also tested this method. For narrow HISA spectra (such as Spectrum 2 in Fig.~\ref{fig_different_HISA_spectra}) the second derivative technique can recover the HISA spectra relatively well. However, for broad spectra (such as Spectrum 1 and Spectrum 5), the HISA spectra were filtered out by the method. Therefore, we do not use this method in our analysis. 

\citet{McCutcheon1978}, \citet{Winnberg1980} and \citet{Andersson1991} used one or several Gaussian profiles to fit the spectra and to derive the off spectra. However, as pointed out by \citet{McCutcheon1978}, this method only works if the absorption feature is very narrow and the shape of the total spectrum is simple and can be represented by a few Gaussians. On the other hand \citet{Denes2018} used a machine learning method, the Autonomous Gaussian Decomposition algorithm (AGD; GAUSSPY) developed by \citet{Lindner2015}, to decompose the emission spectra while masking out the HISA features to derive the off spectra. However, they only need to deal with 47 spectra and it is not clear how well this method would work for our region with $\sim2$ million spectra. It is definitely worthwhile to test the machine learning method in the future, but it is beyond the scope of this paper.

The mean spectrum of the five off-positions shown in Fig.\,\ref{fig_different_off_positions} is shown in gray in Fig.\,\ref{fig_different_HISA_spectra} as well. While the mean off-position represents the \ion{H}{i} spectra of $T_{\rm{off}}$ well in some cases (e.g., Spectrum 2 in Fig.\,\ref{fig_different_HISA_spectra}), but in general it does not (e.g., Spectrum 1 or 5 in Fig.\,\ref{fig_different_HISA_spectra}). There are apparent variations in the \ion{H}{i} spectrum at velocities outside of the HISA feature, such the assumption of a uniform \ion{H}{i} emission background does not seem to be adequate. The mean spectrum method was also discussed by \citet{Myers1978} and \citet{McCutcheon1978}, and they concluded that it is not suitable for HISA studies for the same reason. Hence, we will use the polynomial fit method to extract the HISA feature rather than using a mean off-spectrum.

The noise of the extracted HISA spectra measured at the velocity ranges in $v_{\rm{LSR}} = 40-50$ and $60-70$\,km\,s$^{-1}$ is $\sim8$~K per 1.5~km~s$^{-1}$ for the second order polynomial fit, $\sim5$~K for the forth order fit. Since a forth order polynomial function can always fit the small bumps in the spectra better than the second order polynomial function, it is no surprise the HISA spectra extracted with the forth order fit have smaller noise.

\subsection{HISA extraction}
\label{sect_HISA_extraction}
\begin{figure*}
\centering
 \includegraphics[width=0.8\textwidth]{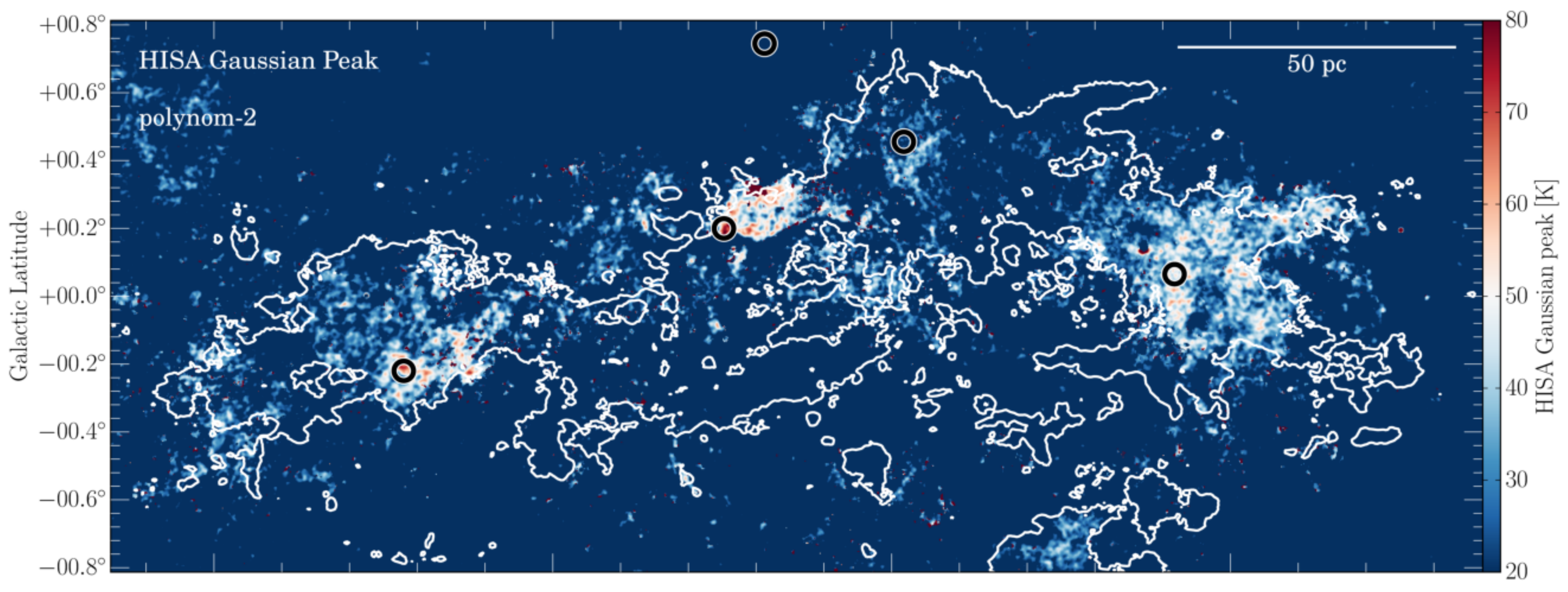}\\
 \includegraphics[width=0.8\textwidth]{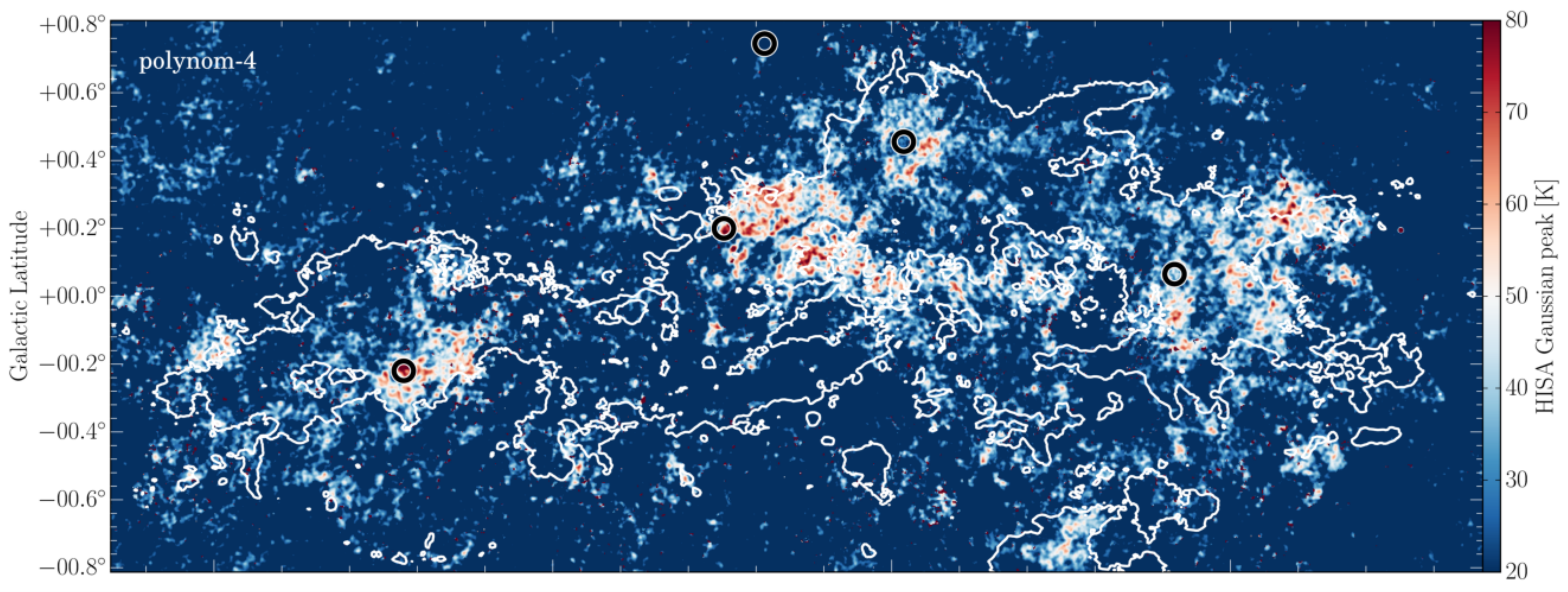}\\
 \includegraphics[width=0.8\textwidth]{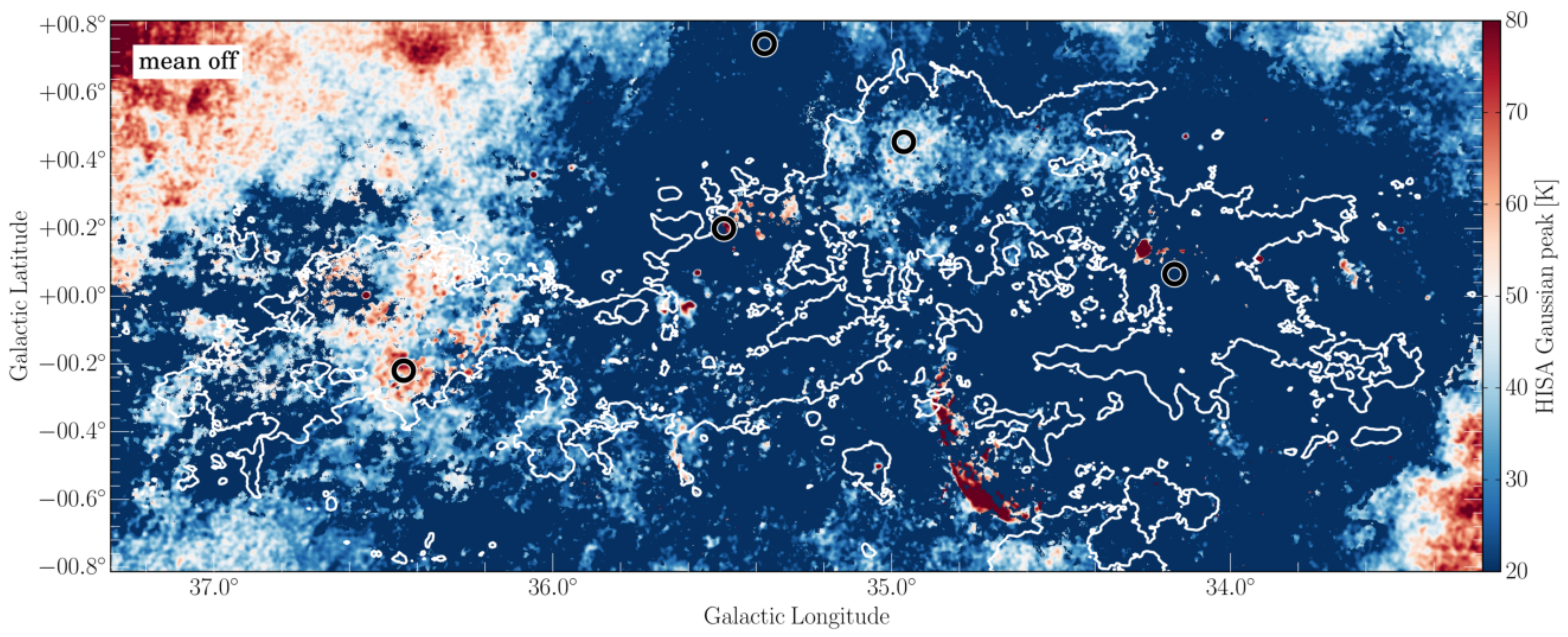}
\caption{Peak value of a Gaussian fit for the absorption depth $T_{\rm{off}}-T_{\rm{on}}$ for different methods of estimating $T_{\rm{off}}$. The top and middle panel show the method using a polynomial fits with second and forth order, respectively. The bottom panel represents the method using a spatially averaged off spectrum shown in Fig.\,\ref{fig_different_off_positions}. The white contours represent the integrated $^{13}$CO emission at levels of 5\,K\,km\,s$^{-1}$ for reference. The circles from left to right marked the same positions as in Fig.~\ref{fig_filament_overview} where Spectra~1 to 5 shown in Fig.\,\ref{fig_different_HISA_spectra} are extracted, respectively.}
  \label{fig_HISA_peak_emission_overlay}
\end{figure*}

With the methods described in Sect.\,\ref{sect_background_estimate_T_off} we can estimate $T_{\rm{off}}$ in Eq.\,\ref{eq_HISA_on_minus_off_solution}. Using this information, we can measure the depth of the absorption feature ($T_{\rm{off}}-T_{\rm{on}}$). To analyze the absorption features, we use a Gaussian curve to fit them. This allows us to study the exact depth of the absorption features and their kinematics. Fits that result in a peak intensity higher than 25~K ($\sim3$ and 5 times the noise level of the HISA spectra extracted with second and forth oder fit, respectively), 1.5~km~s$^{-1}$$<$ full width half maximum (FWHM) linewidth ($=$channel width) $<$20~km~s$^{-1}$ are considered as good fit. The peak values of the fitted Gaussian curves are shown in Fig.\,\ref{fig_HISA_peak_emission_overlay} for different ways of estimating $T_{\rm{off}}$. The absorption depth of the HISA shows values between $\sim$30 and $\sim$80\,K. 

As discussed in Sect.~\ref{sect_background_estimate_T_off}, different methods to estimate $T_{\rm{off}}$ resulted in differences for the peak value. Those based on the spatially averaged off-positions (bottom panel in Fig.\,\ref{fig_HISA_peak_emission_overlay}) show unrealistically large values at the edges as well as in the presence of continuum sources. In the first case, as the Galactic \ion{H}{i} emission drops at the edge, this method falsely identifies the edge regions as absorption features. In the second case, the method picks up the strong \ion{H}{i} absorption towards the bright continuum emission in the background, as the structure of the continuum emission is clearly visible (bottom panel in Fig.\,\ref{fig_HISA_peak_emission_overlay}).

Fitting $T_{\rm off}$ of each \ion{H}{i} spectra with a polynomial function circumvents these problems. Hence, the polynomial fitting method is more appropriate for our analysis and in the following we focus on the determination of $T_{\rm off}$ from HISA-free channels in the $T_{\rm on}$ spectrum. 

We found large differences in $T_{\rm off}$ estimates made with polynomials of second order forth order, in particular towards regions around $l=35\degr$ and $l=33.8\degr$(Fig.\,\ref{fig_HISA_peak_emission_overlay}). Using a forth order polynomial for the background estimate, we found significantly more absorption due to a possible overestimate of the background emission (Sect.~\ref{sect_background_estimate_T_off}). Other regions are not affected significantly by the choice of the fit function, e.g., the regions around spectrum 1 or 2. The uncertainties of the HISA properties introduced by different polynomial function fittings will be discussed in Sect.\,\ref{sect_Uncertainties_in_the_HISA_description}.

\section{Results}

\subsection{Kinematics}
\label{sect_results_kinematics}
In this section, we discuss the kinematic properties of the HISA features. In contrast to the absorption depth, the peak velocity is not significantly affected by the choice of the fit function. Therefore we present here only the velocity structure using the second order polynomial for the determination of $T_{\rm{off}}$. The velocity structure revealed by the fourth order polynomial is similar. 

To compare the kinematics of the HISA feature with those of the $^{13}$CO data, we resampled both the $^{13}$CO and \ion{H}{i} data to the same spatial and velocity resolution (pixel size of 22\arcsec\, and spectral channel width of 1.5~km~s$^{-1}$) and applied Gaussian fitting to the data sets pixel by pixel to determine the peak velocity and FWHM linewidth. Due to limited sensitivity and spectral resolution of our \ion{H}{i} data, we cannot disentangle multiple velocity components. Therefore, we choose to fit both the $^{13}$CO and HISA data with a single Gaussian component for simplicity and consistency. The Gaussian fitting to the HISA spectra is described in Sect.~\ref{sect_HISA_extraction}. For the $^{13}$CO data, fits that result in a peak intensity higher than 2~K ($\sim$10 times noise level of the $^{13}$CO datacube at 1.5~km~s$^{-1}$ channel width), 1.5~km~s$^{-1}<$ FWHM linewidth $<$20~km~s$^{-1}$ are considered as good fit for the $^{13}$CO data. 

The peak velocity maps are presented in Fig.\,\ref{fig_overview_velocity_map}. The $^{13}$CO peak velocity shows that the majority of the filament is at $\sim$54--58~km~s$^{-1}$. The western part of the $^{13}$CO filament around $l=34\degr$ is slightly red-shifted compared to rest of the filament, and has a velocity of $\sim$57-58\,km\,s$^{-1}$. For this region, the peak velocities revealed by the HISA feature are at $\sim$54-55\,km\,s$^{-1}$, which is about 3-4\,km\,s$^{-1}$ lower than the $^{13}$CO velocity. This can also be seen in the right panel of Fig.\,\ref{fig_peak_velocity_histogram}, where we present a histogram of the HISA and $^{13}$CO peak velocities for the eastern and western side of the GMF, respectively. In contrast to this, the eastern side of the filament around $l=36.5\degr$ shows a close correlation of the peak velocities as shown in the left panel of Fig.\,\ref{fig_peak_velocity_histogram}. We will discuss this effect in Sect.\,\ref{sect_discussion_kinematics}. The western and eastern region we refer to here are both under the Galactic coordinate framework, same as here after.

\begin{figure}
\centering
 \includegraphics[width=0.23\textwidth]{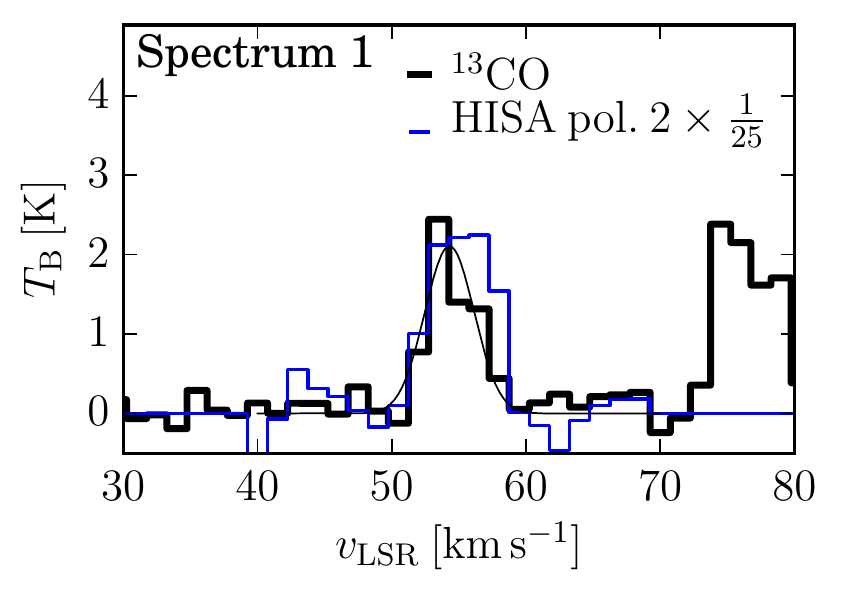}    
 \includegraphics[width=0.25\textwidth]{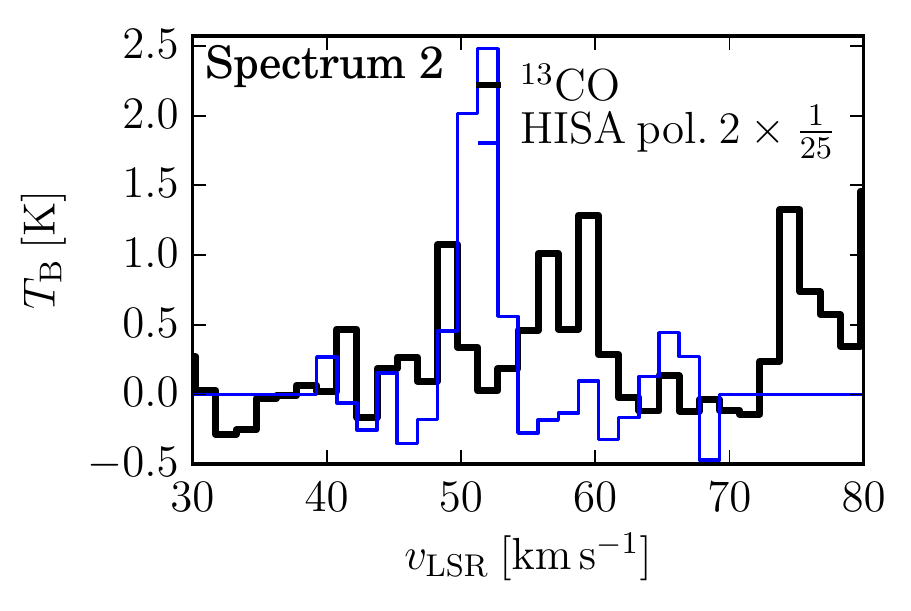}
 \includegraphics[width=0.25\textwidth]{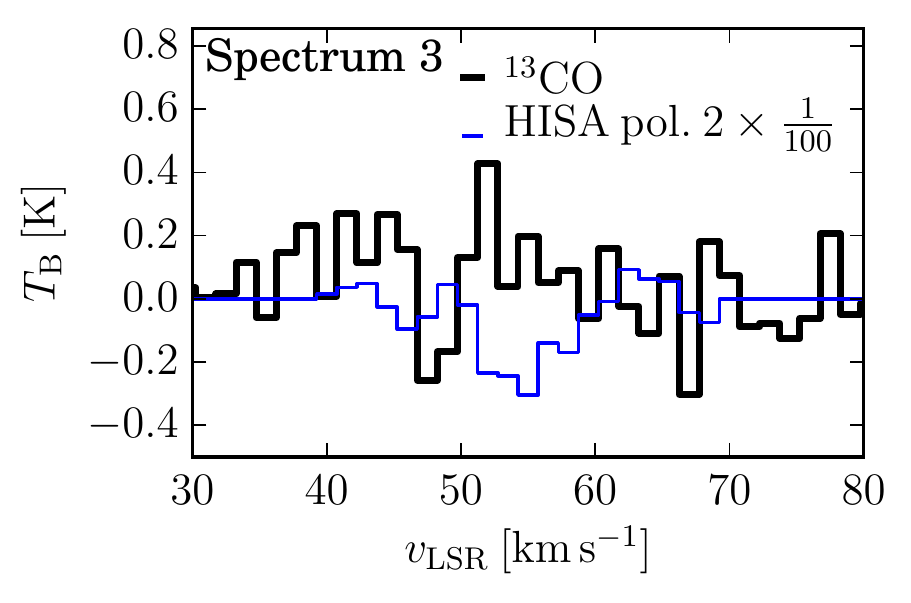}
 \includegraphics[width=0.23\textwidth]{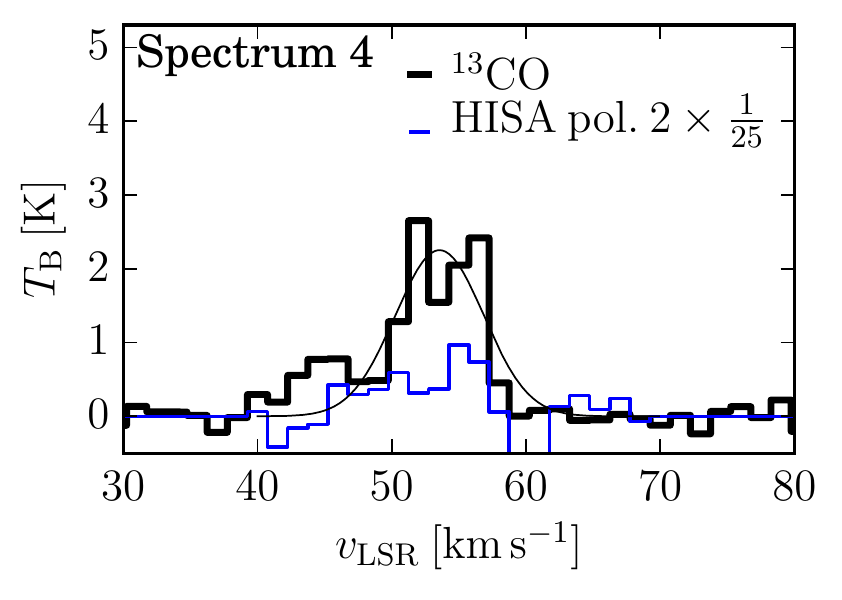}
 \includegraphics[width=0.23\textwidth]{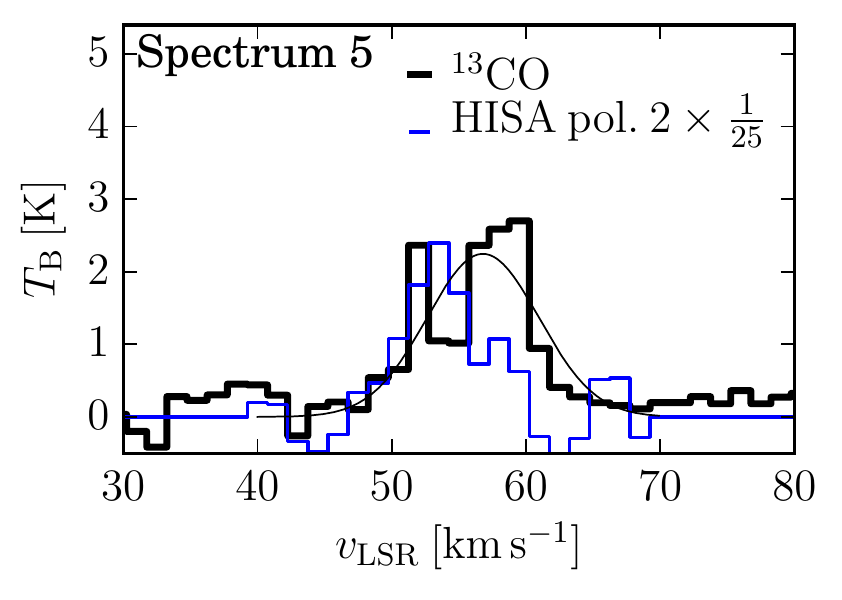}
\caption{Examples of $^{13}$CO spectra extracted from the same positions as in Fig.\ref{fig_different_HISA_spectra} and the Gaussian fitting results if available. The $^{13}$CO spectra are shown in thick black lines. The HISA spectra from the same locations are scaled to 1/25 of its original intensity and shown in thin blue lines. The positions corresponding to each spectrum are shown in Fig.~\ref{fig_filament_overview}. For Spectra~1, 4, and 5, we can fit the spectrum with a Gaussian component to the $^{13}$CO spectrum successfully, we show the Gaussian fitting result with a black solid curve in each panel. For Spectra~2 and 3, the Gaussian fitting results did not meet the criteria we described in Sect.~\ref{sect_results_kinematics} and were ignored.}
  \label{fig_CO_spectra}
\end{figure}

\begin{figure*}
\centering
 \includegraphics[width=\textwidth]{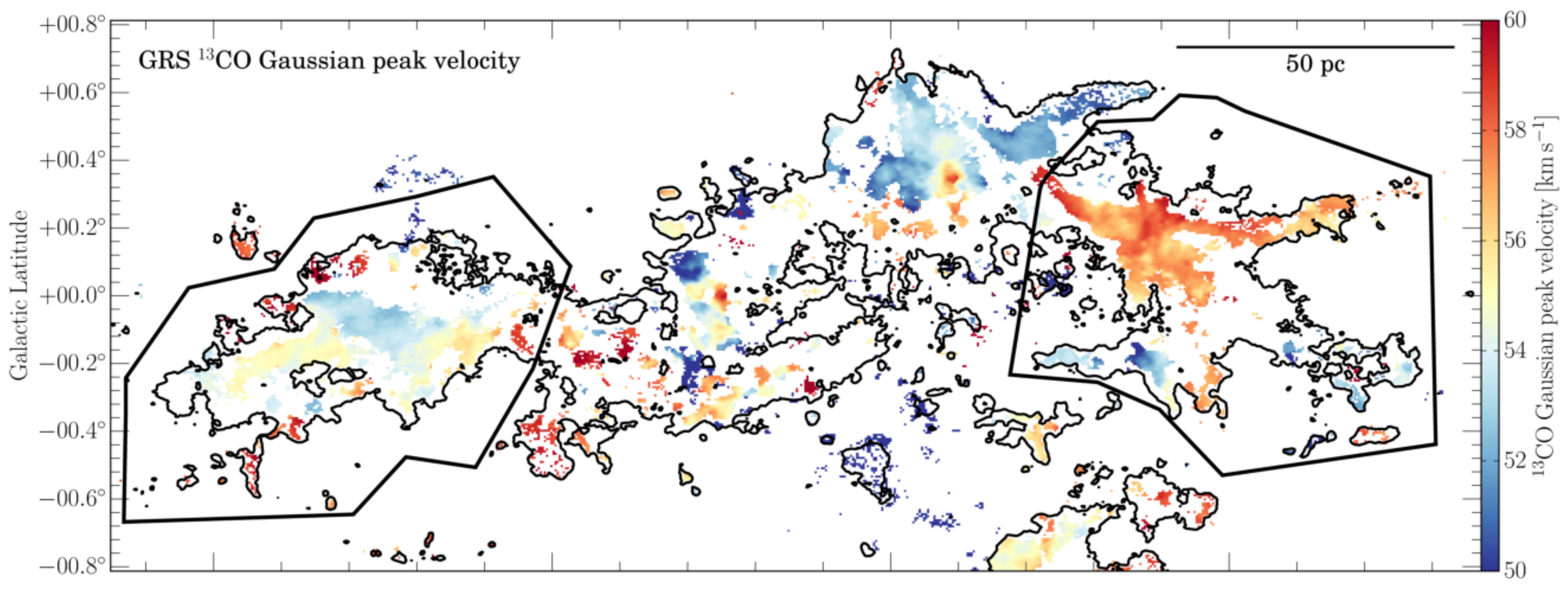}\\
 \includegraphics[width=\textwidth]{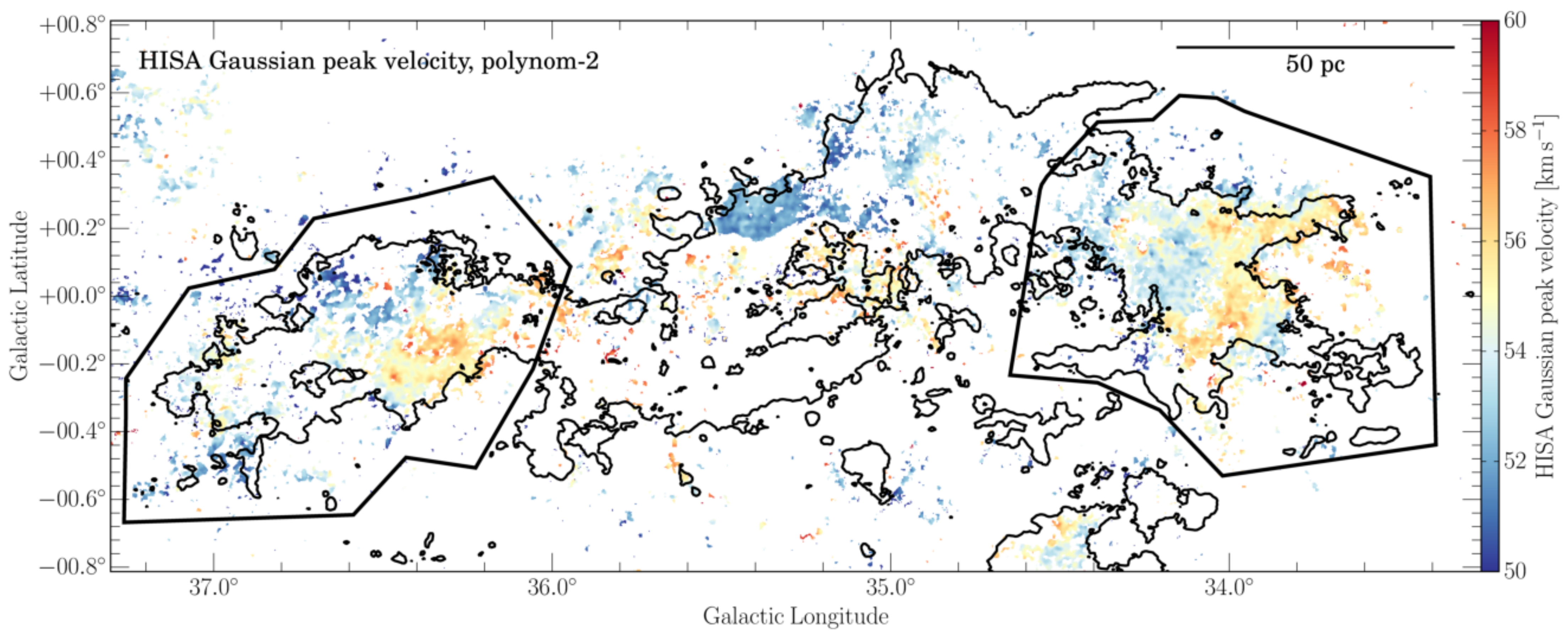}
\caption{Peak velocity of the Gaussian fit of the $^{13}$CO data (top-panel) and the HISA data (bottom-panel), respectively. The HISA are extracted by using a second order polynomial to estimate the background emission. The black contours indicate the integrated $^{13}$CO emission at levels of 5\,K\,km\,s$^{-1}$. The polygons mark the eastern (around $l=36.5\degr$) and western (around $l=34\degr$) subregions, for which we present the histograms of the peak velocities in Fig.~\ref{fig_peak_velocity_histogram}.}
  \label{fig_overview_velocity_map}
\end{figure*}

\begin{figure*}
\centering
 \includegraphics[width=0.33\textwidth]{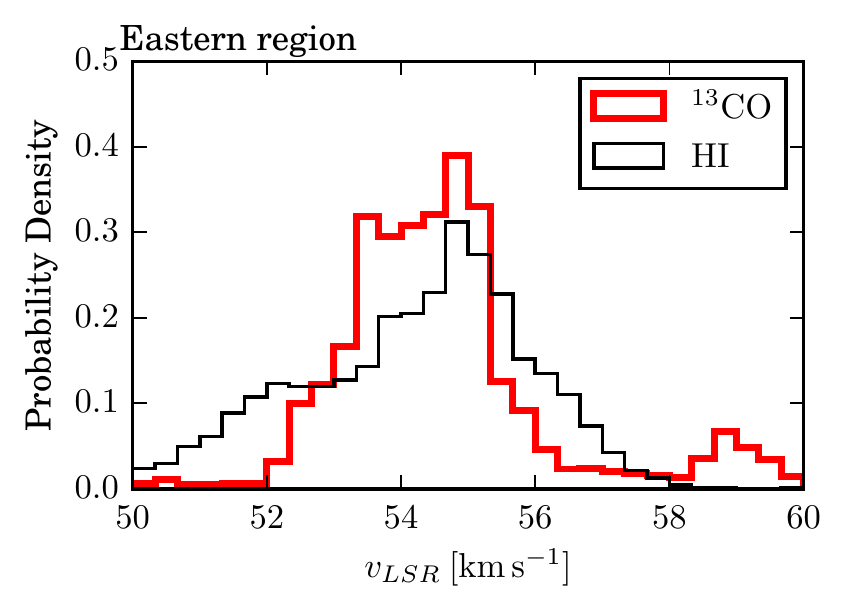}
 \includegraphics[width=0.33\textwidth]{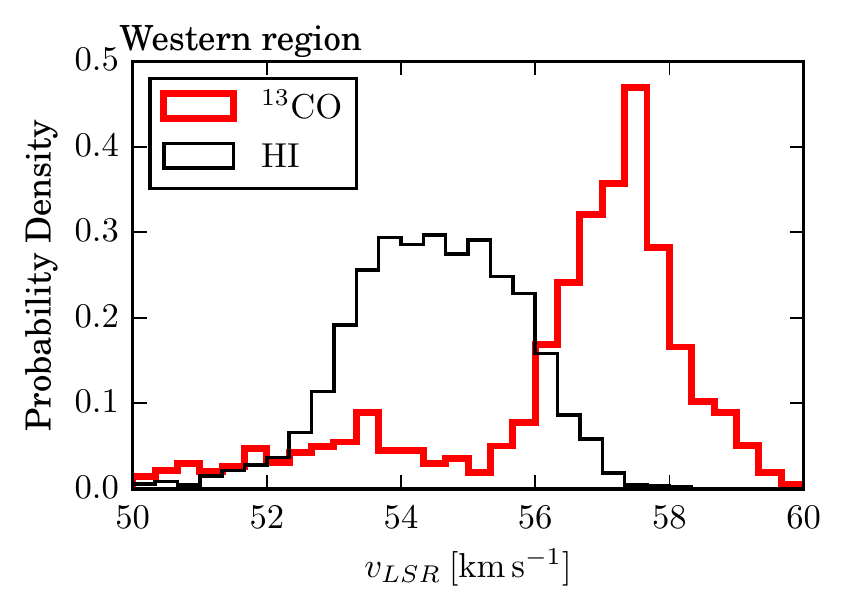}
 \includegraphics[width=0.33\textwidth]{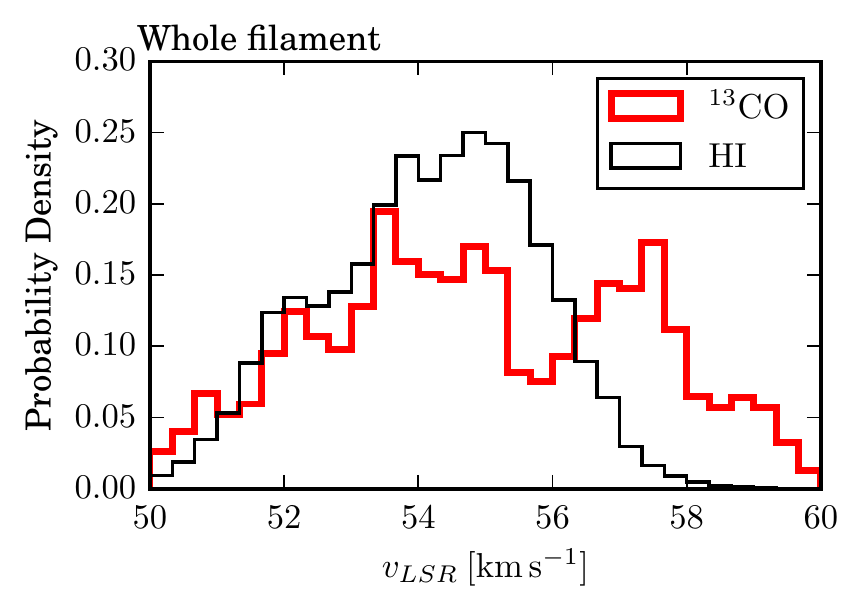} 
\caption{Histograms of the HISA and $^{13}$CO peak velocity, shown in black (thin) and red (thick) respectively. The left and middle panels show all extracted pixels within the marked eastern (around $l=36.5\degr$) and western (around $l=34\degr$) polygon in Fig.\,\ref{fig_overview_velocity_map}. The right panel shows all extracted pixels for the whole filament (red polygon in Fig.\,\ref{fig_overview_velocity_map}).}
  \label{fig_peak_velocity_histogram}
\end{figure*}

\begin{figure*}
\centering
 \includegraphics[width=\textwidth]{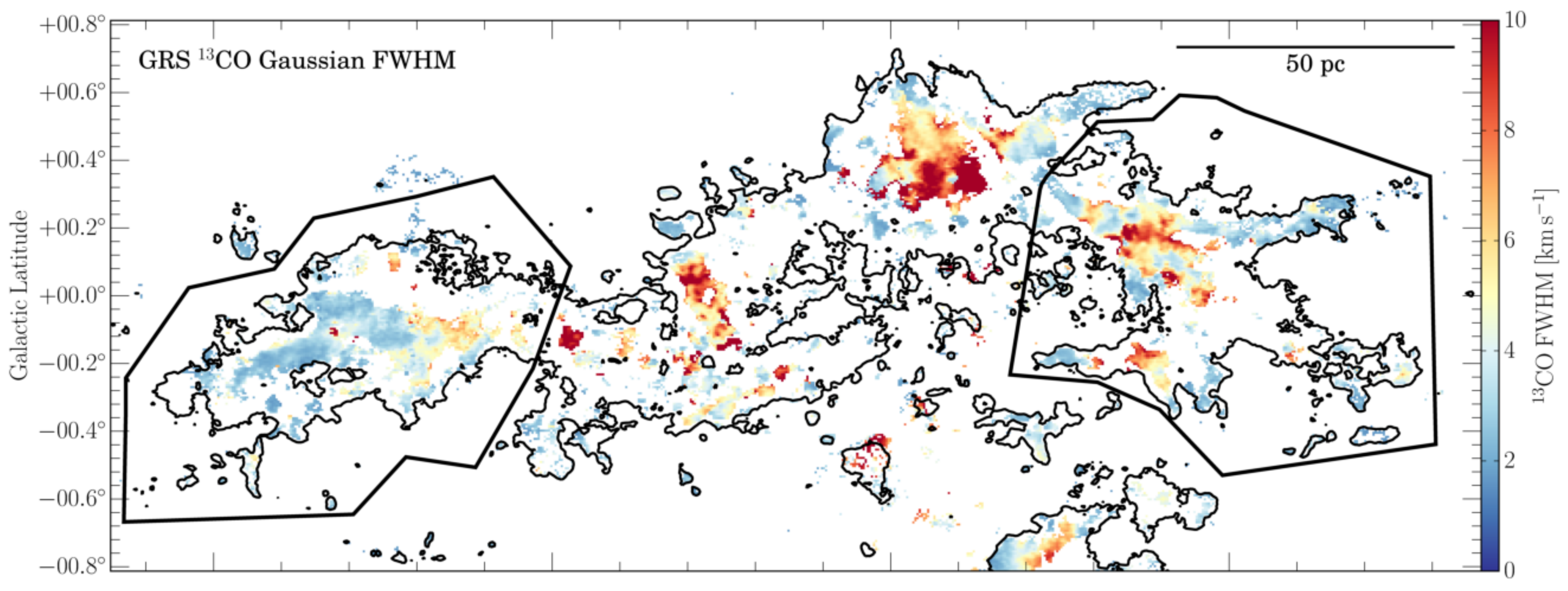}\\
 \includegraphics[width=\textwidth]{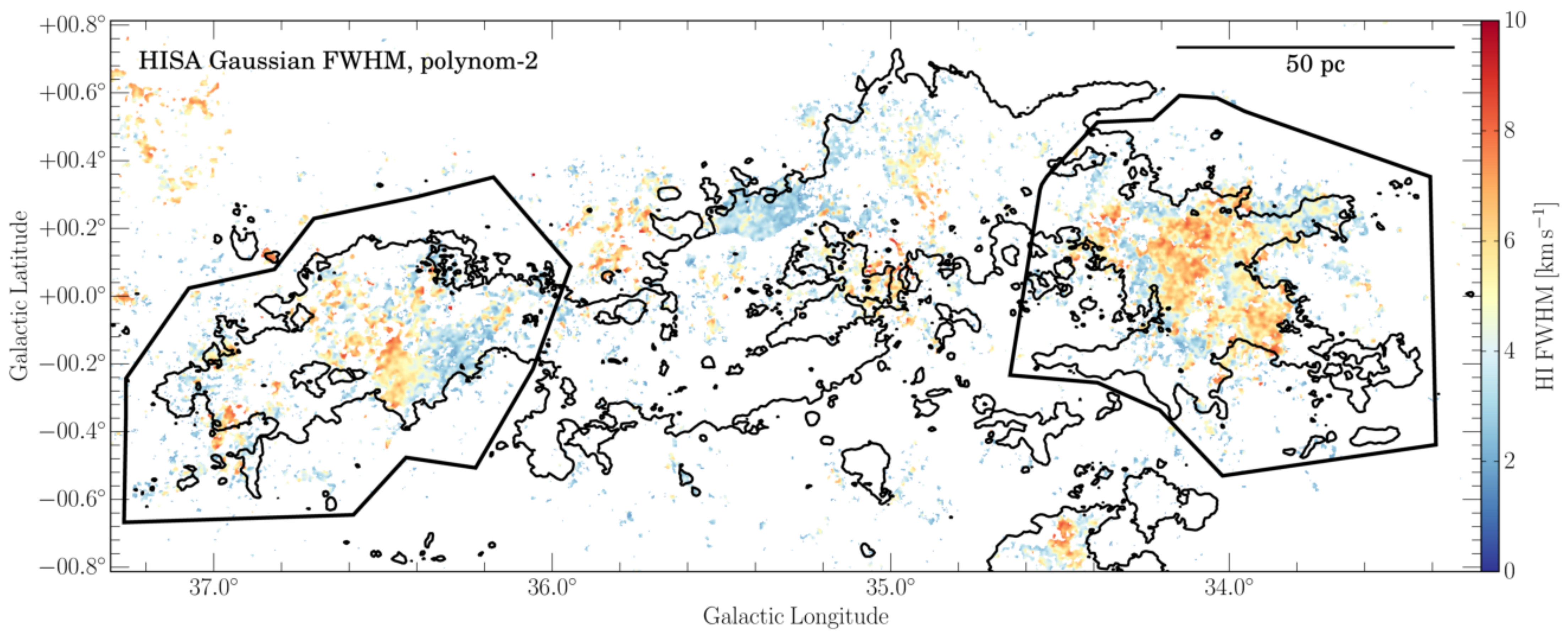}
\caption{ Full width at half maximum (FWHM) of the $^{13}$CO emission (top-panel) and the HISA feature (bottom panel), respectively. The linewidth is determined by using a single Gaussian fit. The contours indicate the integrated $^{13}$CO emission at levels of 5\,K\,km\,s$^{-1}$ for reference. The polygons mark the eastern (around $l=36.5\degr$) and western (around $l=34\degr$) subregions, for which we present the histograms of the FWHM linewidths in Fig.~\ref{fig_fwhm_histogram}.}
  \label{fig_overview_FWHM_map}
\end{figure*}

\begin{figure*}
\centering
 \includegraphics[width=0.33\textwidth]{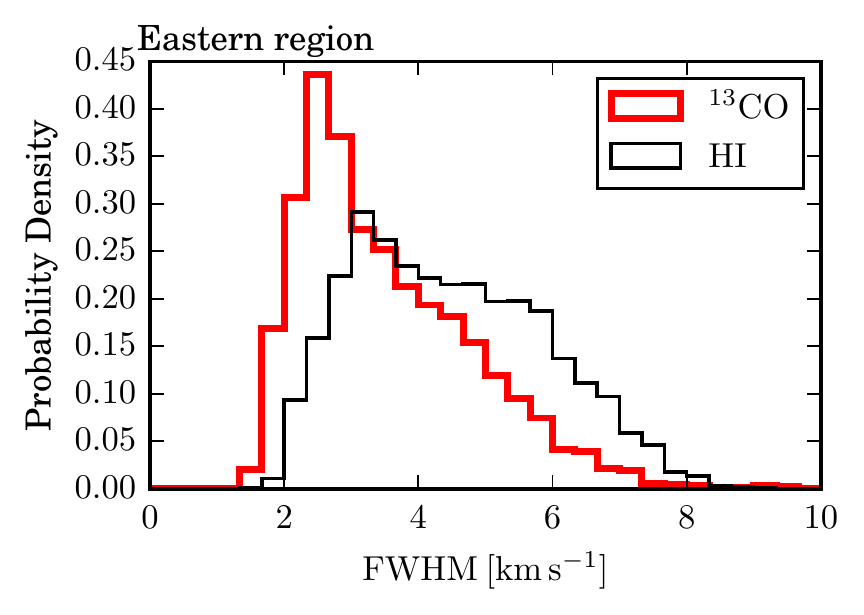}
 \includegraphics[width=0.33\textwidth]{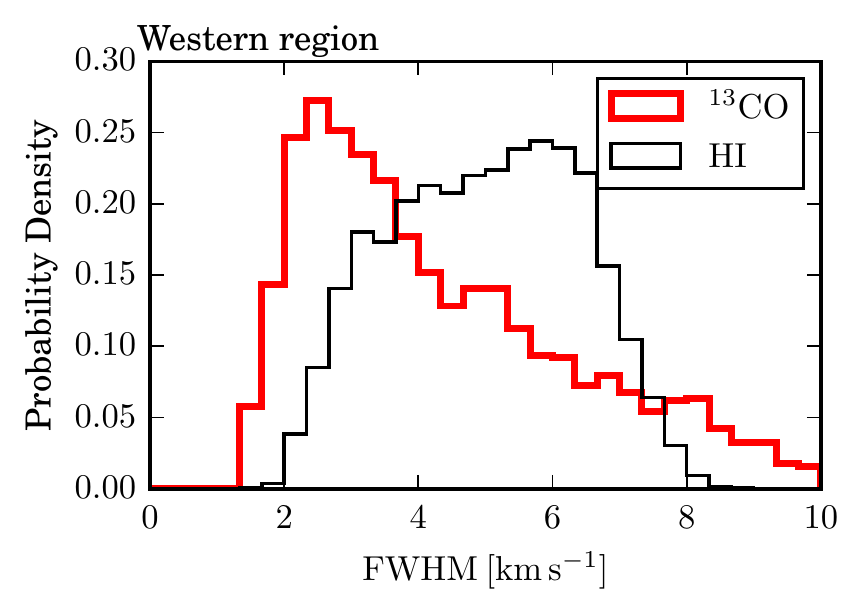}
 \includegraphics[width=0.33\textwidth]{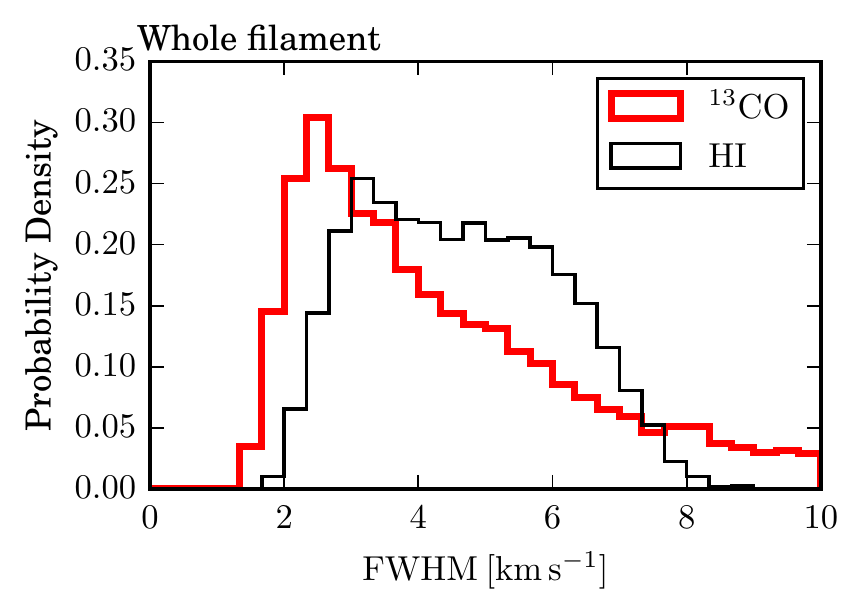}
\caption{Histograms of the FWHM linewidth of the HISA and $^{13}$CO emission in black (thin) and red (thick), respectively. The left and middle panels show all extracted pixels within the marked eastern (around $l=36.5\degr$) and western (around $l=34\degr$) polygon in Fig.\,\ref{fig_overview_velocity_map}. The right panel shows all extracted pixels for the whole filament (red polygon in Fig.\,\ref{fig_overview_velocity_map}).}
  \label{fig_fwhm_histogram}
\end{figure*}

The linewidths of both $^{13}$CO and HISA are shown in Fig.\,\ref{fig_overview_FWHM_map}. The linewidth for the $^{13}$CO emission shows extremely high values of more than 10\,km\,s$^{-1}$ for the central region of the filament around $l=35\degr$. However, these values have to be treated cautiously as the $^{13}$CO emission exhibits multiple lines in this region and we only use a single Gaussian function to fit them. On the eastern side of the filament around $l = 36.5\degr$ we find mostly single components for the $^{13}$CO emission and the linewidth is $\Delta v \sim 2$ to 4~km~s$^{-1}$. The linewidth of the HISA feature shows values $\Delta v \sim 3 - 6$\,km\,s$^{-1}$ for the whole filament. This can also be seen in the left panel of Fig.\,\ref{fig_fwhm_histogram}, where a histogram of the linewidths is shown. The linewidth distribution of the $^{13}$CO emission is systematically higher in the western region of the filament, whereas the linewidth of the HISA feature is similar to the eastern region. This result has to be treated cautiously as we see multiple components for the $^{13}$CO line within the western region, which increases the linewidth. 

\begin{figure}
\centering
 \includegraphics[width=0.45\textwidth]{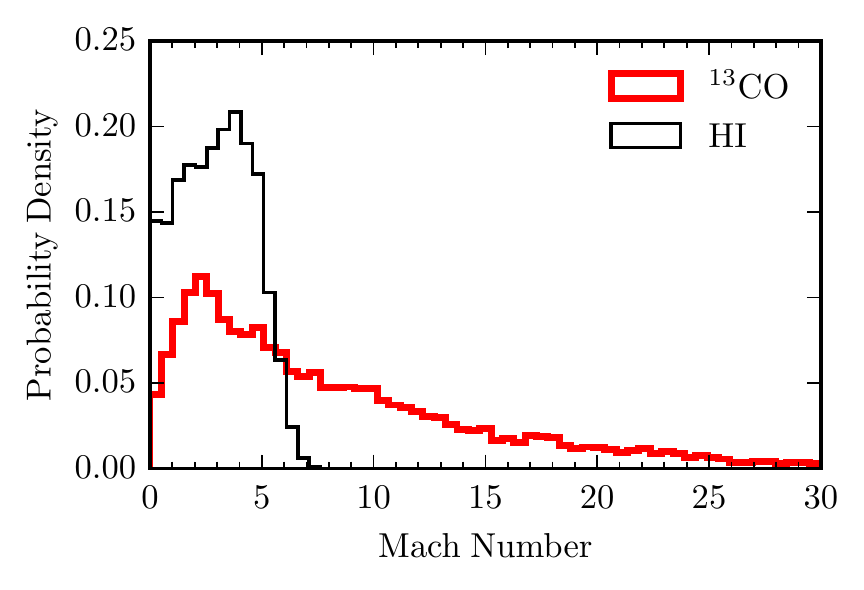}
 \caption{Histograms of the Mach number of the HISA and $^{13}$CO emission, shown in black and red, respectively.}
  \label{fig_mach_histogram}
\end{figure}

To estimate the contribution of the non-thermal component in the HISA features and the molecular emission, we assume the relation $\sigma_{\rm nth}=\sqrt{\sigma_{\rm obs}^2-\sigma_{\rm th}^2-\sigma_{\rm res}^2}$, where $\sigma_{\rm obs}$ is the measured velocity dispersion, $\sigma_{\rm th}$ is the radial component of the thermal velocity dispersion, and $\sigma_{\rm res}$ is the velocity dispersion introduced by the channel width of our data (1.5~km~s$^{-1}$). Assuming a Gaussian line profile with the FWHM linewidth $\Delta v$ obtained from the aforementioned fittings we get $\sigma_{\rm obs}=\Delta v/\sqrt{8 \rm{ln}2}$, and $\sigma_{\rm res}=1.5/\sqrt{8 \rm{ln}2}$~km~s$^{-1}$. Assuming a Maxwell-Boltzmann velocity distribution, $\sigma_{\rm th} = \sqrt{k_{\rm B} T_{\rm k}/(\mu m_{\rm H})}$, where $k_{\rm B}$ is the Boltzmann constant, $\mu$ is the molecular weight, $m_{\rm H}$ is the mass of the hydrogen atom, and $T_{\rm k}$ is the kinetic temperature. For HISA, we assume $T_{\rm k}=T_{\rm HISA}=40$~K. The peak excitation temperature of $^{13}$CO we derived for the filament is $\sim25$~K (see Sect.~\ref{sect_H2_column_density_estimate}) which agrees to what \citet{Roman-Duval2010} found for the GRS molecular clouds. If we assume that the brightest $^{13}$CO emission is coming from regions where the line is optically thick and thermalized, then this excitation temperature will be comparable to the actual gas kinetic temperature, which must therefore have a value close to 20~K. Furthermore, simulations of molecular clouds in a variety of different radiation fields show that the CO mass-weighted temperature of the gas is typically in the range 10--30~K, with very little dependence on the local environment \citep{Penaloza2018}. Therefore, we assume a uniform $T_{\rm k}$ of 20~K for $^{13}$CO.  Assuming spatial isotropy, the Mach number is estimated to be $\sqrt{3} \sigma_{\rm nth}/c_{\rm s}$, where $c_{\rm s}$ is the sound speed estimated using a mean molecular weight $\mu=2.34$ for the molecular cloud and $\mu=1.27$ for the \ion{H}{i} cloud \citep{Allen1973, Cox2000}. The distribution of the Mach number for HISA and $^{13}$CO across the whole filament (Fig.~\ref{fig_mach_histogram}) shows that the $^{13}$CO emission is dominated by supersonic motions, whereas HISA features have a much smaller Mach number in general with a significant fraction of the HISA features, i.e., a significant fraction of the CNM being at subsonic and transonic velocities. If the HISA and $^{13}$CO lack spatial isotropy, we could be overestimating the Mach number by a maximum factor of $\sqrt{3}$.

\subsection{Column density}
\label{HISA_column_density}

\subsubsection{H$_2$ column density}
\label{sect_H2_column_density_estimate}

We use the $^{13}$CO(1--0) data from the Galactic Ring Survey \citep[GRS, ][]{Jackson2006} to derive the column density and kinematic properties of the molecular gas of the filament. The data have an angular resolution $\Theta=46\arcsec$ and a velocity resolution of $\Delta v = 0.21$\,km\,s$^{-1}$. We can estimate the column density of the $^{13}$CO molecule including opacity correction with the equation \citep{wilson2010}:
\begin{equation}
N(^{13}{\rm{CO}})  = 3.0\times10^{14} \: \frac{\tau_{13}}{1-{\rm e}^{-\tau_{13}}} \: \frac{\int T_{\rm{MB}}(v)\, dv}{1-{\rm e}^{-5.3/T_{\rm{ex}}}},
\label{eq_column_density_13CO}
\end{equation}
where $N(^{13}\rm{CO})$ is the column density of the $^{13}$CO in units of cm$^{-2}$, $dv$ is the velocity in km\,s$^{-1}$, $\tau_{13}$ is the $^{13}$CO opacity, and $T_{\rm{MB}}$ is the main beam brightness temperature and $T_{\rm{ex}}$ is the excitation temperature, both in K. We do not have a direct measurement for the excitation temperature, and we assume that the excitation temperatures of the $^{12}$CO and $^{13}$CO are the same. Assuming that the $^{12}$CO line is optically thick, we used the $^{12}$CO(1--0) data from the FOREST unbiased Galactic plane imaging survey with the Nobeyama 45~m telescope (FUGIN; \citealt{Umemoto2017}) to estimate $T_{\rm{ex}}$ following the formula \citep{wilson2010}:
\begin{equation}
T_{\rm{ex}}=\frac{5.5}{{\rm ln}\left(1+\frac{5.5}{T_{\textrm{mb}}(^{12}\rm{CO})+0.82}\right)},
\label{eq_tex}
\end{equation}
where $T_{\textrm{mb}}(^{12}\rm{CO})$ is the peak main-beam brightness temperature of $^{12}\rm{CO}$(1--0) line. We calculated $T_{\rm{ex}}$ for regions where $T_{\textrm{mb}}(^{12}\rm{CO})$ is above the 5$\sigma$ level (2~K), which results in a $T_{\rm{ex}}$ between $\sim$5 to 25~K. For regions where $T_{\textrm{mb}}(^{12}\rm{CO})$ is below the 5$\sigma$ level (2~K), an upper limit of 5~K for $T_{\rm{ex}}$ is applied. Following Eq.~B.6 in \citet{Schneider2016}, we can derive $\tau_{13}$ from $T_{\rm{ex}}$ and $T_{\textrm{mb}}(^{13}\rm{CO})$. For regions where $T_{\textrm{mb}}(^{13}\rm{CO})$ is above 5$\sigma$ level (1.05~K), $\tau_{13}$ is estimated to be between $\sim$0.1 and 3. For regions where $T_{\textrm{mb}}(^{13}\rm{CO})$ is below 5$\sigma$ level (1.05~K), an upper limit of 0.1 for $\tau_{13}$ is applied. The majority of the region along the filament has a $\tau_{13}\lesssim1.0$, only the region around $(l=34.30,~ b=0.18)$ and a few pixels around $(l=35.55,~b=0.0)$ have a $\tau_{13}>2.0$.

For the Galactocentric distance of 5.9~kpc of G38a, the fractional abundance of $^{13}$CO relative to H$_2$ is estimated to be 2.9$\times10^{-6}$ following the relations reported by \citet{Giannetti2014}. With this abundance, we converted the $N(^{13}\rm{CO})$ to $N(\rm{H_2})$. \citet{Zhang2019} employed a similar method (uniform CO abundance, $T_{\rm mb}$($^{12}$CO) $\rightarrow$ $T_{\rm ex}$) to estimate the column density and mass of a sample of GMFs (they estimated the mass of GMF38a to be $\sim3.8--11.0\times 10^5$~$M_\odot$), and discussed in detail the uncertainties brought in by $T_{\rm ex}$ and the $^{13}$CO abundance. According to their results, the 1$\sigma$ uncertainty of the column density estimated from $^{13}$CO is $\sim50\%$. Simulations show that the abundance of $^{13}$CO could vary and we could underestimate the column density for the low column density part by $\sim40\%$ \citep[$N(^{13}{\rm CO})<10^{16}$, or $N({\rm H_2})<3.4\times10^{21}$ in this paper,][]{Szucs2014, Szucs2016}.

\subsubsection{CNM column density from HISA measurements}
Besides the kinematics, the column density of \ion{H}{i} is also a critical cloud parameter. To estimate the column density of the HISA feature we use the equation given by \citet{wilson2010}:
\begin{equation}
N_{\rm H} = 1.8224\times10^{18}\: T_{\rm S} \int \tau(v)\, dv ,
\label{eq_column_density_hi}
\end{equation}
where $T_{\rm{S}}$ and $\tau$ are the spin temperature and optical depth, respectively. For the CNM traced by HISA, $T_{\rm{S}}=T_{\rm{HISA}}$ and $\tau=\tau_{\rm HISA}$ (Eq.~\ref{eq_HISA_on_minus_off_solution}). However, as mentioned in Sect.\,\ref{sec_HI_self_absorption} we measure the spin temperature and the optical depth together and disentangling them is difficult. Hence, we assume a constant spin temperature over the cloud and calculate the optical depth using Eq.\,\ref{eq_HISA_on_minus_off_solution}. As we have no measurement for $p$, this value is difficult to estimate. Considering GMF38a is at the near side of the Milky way, we assume a value of $p=0.9$ in the following and discuss the corresponding uncertainties in Sect.~\ref{sect_p} and \ref{sect_Uncertainties_in_the_HISA_description}. We will discuss the uncertainty of column density brought in by the method we choose to estimate the background temperature $T_{\rm{off}}$ in Sect.\,\ref{sect_Uncertainties_in_the_HISA_description}

We integrate between 50 to 60~km$^{-1}$ and derive the column density map shown in Fig.\,\ref{fig_column_density}, assuming for the HISA feature a spin temperature $T_{\rm S} = 40$\,K, $p=0.9$ and using a second order polynomial to estimate the background temperature $T_{\rm{off}}$. Larger spin temperatures do not change the structure of the column density map significantly, but will increase its value everywhere. 

The column density peaks in the $^{13}$CO map do not coincide well with the column density peaks of the atomic hydrogen. As shown in Fig.\,\ref{fig_filament_overview}, the highest peak in the $^{13}$CO (around $l=34\degr$) coincides with a strong continuum source and hence makes the determination of the HISA feature at this position impossible. However, we use this continuum source to constrain the optical depth, which we present in Sect.\,\ref{sect_HI_optical_depth_continuum_source}. The highest column density peak for the atomic hydrogen can be found in the eastern area of the filament ($l=36.5\degr$). In this region, the H$_2$ is diffuse and its column density is low. Another CNM column density peak can be found in the center of the filament around $l=35.4\degr$, $b=+0.3\degr$. This CNM feature has almost a round shape and only a very weak counterpart in the $^{13}$CO emission. 

Assuming a typical CNM thermal pressure of $P_{\rm CNM}/k\sim$4000~K~cm$^{-3}$ \citep{Heiles1997, Jenkins2011, Goldsmith2013}, and $T_{\rm K}=T_{\rm HISA}=40$~K, we can estimate the size of the CNM along the line of sight by dividing the column density by $P_{\rm CNM}/k/T_{\rm K}$. Depending on the column density, the line of sight size is estimated to be $\sim$0.5--3~pc, which is much smaller than the width and length of the filament ($\sim25$~pc and $\sim$230~pc, respectively). Arguing the other way round, if the line of sight size of the CNM is similar to the width of the filament ($\sim25$~pc), the thermal pressure would be significantly lower, and the CNM could be under-pressured or the pressure may be dominated by some other (magnetic/turbulent) component.

\subsubsection{Atomic gas column density from \ion{H}{i} emission}
Since HISA traces only the CNM, it is likely that this component is surrounded or even mixed with a warm component. We also derive the column density traced by the \ion{H}{i} emission (between 50 and 60~km~s$^{-1}$, $32.65\degr<l<37.27\degr$ and $|b|\leq1.25\degr$), which traces both the CNM and WNM. Following the method described in \citet{Bihr2015b}, we derived the mean \ion{H}{i} optical depth map from the strong continuum sources in the background (G33.498+0.194, G33.810-0.189, G33.915+0.110, G34.133+0.471, G35.053--0.518, G35.574+0.068, G35.947+0.379, G36.056+0.357 and G36.551+0.002; \citealt{Wang2018}). Following the method described in \citet{Bihr2015b}, the optical depth measured towards the continuum source is:
\begin{equation}
\tau = -{\rm ln}\left(\frac{T_{\rm{on,\ cont}} - T_{\rm off,\ cont}}{T_{\rm{cont}}}\right),
\label{eq_tau}
\end{equation}
where $T_{\rm on,\ cont}$ is the on-continuum-source brightness temperature, and $T_{\rm off,\ cont}$ is the off-continuum-source temperature. Since we use the THOR C array data to calculate $\tau$, the smooth, large-scale structure is mostly filtered out \citep{beuther2016}. We can neglect the off emission $T_{\rm off,\ cont}$ and simplify Eq.~\ref{eq_tau} to:
\begin{equation}
\tau_{\rm{simplified}} = -{\rm ln}\left(\frac{T_{\rm{on, cont}}}{T_{\rm{cont}}}\right) .
\label{eq_tau_simplified}
\end{equation}
For channels with a $T_{\rm{on,\ cont}}$ value smaller than 3 times the rms, we use the 3$\sigma$ value to get a lower limit of $\tau$. The mean optical depth varies between 1.1 to 1.9 from 50 to 60~km~s$^{-1}$. The optical depth corrected spin temperature is $T_{\rm S}=T_{\rm B}/(1-{\rm e}^{-\tau})$, where $T_{\rm B}$ is the brightness temperature of the \ion{H}{i} emission. The optical depth corrected atomic hydrogen column density is calculated with Eq.~\ref{eq_column_density_hi}. Since the absorption features towards these continuum sources between 50 to 60~km~s$^{-1}$ often saturates, the optical depth we derived is a lower limit as shown in Fig.~\ref{fig_tau_spectra} and we are underestimating the \ion{H}{i} column density. 

\begin{figure*}
\centering
 \includegraphics[width=0.8\textwidth]{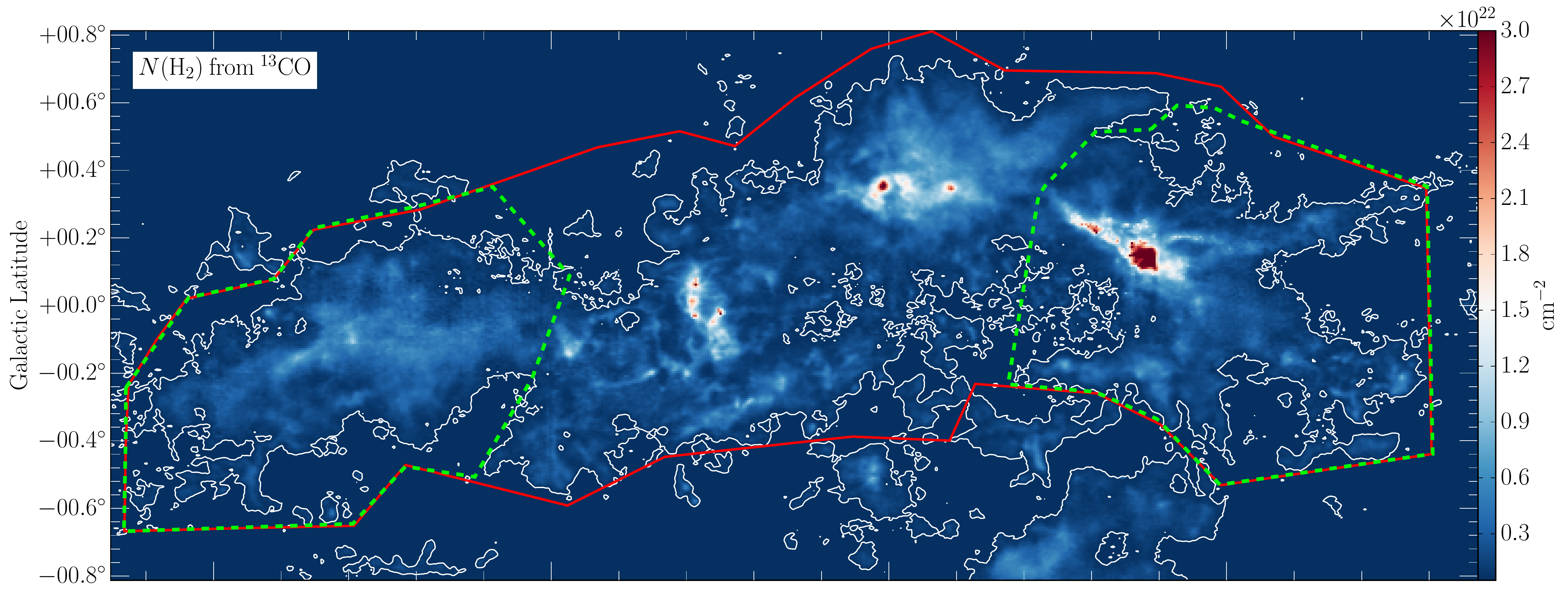}
 \includegraphics[width=0.8\textwidth]{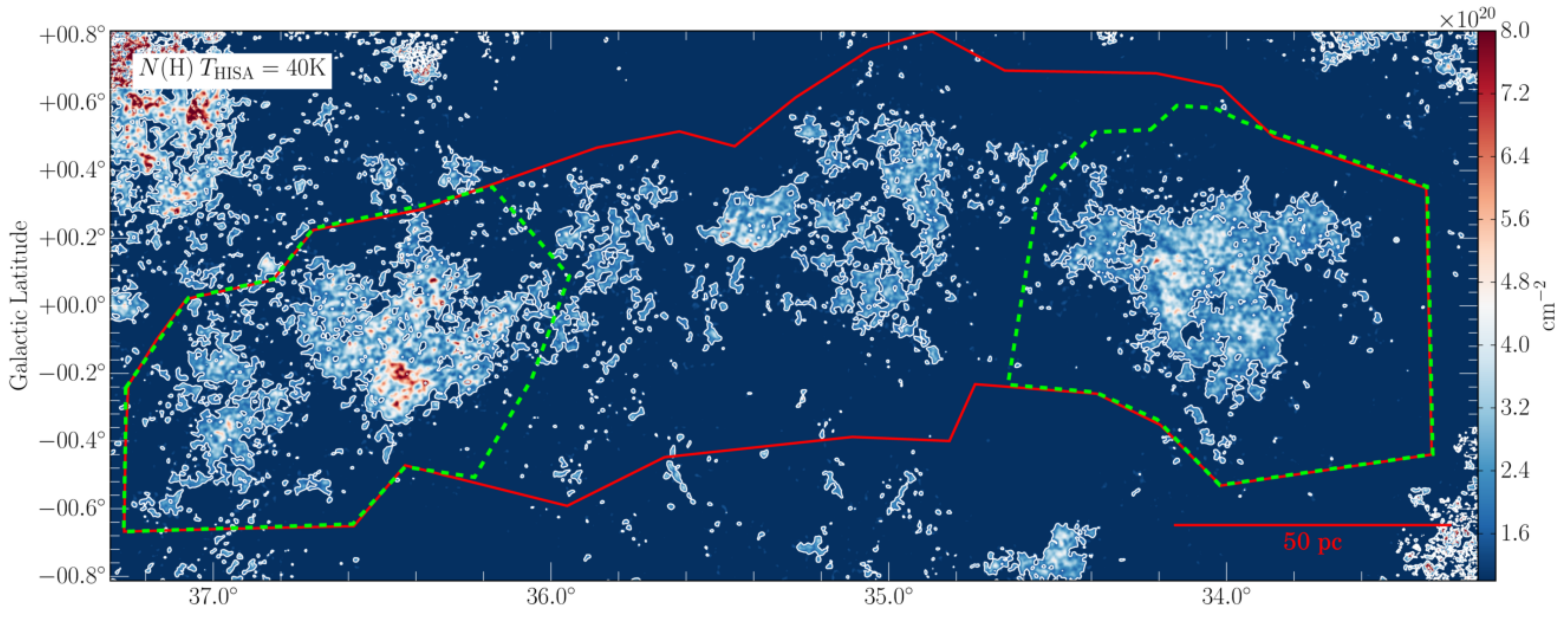}
 \includegraphics[width=0.8\textwidth]{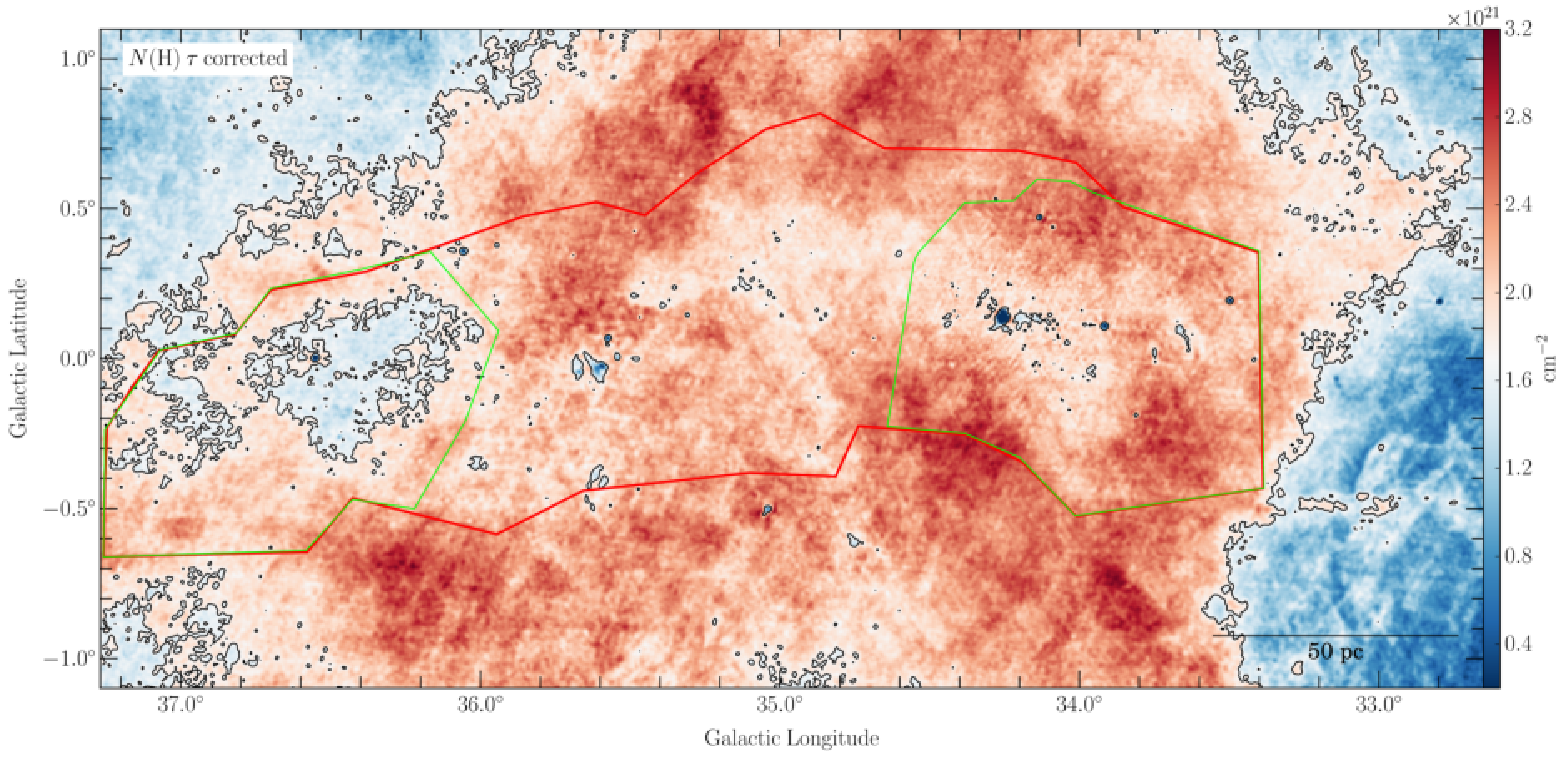} 
\caption{{\it Top-panel:} Column density map of the molecular cloud derived from $^{13}$CO emission (see Sect.\,\ref{sect_H2_column_density_estimate}). {\it Middle-panel}: The column density map of the CNM derived from HISA, assuming $T_{\rm S} = 40$\,K, $p=0.9$ and using a second order polynomial to estimate the background temperature $T_{\rm{off}}$. {\it Bottom-panel}: The optical depth corrected column density map of atomic hydrogen derived from \ion{H}{i} emission. The contours indicate column density threshold used for the N-PDFs (see Sect.\,\ref{sect_discussion_column_density_pdf}). For the H$_2$, HISA and \ion{H}{i} emission they have levels of $6\times10^{20}$, $1.5\times10^{20}$, and 1.7$\times10^{21}$\,cm$^{-2}$, respectively. The red and green dashed polygons mark the region for the mass estimates and the column density PDF measurements shown in Figs.\,\ref{fig_column_density_pdf_overview} and \ref{fig_column_density_pdf_left_and_right}, respectively.}
  \label{fig_column_density}
\end{figure*}

The column density we derived from \ion{H}{i} emission is a combined result from the far (10.3~kpc) and near side (3.4~kpc) due to the kinematic distance ambiguity. Assuming the atomic gas in the Galactic plane is approximately axisymmetric with respect to the Galactic center \citep{Kalberla2008}, the atomic gas at near side and far side that are at the same Galactocentric distance share the same density distribution in the vertical direction. We used the average vertical density profile described by \citet{Lockman1984} (see Eq.~5 and Table.~1 in their paper) to estimate how much gas is at near distance for each line of sight and derived the column density map of \ion{H}{i} emission at 3.4~kpc shown in Fig.~\ref{fig_column_density}.

\begin{figure}
\centering
 \includegraphics[width=0.5\textwidth]{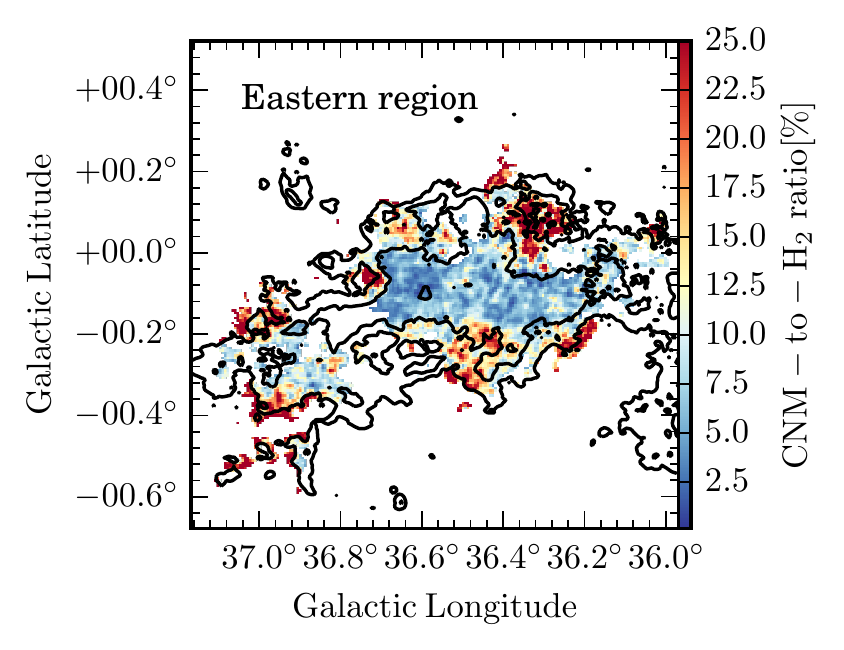}
 \includegraphics[width=0.5\textwidth]{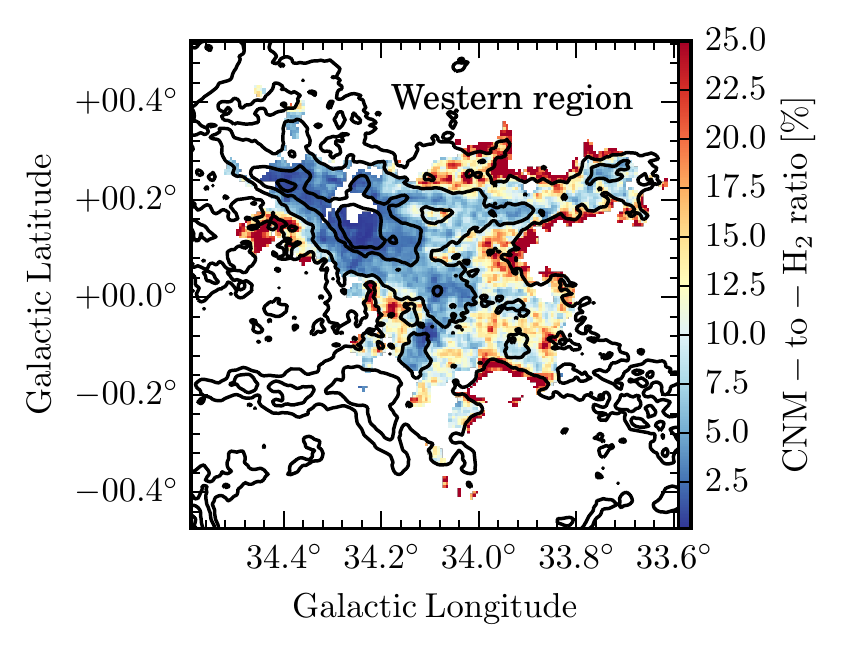}
\caption{Ratio of the CNM and H$_2$ column densities in percentage of the Eastern region ({\it top-panel}) and Western region ({\it bottom-panel}). The contours in both panels indicate the integrated $^{13}$CO emission at levels of 5, 10, 20, and 30\,K\,km\,s$^{-1}$.}
  \label{fig_cnm_ratio}
\end{figure}

By comparing the column density maps of CNM and H$_2$, we produce the CNM-to-H$_2$ ratio map and show it for the Eastern and Western regions in Fig.~\ref{fig_cnm_ratio}. We masked out regions outside the column density threshold contours shown in Fig.~\ref{fig_column_density}. In both subregions, the CNM-to-H$_2$ ratio varies between $\sim0.5-25\%$ with a median value of $\sim 9\%$. Fig.~\ref{fig_cnm_ratio} shows that the outer layers of the filament have high CNM-to-H$_2$ ratio, while the inner regions show lower CNM-to-H$_2$ ratio. This change from the outside to the inside of the cloud can be interpreted as signature of the conversion of atomic to molecular gas with increasing density. \citet{Zuo2018} studied HINSA towards nearby clouds and found a much lower ratio between 0.2 and 2\%. 

\begin{figure}
\centering
 \includegraphics[width=0.5\textwidth]{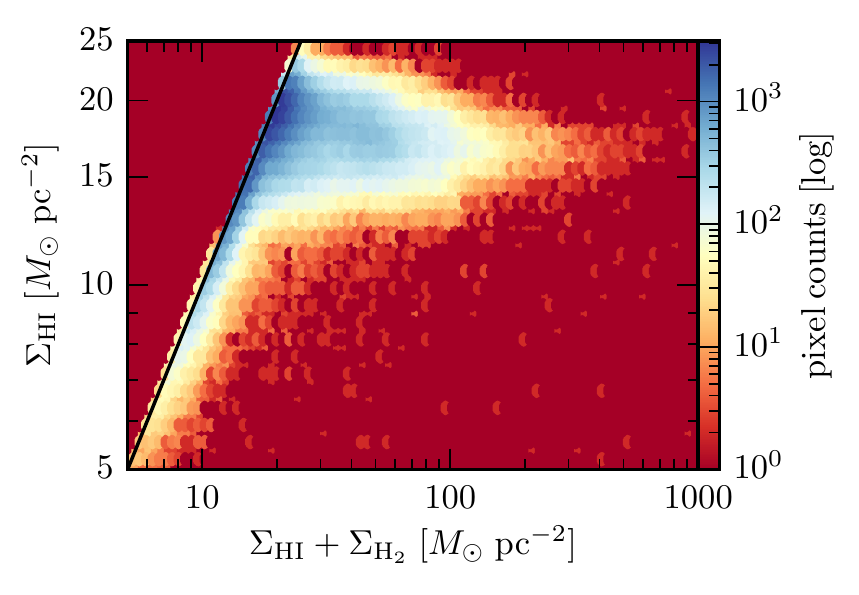}
\caption{Surface density of the atomic hydrogen ($v_{\rm lsr}=50-60$~km~s$^{-1}$) as a function of the total gas surface density ($\Sigma_{\rm HI} + \Sigma_{\rm H_2}$). The solid line represents a one-to-one relation.}
  \label{fig_krumholtz}
\end{figure}

Figure~\ref{fig_krumholtz} shows the surface density comparison between the atomic hydrogen (CNM+WNM) and the total gas (CNM+WNM+H$_2$). The surface density of the atomic hydrogen rises up to $\sim14-23~M_\odot$~pc$^{-2}$ ($\sim1.8-2.9\times10^{21}$~cm$^{-2}$) and then saturates to an almost flat distribution. This turnover is at lower values than that found by \citet{Bihr2015b} towards W43 (50--80~M$_\odot$~pc$^{-2}$) but it is still higher than the 10~M$_\odot$~pc$^{-2}$ observed towards nearby clouds \citep{Lee2015} and predicted by models \citep[e.g.,][]{Krumholz2008, Krumholz2009, Sternberg2014}. Such higher than predicted \ion{H}{i} column densities can be explained by the clumpy nature of the ISM with several \ion{H}{i}-to-H2 transitions along the line of sight \citep{Bialy2017b}.

\subsection{Mass estimate}
\label{sect_mass_estimate}
\begin{table*}[h]
\caption{Mass estimates of the GMF.}
\centering
\begin{threeparttable}
\label{table_mass_estimates}
\begin{tabular}{l c c c c c } 
\hline\hline
Region & $M$(H$_2$)                 & $M$(CNM) & $M$(CNM) & $M$(CNM)/M(H$_2$) & $M$(\ion{H}{i} emission)\\
           &                                         & (20\,K) & (40\,K) & (40\,K) & \\

           & [$M_{\odot}$] & [$M_{\odot}$] &  [$M_{\odot}$] &  &[$M_{\odot}$]   \\
\hline\vspace{-0.35cm}\\
Full filament &3.6$\times$10$^5$& $3.4\times 10^{3}$& $9.5\times 10^{3}$ &  3\% &  2.3$\times 10^{5}$ (6.1$\times 10^5$$^*$)\\
Eastern region & 7.9$\times$10$^4$ &$1.1\times 10^{3}$ & $3.4\times 10^{3}$& 4\%& 5.0$\times 10^{4}$\\
Western region & 1.1$\times$10$^5$ & $1.1\times 10^{3}$&$ 2.9\times 10^{3}$ & 3\%& 7.1$\times 10^{4}$\\

\hline   
\end{tabular} 
\begin{tablenotes}
\scriptsize
\item $*$This mass is calculated including all \ion{H}{i} emission between 50 and 60 km~s$^{-1}$, $32.65\degr<l<37.27\degr$, $|b|\leq1.1\degr$ (bottom-panel in Fig.~\ref{fig_column_density}).
\end{tablenotes}
\end{threeparttable}
\end{table*}

As we know the column density and the distance to the cloud ($\sim$3.4\,kpc), we can directly estimate the mass of \ion{H}{i} and H$_2$ gas. We do so for three different regions, the ``full filament'' (red polygon in Fig.\,\ref{fig_column_density}), the ``eastern region'' (eastern green dashed polygons in Fig.\,\ref{fig_column_density}) and the ``western region'' (western green dashed polygons in Fig.\,\ref{fig_column_density}). Table \ref{table_mass_estimates} summarizes the mass measurements. The molecular hydrogen mass for the entire filament is $\sim$3.6$\times$10$^5$\,$M_{\odot}$ and the CNM mass traced by the HISA is significantly less, showing values of 3.4$\times$10$^3$ to 9.5$\times$10$^3$~$M_{\odot}$, depending on the assumed spin temperature. 

Furthermore, we studied the cold atomic to the molecular hydrogen mass ratio $M$(CNM)/$M$(H$_2$). For the entire filament, this value is between 1\% and 3\%, again depending on the assumed spin temperature. However, this ratio has to be treated cautiously. The HISA extraction method does not work reliably in the center of the filament due to strong continuum emission and we might miss some \ion{H}{i} mass. The mass ratio for the smaller regions show slightly higher values (3\% to 4\%). The H$_2$ column density shows significantly higher values for the western region in comparison to the eastern region. In contrast to this, the \ion{H}{i} column density reveals a prominent peak on the eastern side and hence the $M$(CNM)/$M$(H$_2$) ratio is lower for the western than that for the eastern region. Considering that $^{13}$CO does not trace all the molecular hydrogen gas \citep{Pineda2008, Goodman2009, Gong2018}, the $M$(CNM)/$M$(H$_2$) ratio we derive could be just an upper limit.

We further estimate the mass of the atomic component traced by \ion{H}{i} emission. Within the same area (polygons in Fig.~\ref{fig_column_density}, the atomic gas traced by \ion{H}{i} emission has lower mass as the molecular gas (Table~\ref{table_mass_estimates}). However, since there is no clear boundary of the GMF shown in the \ion{H}{i} column density map (see bottom-panel in Fig.~\ref{fig_column_density}), we can assume all \ion{H}{i} emission between 50 to 60~km~s$^{-1}$ within $32.65\degr<l<37.27\degr$ and $|b|\leq1.1\degr$ (same latitude range as the GRS coverage) is associated with the filament to obtain the mass. The mass is estimated to be 6.1$\times 10^5$~$M_{\odot}$, which is about $\sim$60 times larger than the mass estimated for the CNM traced by HISA (40~K). The mass of the atomic hydrogen is about 60\% larger than the molecular hydrogen mass (3.6$\times$10$^5$~$M_{\odot}$), which makes sense since the molecular cloud is surrounded by a large reservoir of atomic gas.

\section{Discussion}
\label{sect_discuss}
\subsection{Kinematics}
\label{sect_discussion_kinematics}
For nearby galaxies, the ratios of the CO to \ion{H}{i} linewidth is around $\sigma_{\rm HI}/\sigma_{\rm CO} = 1-1.4$ \citep{Caldu-Primo2013,Mogotsi2016}. The linewidth values found in these studies are approximately $\sigma \sim 6-12$\,km\,s$^{-1}$, which corresponds to $\Delta v_{_{\rm{FWHM}}}  \sim 14-28$\,km\,s$^{-1}$ for both the \ion{H}{i} and CO lines. These measurements are done for the CO and \ion{H}{i} emission over large regions ($\sim$0.5\,kpc), which can increase the linewidth due to superposition of different velocity components in the supersonically turbulent ISM \citep[e.g., ][]{Klessen2016}. Our Galactic HISA measurements of GMF38a show significantly smaller values for the linewidth of $\Delta v_{_{\rm{FWHM}}} \sim 2-8$\,km\,s$^{-1}$ for CNM. The reason is that we observe a much smaller region and we are able to separate multiple components. Furthermore, they observe \ion{H}{i} emission and the linewidth is dominated by WNM, whereas we observe cold \ion{H}{i} absorption features produced by CNM.

\begin{figure}
\centering
 \includegraphics[width=0.5\textwidth]{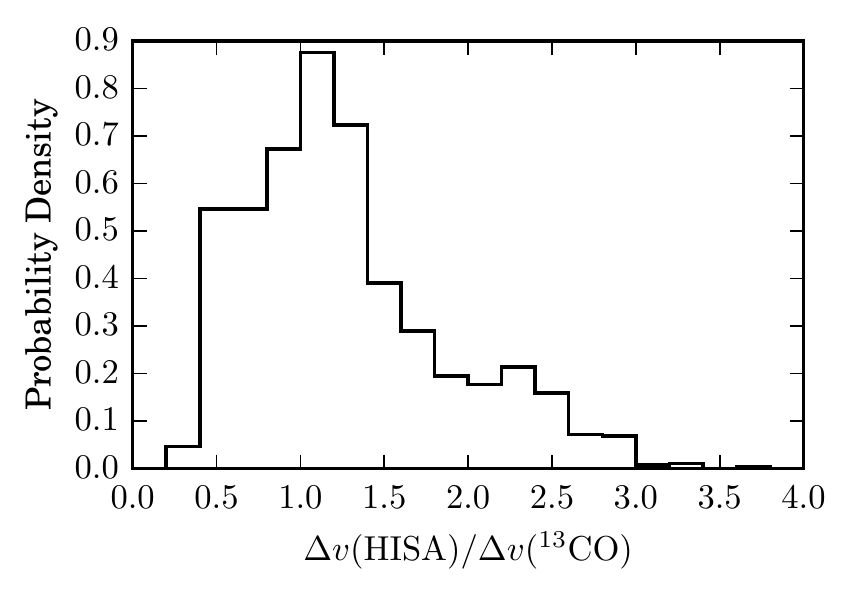}
\caption{Ratio of HISA and $^{13}$CO linewidths in the eastern region indicated in Fig.\,\ref{fig_overview_FWHM_map}.}
  \label{fig_fwhm_ratio_left}
\end{figure}

To study the linewidth ratio in detail, we determine this ratio for the eastern region, which is indicated in Fig.\,\ref{fig_overview_FWHM_map}. We focus on this region as it is not significantly affected by multiple component line spectra. A histogram of the \ion{H}{i}/$^{13}$CO ratio is shown in Fig.~\ref{fig_fwhm_ratio_left}. The mean values for the linewidths are $\Delta v_{_{\rm{FWHM}}} ({\rm ^{13}CO}) = 3.6$\,km\,s$^{-1}$ and $\Delta v_{_{\rm{FWHM}}} (\ion{H}{i}) = 4.5$\,km\,s$^{-1}$.

\subsection{The percentage of background emission -- $p$}
\label{sect_p}
The percentage of background emission, parameterized with $p$, is difficult to estimate. Different assumptions can be found in the literature. For example, \citet{McClure-Griffiths2006} and \citet{Denes2018} assume that $p=1$ and 0.9 for the observed Riegel-Crutcher cloud, respectively, as the corresponding distance is small ($\sim$125\,pc). \citet{Rebolledo2017} studied the HISA features in the Gum~31 molecular complex with a simple two component assumption ($p=1$). \citet{Li2003} studied the \ion{H}{i} narrow self-absorption towards dark clouds in the Taurus/Perseus region. Since these clouds are located at high Galactic latitude away from the Galactic mid-plane, they can assume a simple Gaussian Galactic \ion{H}{i} disk model \citep{Lockman1984} and estimate the factor $p$ with the complementary error function. 

Since the background and foreground \ion{H}{i} emission occurs from warm and diffuse \ion{H}{i} clouds, we do not expect fluctuations of this emission on small scales. Hence, the assumption of a constant $p$ for the entire filament is reasonable. Furthermore, we can obtain a lower limit for $p$. As shown in Fig.\,\ref{fig_Ts_vs_tau}, low values of $p\lesssim0.4$ are not feasible as the spin temperature would become smaller than the temperature of the CMB. As we further discussed in Sect.~\ref{sect_maximum_spin_temperature}, low values of $p\lesssim0.7$ would also result an unrealistic low spin temperature. 

For Galactocentric radius $7\lesssim R \lesssim 35$~kpc, \citet{Kalberla2008} reported the average mid-plane volume density distribution of the atomic gas follows $n(R)\sim n_0~{\rm e}^{-(R-R_\odot)/R_n}$ with $n_0=0.9$~cm$^{-3}$, and $R_n$=3.15~kpc, $R_\odot=$8.5~kpc (IAU recommendations). Assuming $n(R<7~{\rm kpc})=n(7~{\rm kpc})$, we integrate the density along the line of sight of GMF38a ($l=35.5$\degr, $b=0$\degr) and obtain the amount of gas in the foreground and background of GMF38a. Assuming optically thin emission, $p$ is estimated to be 0.91, which agrees with our adopted value of 0.9. 


\subsection{Optical depth measurement toward a strong continuum source}
\label{sect_HI_optical_depth_continuum_source}
\begin{figure}
\centering
 \includegraphics[width=0.5\textwidth]{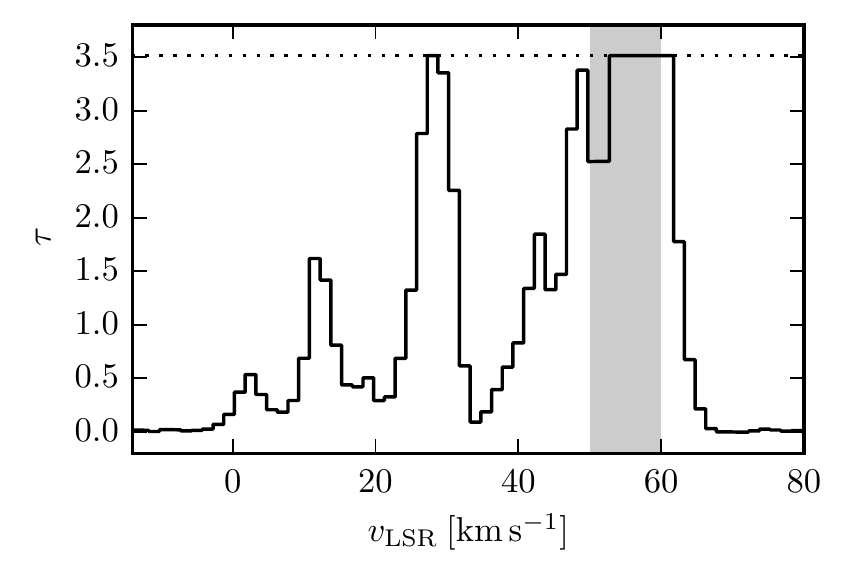}
 \caption{Spectrum of the \ion{H}{i} optical depth towards the UC\ion{H}{ii} region G34.256+0.146 using the THOR data. For some channels, the absorption spectra saturates and the measured optical depth is a lower limit of $\tau = 3.5$, which is indicated by the dotted line. The gray shaded area indicates the velocity range of the HISA feature ($v_{\rm{LSR}} = 50-60$\,km\,s$^{-1}$).}
  \label{fig_tau_spectra}
\end{figure}

We can use strong continuum sources to estimate the optical depth of the CNM. The UC\ion{H}{ii} region G34.256+0.146 (see Fig.\,\ref{fig_filament_overview}) is an ideal candidate to perform this task as it is very bright ($T_{\rm{cont}}(max) \sim 1300$\,K) and slightly extended ($d\sim 70\arcsec$). We use the THOR data to extract the absorption spectrum towards the continuum peak and determine the optical depth and the lower-limit of the optical depth using Eq.~\ref{eq_tau_simplified}. The optical depth spectrum is shown in Fig.\,\ref{fig_tau_spectra}. In the velocity range of the HISA feature, the absorption spectrum saturates and the determined optical depth $\tau = 3.5$ is a lower limit. Furthermore, since the UC\ion{H}{ii} region is at the same distance as the filament \citep{anderson2014} and associated with the filament, there could be CNM that are behind the UC\ion{H}{ii} region and are not traced by the absorption spectrum. Therefore, the determined optical depth represents a lower limit.

As explained in Sect.\,\ref{sec_HI_self_absorption}, the general HISA extraction method measures the optical depth together with the spin temperature and we are not able to disentangle them. However, the additional information from the absorption spectra towards the strong continuum source and the corresponding optical depth measurement allows us to overcome this problem. Figure \ref{fig_Ts_vs_tau} presents the optical depth as a function of the spin temperature for different values of $p$. The lower limit of the optical depth measurement is shown at $\tau = 3.5$ using a black horizontal line. Assuming $p=0.9$ reveals a spin temperature of $T_{\rm{HISA}} \sim 55$\,K. This is a bit higher than the assumed spin temperature for the column density determination presented in Fig.\ref{fig_column_density} (40~K). Since we measure the spin temperature close to an UC\ion{H}{ii} region, we expect rather high values. 

A problem for HISA studies is that low level of background emission can be interpreted as an absorption feature. Studying very narrow absorption features, HINSA \citep[e.g.,][]{Li2003, Goldsmith2005, Goldsmith2007, Krco2010}, can avoid this problem, since the steep absorption profiles of these HINSA features cannot be induced by two broad emission profiles on each side of the absorption feature. The broad HISA features we identified could be caused by two emission components. Fortunately, the optical depth information from the spectrum toward the strong continuum source helps to solve this problem. Since the optical depth is high ($\tau > 3.5$) for the velocity range of the HISA feature (Fig.\,\ref{fig_tau_spectra}), we are confident that we actually indeed observe a HISA feature rather than missing \ion{H}{i} emission. Furthermore, the correlation of the HISA feature with the $^{13}$CO emission is another indicator of cold, dense \ion{H}{i}.

\subsection{Maximum spin temperature}
\label{sect_maximum_spin_temperature}
\begin{figure}
\centering
 \includegraphics[width=0.5\textwidth]{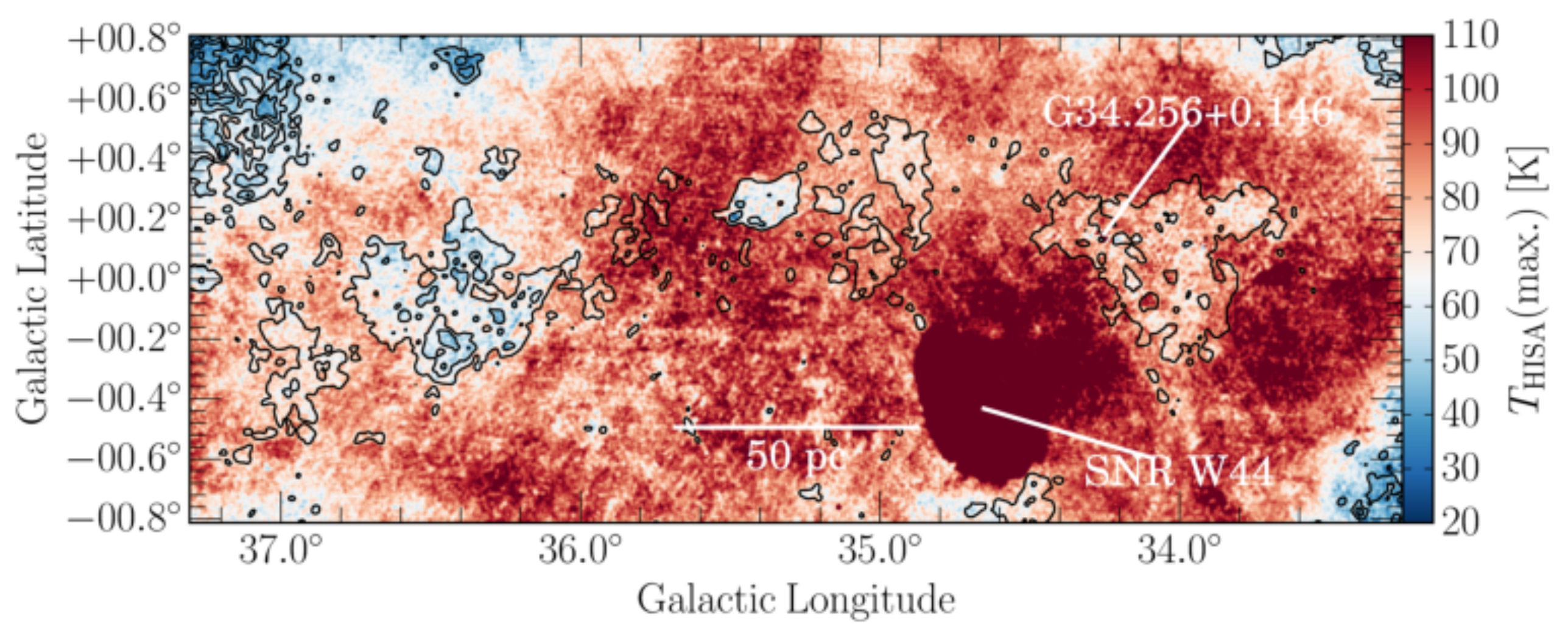}
 \caption{Maximum spin temperature (Eq.\,\ref{eq_T_s_max}) of the absorption features assuming $p=0.9$. The black contours show the HISA column density map assuming $T_{\rm HISA}=$40~K, $p=0.9$, and the contour levels are 1.5, 3.5 and 5.5$\times10^{20}$~cm$^{-2}$ with a smoothing of 3 pixels. Regions with strong continuum emission are not reliable, such as SNR W44 and G34.256+0.146 (see Fig.\,\ref{fig_filament_overview}).}
  \label{fig_maximum_spin_temperature}
\end{figure}

As explained in Sect.\,\ref{sect_HI_optical_depth_continuum_source}, we can use strong continuum sources to measure the optical depth and therefore disentangle the spin temperature and the optical depth. However, this is only possible in a selected number of locations, i.e., in the vicinity of a strong continuum source. In general we can only give the spin temperature as a function of the optical depth. Fig.\,\ref{fig_Ts_vs_tau} shows that the function becomes very steep for certain spin temperatures. Hence, this shows that the maximum spin temperature will be reached for the case of large optical depth. This can also be seen by solving Eq.\,\ref{eq_HISA_on_minus_off_solution} for $T_{\rm{HISA}}$:
\begin{equation}
T_{\rm{HISA}} = \frac{T_{\rm{on-off}}} {1-{\rm e}^{-\tau}} + p \: T_{\rm{off}} + T_{\rm{cont}}.
\label{eq_HISA_on_minus_off_solution_T_s}
\end{equation}
Since $T_{\rm{on-off}}$ is always negative, $T_{\rm{HISA}}$ reaches an upper limit for a given $T_{\rm{off}}$ and $T_{\rm{cont}}$ when $\frac{T_{\rm{on-off}}} {1-{\rm e}^{-\tau}}$ has its minimum value, which occurs for $\tau \rightarrow \infty$. This means the maximal $T_{\rm{HISA}}$ is:
\begin{equation}
\begin{split}
T_{\rm{HISA}}(max.) &= T_{\rm{on-off}} + p \: T_{\rm{off}} + T_{\rm{cont}}.
\end{split}
\label{eq_T_s_max}
\end{equation}
This equation depends on the assumption of the ratio of foreground and background emission, which is described by the factor $p$. For $p=1$, the upper limit of the spin temperature reaches a maximum. We can use this information to calculate the upper limit of the spin temperature for each pixel in our map. However, we do not assume $p=1$, but rather a more realistic value of $p=0.9$. The result is given in Fig.\ref{fig_maximum_spin_temperature}. We focus the discussion on regions offset from strong continuum sources, since for strong continuum sources the attenuation of the continuum exceeds by far the contribution of  self-absorption to the \ion{H}{i} absorption spectrum. As expected, we see a clear anti-correlation of the upper limit for the spin temperature and the column density of the HISA feature. A weaker absorption feature ($T_{\rm{on-off}}$ is always negative) will result in a higher $T_{\rm{HISA}}$, and a lower column density. The lowest values are found for the compact HISA feature in the center around $l=35.5\degr$ with values around $T_{\rm{HISA}}(max.)\sim25$\,K. Similar values can be found for the eastern region of the filament, whereas the western side of the filament shows in general higher values around $T_{\rm{HISA}}(max.)\sim 75$\,K. As this is only an upper limit for the spin temperature, we cannot directly infer the actual temperature. However, it is plausible that the \ion{H}{i} spin temperature is higher on the western side of the cloud due to star formation activity and feedback processes, such as the prominent UC\ion{H}{ii} region. We will discuss this aspect further in Sect.\,\ref{sect_evolutionary_stages}.

For the \ion{H}{i} column density determination in Sect.\,\ref{HISA_column_density}, we assumed a spin temperature of $T_{\rm{HISA}} = 40$\,K. As seen in Fig.\,\ref{fig_maximum_spin_temperature}, this is higher than the upper limit of the spin temperature for certain cold regions. Hence for a few channels we cannot determine a column density for these regions, we exclude these pixels in these channels from our column density calculation. Since we observe only small regions with $T_{\rm{HISA}}(max.)<40$\,K in only a few velocity channels, the column density calculation is not affected significantly. However, assuming a larger value for the spin temperature increases this effect and larger regions are affected, which would make the determined column density unreliable.

Furthermore, as indicated in Equation~\ref{eq_T_s_max}, the factor $p$ also affects the value of $T_{\rm{HISA}}(max.)$. If we take $p=0.7$, the $T_{\rm{HISA}}(max.)$ for the whole filament would drop $\sim20$~K. Regions around $l=35.5\degr$ and in the eastern part of the filament would have a $T_{\rm{HISA}}(max.)<10$~K, which is highly unlikely, since simulations find very little \ion{H}{i} has a temperature $\lesssim$20~K \citep{Glover2016}. Furthermore, previous HISA study towards nearby molecular clouds reveals a $T_{\rm S}$ of 20-80~K \citep{Gibson2000, Gibson2005b, Denes2018}. Therefore, our assumption of $T_{\rm{HISA}}=40$~K and $p=0.9$ in the previous sections is reasonable.


\subsection{Column density probability density functions (N-PDFs)}
\label{sect_discussion_column_density_pdf}

\begin{figure*}
\centering
 \includegraphics[width=0.3\textwidth]{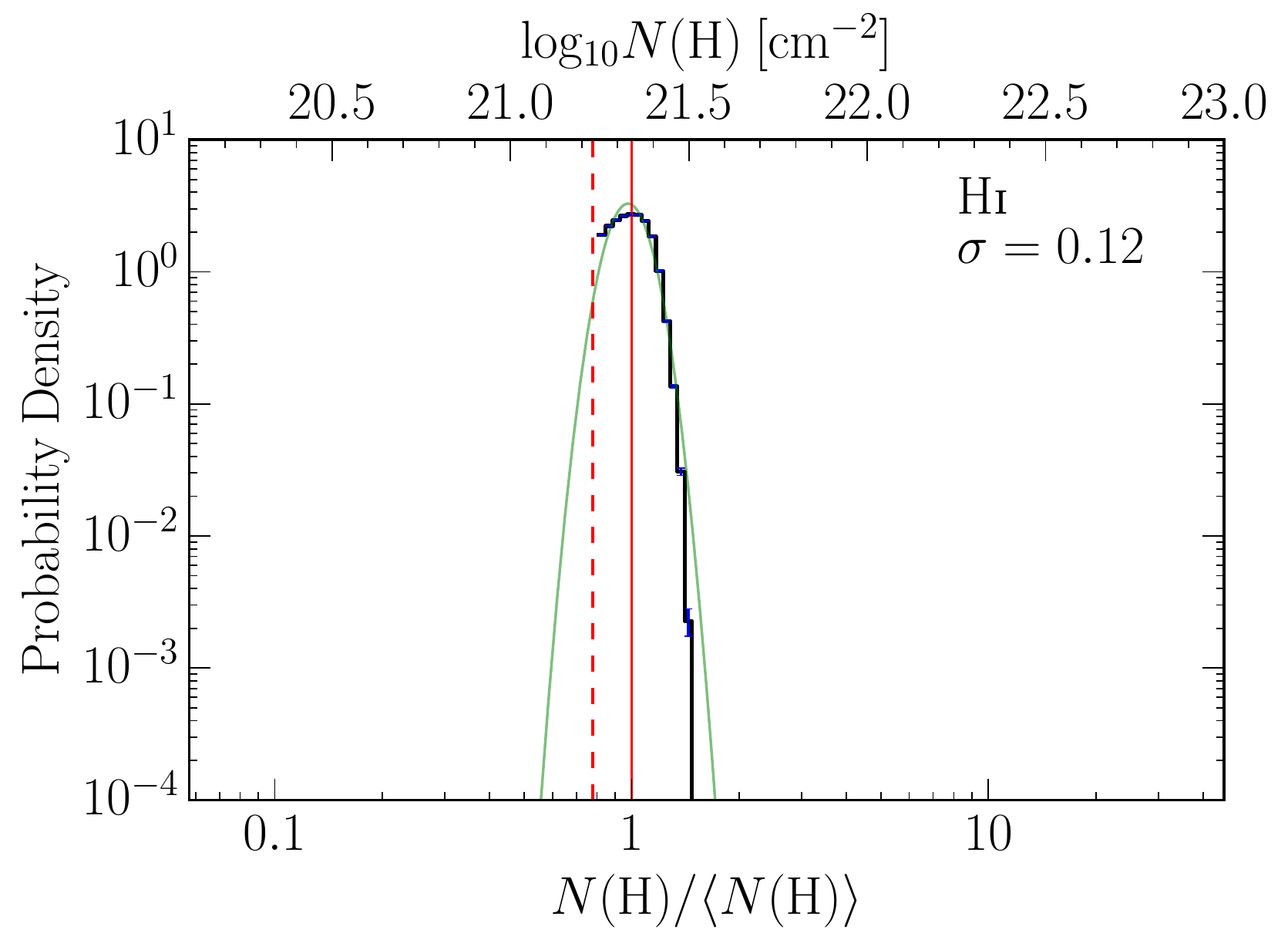}
 \includegraphics[width=0.3\textwidth]{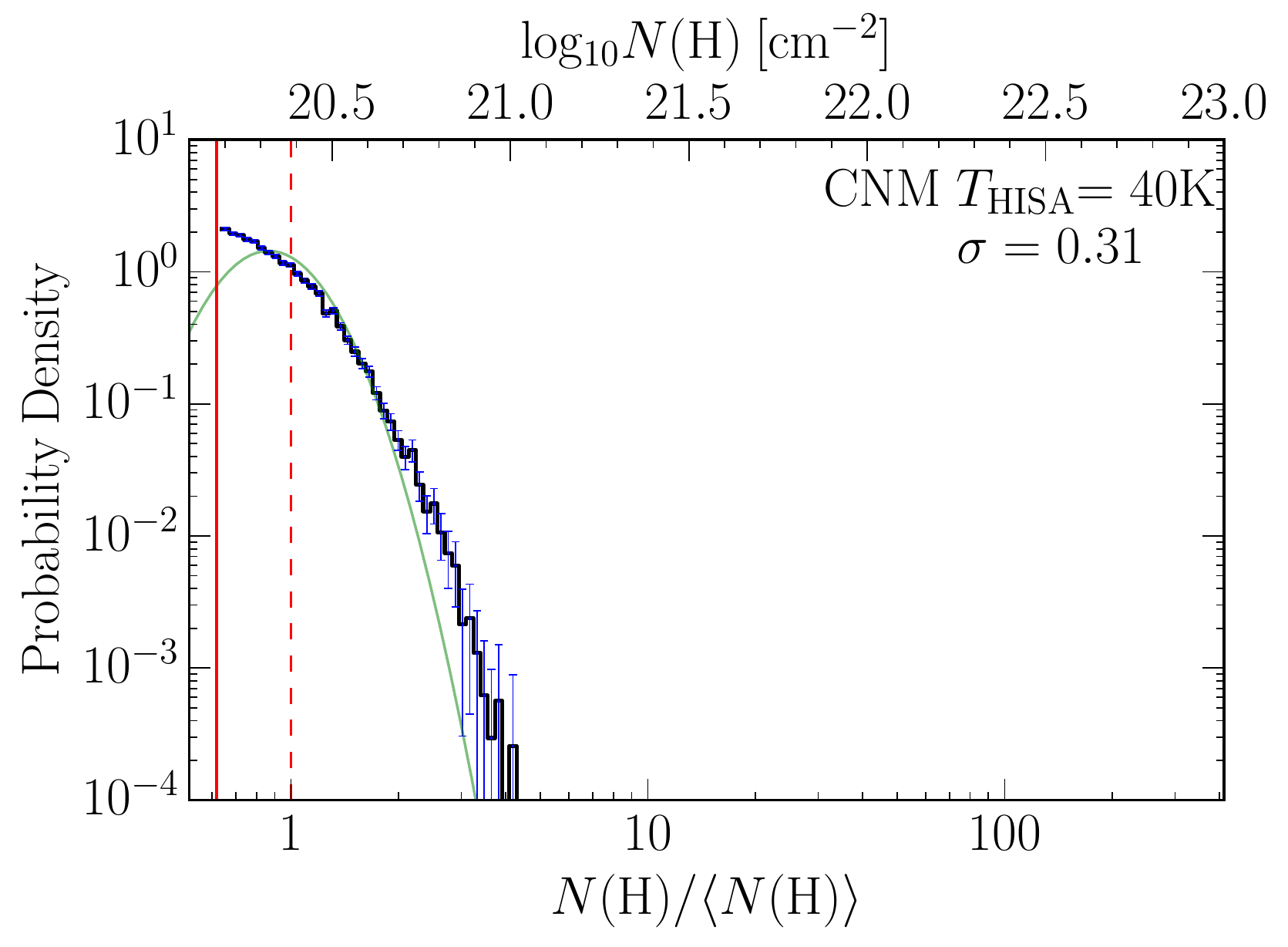}
 \includegraphics[width=0.3\textwidth]{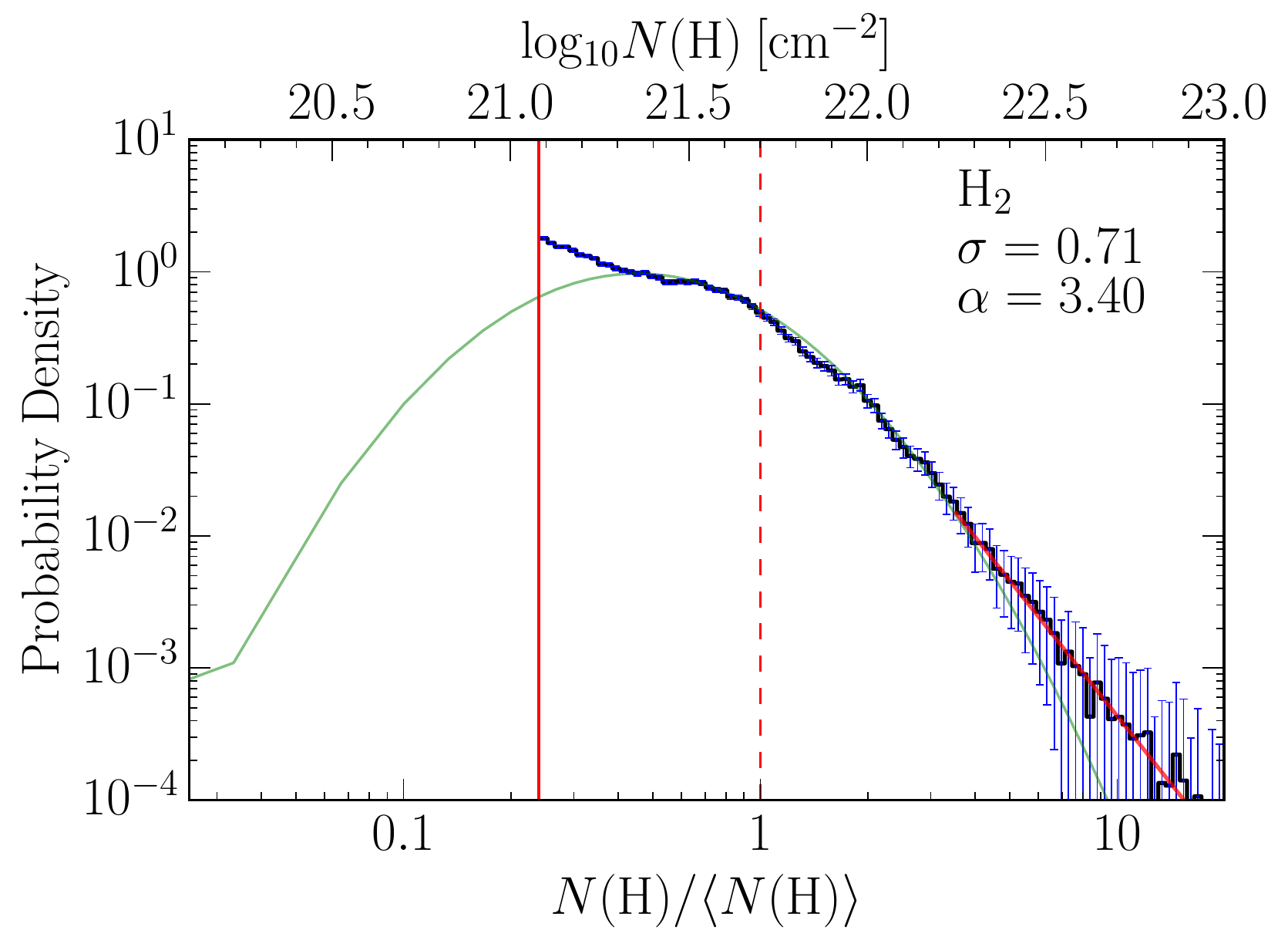} 
\caption{The N-PDFs of the gas traced by \ion{H}{i} (left panel) emission, HISA (middle panel), and $^{13}$CO (right panel) in units of hydrogen atoms per square cm. The light green lines show the log-normal fit in each panel. $\sigma$ is the width of the log-normal distribution. The dashed vertical lines in each panel mark the column density threshold, and the solid vertical lines mark the mean column densities. The red line in the $^{13}$CO panel shows the power-law fit with an index $-\alpha$ to the high column density tail.}
  \label{fig_column_density_pdf_overview}
\end{figure*}

The column density maps derived in Sect.\,\ref{HISA_column_density} can be utilized to determine the probability density functions of column densities (N-PDFs). We resampled all the column density maps into the same spatial resolution and constructed the N-PDFs. Fig.\,\ref{fig_column_density_pdf_overview} presents the N-PDFs in units of hydrogen atoms per square cm for the entire filament traced by HISA, \ion{H}{i} emission and $^{13}$CO. For the HISA feature, we assumed a spin temperature of $T_{\rm{HISA}}=40$~K, $p=0.9$, and used a second order polynomial to estimate the background emission, which are the same assumptions used to produce Fig.\,\ref{fig_column_density}. We derived the H$_2$ column density from $^{13}$CO (see Sect.~\ref{sect_H2_column_density_estimate}), and converted it into the unit of hydrogen atoms per square centimeter for easy comparison with HISA and \ion{H}{i} emission. We calculated the column density traced by \ion{H}{i} emission with optical depth correction (see Sect.~\ref{sect_mass_estimate}). 


In the following, we first consider the quantification of the shapes of the observed N-PDFs and then the interpretation of the observed shapes. It has been argued that it may be necessary to consider ``completeness'' of the N-PDFs when defining the column density range that can be studied \citep[][see also \citealt{Brunt2015, Lombardi2015, Alves2017}]{Kainulainen2013}, although, it is unclear if such a requirement is meaningful when studying possibly turbulence-dominated gas \citep{Kortgen2019}. In our analysis, we define the completeness for the HISA and $^{13}$CO data, but do not apply such requirement for the \ion{H}{I} emission data that we expect to be clearly turbulence-dominated. We define the completeness with the help of the lowest ``closed contours'' in the column density maps. These lowest closed contours are $1.5\times10^{20}$ cm$^{-2}$ for the HISA data and $1.2\times10^{21}$ cm$^{-2}$ for the $^{13}$CO data and lead to regions that are marked with red polygons in Fig.~\ref{fig_column_density}. For \ion{H}{I} emission, we include all the data within $32.65\degr<l<37.27\degr$, $b<1.1\degr$ and use the column density level of $1.7\times10^{21}$ cm$^{-2}$ to define the range over which the range is analyzed. We normalize all N-PDFs by the mean column density of the tracer (shown in Table~\ref{table_npdf}). 

We first note that the N-PDFs of all components appear curved in log-log representation, even when only taking into account the data above the completeness levels. Therefore, we do not make an effort to quantify the N-PDFs with single power-law functions. Given the appearance of the N-PDFs, we decided to quantify their shapes through fits of log-normal functions. The N-PDF from \ion{H}{i} emission shows a clear log-normal shape. We cannot identify a clear peak in the N-PDFs of HISA and $^{13}$CO, but proceed with the log-normal fit nevertheless. The best fit for HISA is not well constrained and shows some excess over the fit at the high column density side. The best fit for $^{13}$CO agrees well with a log-normal function over a wide range at low column densities, but shows an excess over it at higher columns; a power-law could also describe the high column densities relatively well. We fitted the power-law tail of the $^{13}$CO N-PDF with a power-law ($p(x)\propto x^{-\alpha}$) from the optimal column density threshold $N_{\rm min}$, which results in the minimum Kolmogorov-Smirnov distance between the fit and the N-PDF. The fit was performed with the python package {\it Powerlaw} \citep{Alstott2014}. The fitted parameters of the log-normal functions and power-laws, when applicable, are listed in Table~\ref{table_npdf}. 

%
%
The N-PDF from the \ion{H}{i} emission peaks at 2.2$\times 10^{21}$~cm$^{-2}$, corresponding to $A_V=1$ \citep{Guever2009}. Studies towards nearby clouds have found that the N-PDFs from \ion{H}{i} emission there peak around $\sim 1-2 \times 10^{21}$~cm$^{-2}$ \citep{Imara2016}, slight lower than we find here for our target. The widths of the fitted log-normal models vary among the tracers; the \ion{H}{i} emission has the smallest width, followed by HISA and the $^{13}$CO that has the widest width (see Table~\ref{table_npdf}). The narrowness of the \ion{H}{I} emission N-PDF indicates that \ion{H}{i} emission is relatively smoothly distributed without large variations and substructure (Fig.~\ref{fig_column_density}). The width of \ion{H}{I} emission N-PDF is similar to those seen towards nearby clouds \citep{Burkhart2015, Imara2016}. We note that the width of the HISA N-PDF is much larger than the error introduced by our HISA extraction method (see Appendix~\ref{app_pdf_width}); we consider it is robustly between the widths of \ion{H}{I} emission and $^{13}$CO. 



\begin{figure}
\centering
 \includegraphics[width=0.4\textwidth]{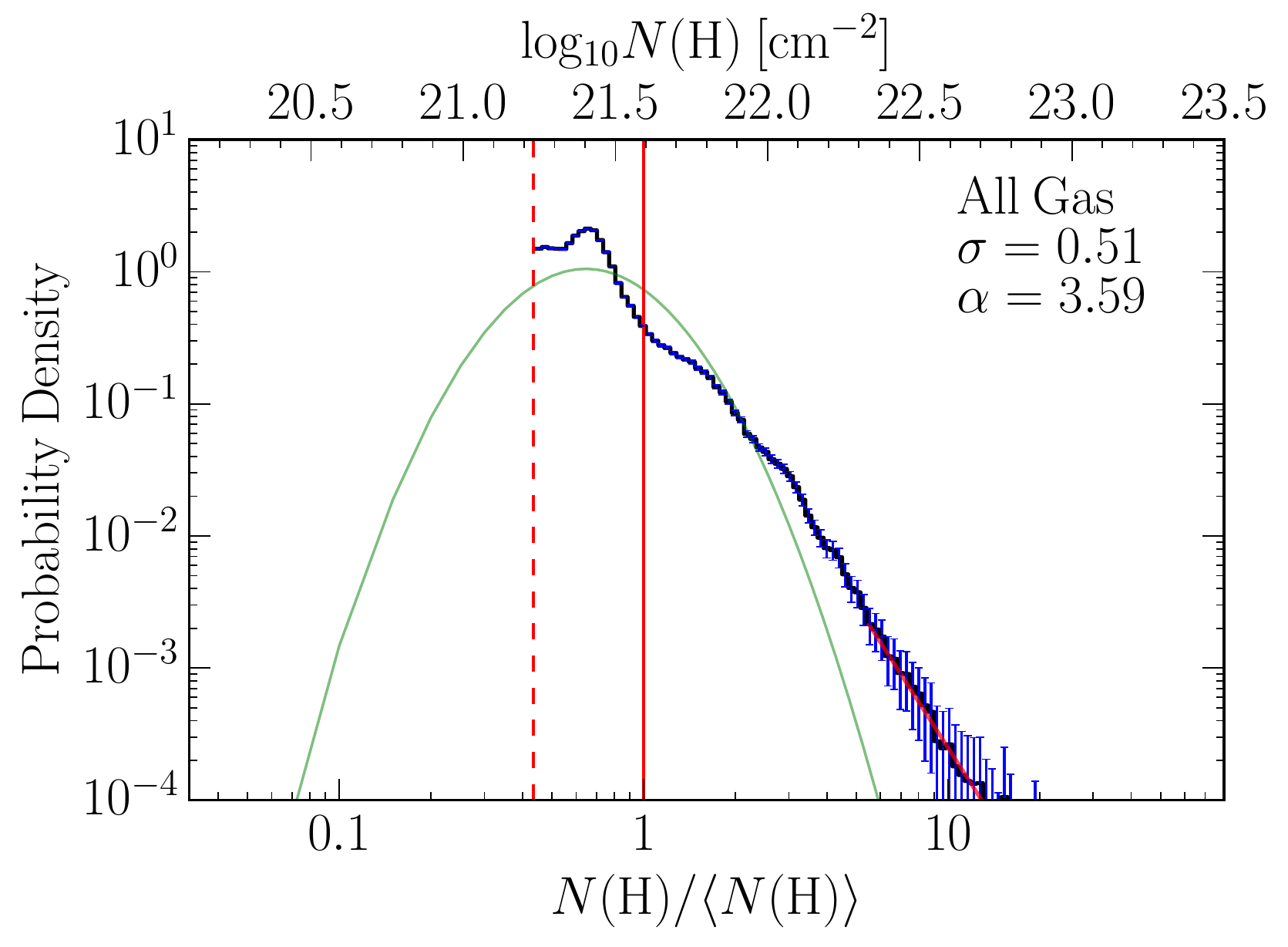}
\caption{Probability distribution function of the combination of atomic and molecular gas over the entire region in Fig.\,\ref{fig_column_density}. The green line shows the log-normal fit. The dashed vertical line mark the column density threshold (the contour levels in Fig.~\ref{fig_column_density}), and the vertical solid line mark the mean column density. The red line shows the power-law fit with an index $-\alpha$ to the high column density tail.}
  \label{fig_column_density_pdf_all}
\end{figure}

We also examine the N-PDF of ``all gas'', derived by adding together the column density maps from all three tracers (Fig.~\ref{fig_column_density_pdf_all}). We note that ``all gas'' does not, in fact, trace all the gas, since there is some amount of H$_2$ that $^{13}$CO does not trace \citep{Pineda2008,Goodman2009}, aka., CO-dark gas. This fraction could be as high as 26\% to 79\% of the total H$_2$ gas \citep{Gong2018}. The N-PDF of ``all gas'' (Fig.~\ref{fig_column_density_pdf_all}) can not be fitted with single log-normal or power-law function. Similar to what we applied to the $^{13}$CO, we fitted the N-PDF with a log-normal function and the high-column density side with a power-law, but obviously, a single power law cannot account for the peak in the observed distribution. 


%

\begin{table*}[h]
\caption{Results of the fits to the N-PDFs.}
\centering
\begin{threeparttable}
\label{table_npdf}
\begin{tabular}{l c c c c c} 
\hline\hline
Component & $N$ Threshold&$\langle N$(H)$\rangle$ & Log-normal width & power-law index& $N_{\rm min}$\\
           & log$_{10}{(N{\rm(H)})}$  & log$_{10}{(N{\rm(H)})}$ &$\sigma$ &$\alpha$ &log$_{10}{(N{\rm(H)})}$ \\
\hline\vspace{-0.35cm}\\
CNM & 20.18 &20.39& 0.31 & --&--\\
\ion{H}{i} emission&21.23 & 21.34&  0.12 & --&--\\
H$_2$ &21.08& 21.70& 0.71 & 3.40& 22.25\\
All gas&21.23 &  21.59& 0.51 & 3.59& 22.33\\
CNM West& 20.18 &20.42& 0.27 & --&--\\
CNM East& 20.18 &20.38& 0.35 & --&--\\
H$_2$ West &21.64& 22.03& 0.71 & 2.61& 21.64\\
H$_2$ East &21.64& 21.86& 0.71 & --& --\\
\hline   
\end{tabular} 
\end{threeparttable}
\end{table*}


%

\begin{figure}
\centering
 \includegraphics[width=0.4\textwidth]{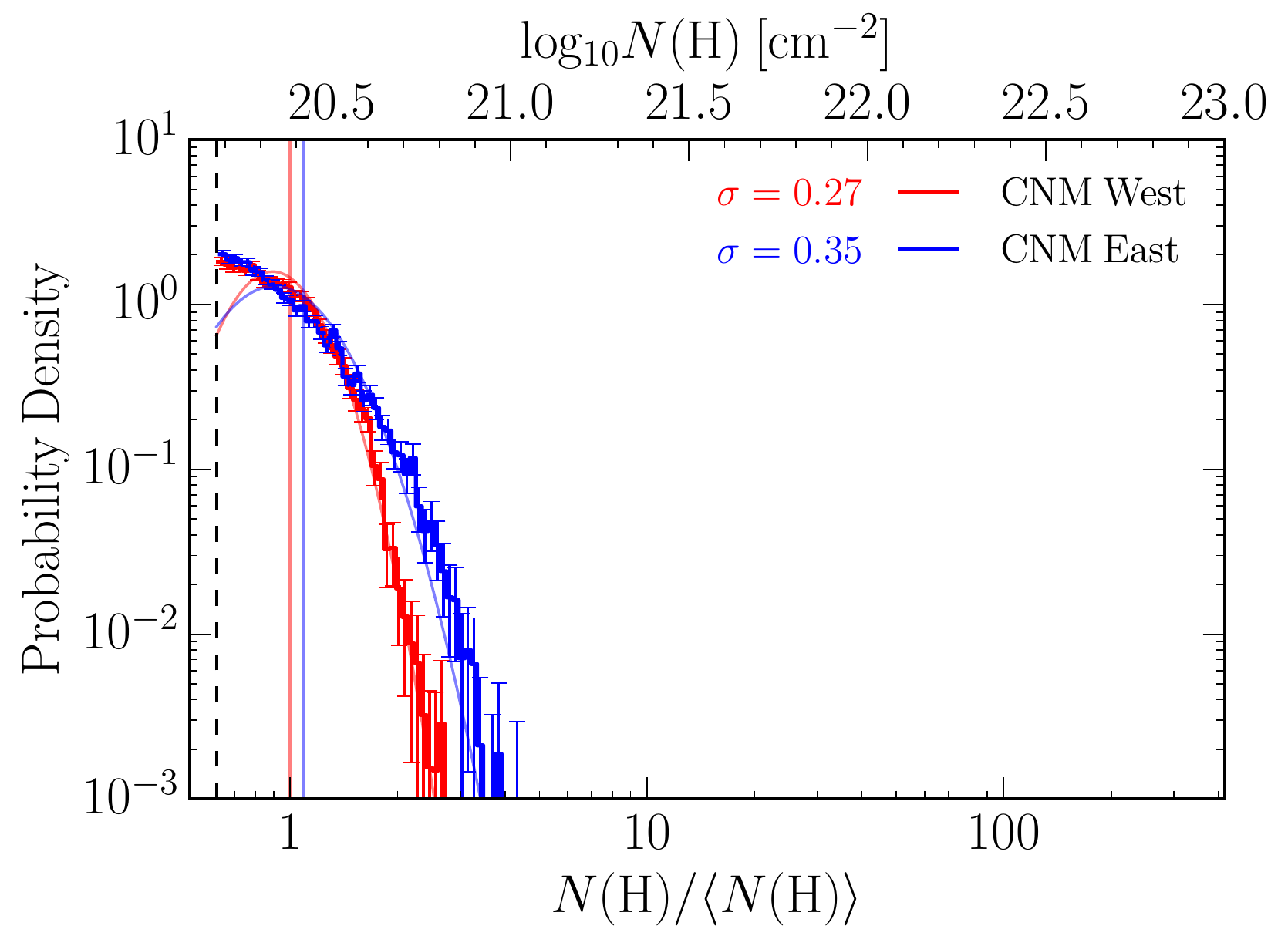}
 \includegraphics[width=0.4\textwidth]{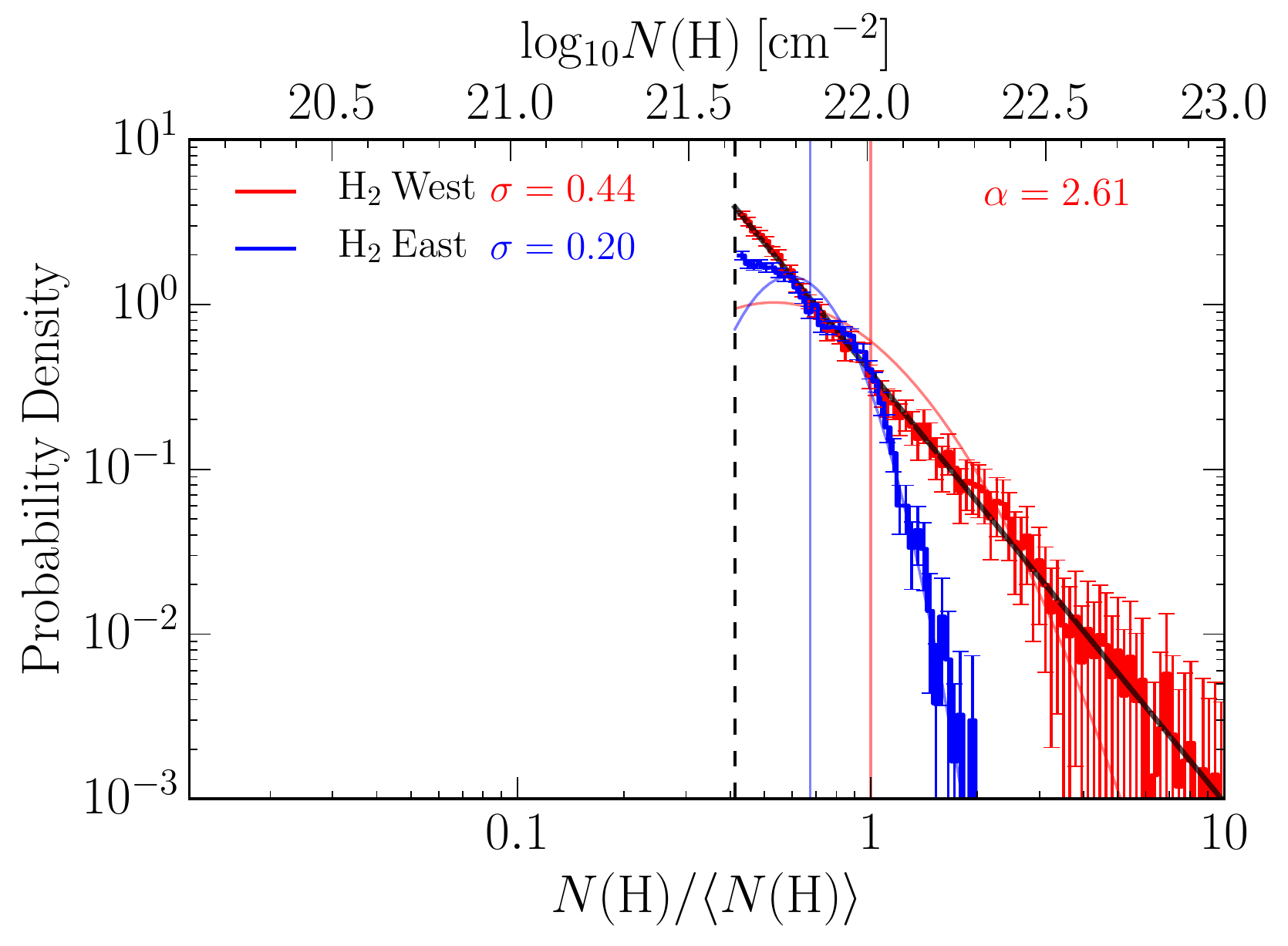}
\caption{The N-PDFs for the two subregions within the filament marked with dashed polygons in Fig.~\ref{fig_column_density}. The top and bottom panel show the N-PDF of atomic and molecular hydrogen derived with HISA and $^{13}$CO measurements, respectively. The lighter lines show the log-normal (top and bottom panel) and power-law (bottom panel) fits. The black dashed lines in each panel mark the column density threshold, and the solid vertical lines indicate the mean column densities. The bottom x-axes in both panels are the relative column density for the N-PDF of the western subregion, i.e., the red histogram in both panels. Other N-PDFs are shifted to the correct absolute column density for comparison. The thick black line in the molecular component panel shows the power-law fit with an index $-\alpha$ to the column density distribution of the western region.}
  \label{fig_column_density_pdf_left_and_right}
\end{figure}

We next discuss the N-PDFs of the two subregions of the filament. The N-PDFs for the eastern and western regions, indicated with dashed polygons in Fig.~\ref{fig_column_density}, are shown in Fig.~\ref{fig_column_density_pdf_left_and_right}. The N-PDFs of the cold dense \ion{H}{i} traced by HISA for both subregions still can be described by a single log-normal function with a similar width as that found over the whole filament. The N-PDFs of the molecular gas show very different shapes for two subregions. While the N-PDF for the eastern subregion can be described by a log-normal function, the N-PDF for the western subregion shows a clear power-law shape. 

Recall that we pointed out in Sect.~\ref{sect_intro} that theoretical studies predicted that the shapes of the N-PDFs are depend on the physical processes acting within the cloud, a log-normal N-PDF indicates that turbulent dominates and a power-law indicates that gravity dominates \citep[e.g.,][]{Federrath2010, Ballesteros-Paredes2011, Kritsuk2011, Federrath2013, Burkhart2015B}. A possible explanation for the different N-PDF for different subregions is that the western subregion shows ongoing high-mass star formation activities, i.e., UC\ion{H}{ii} region, indicating that the molecular cloud is dominated by gravitational collapse. 


One possible reason we do not see a power-law tail in the N-PDF of the molecular gas in the eastern subregion is that the $^{13}$CO could be frozen onto the dust grains in the densest and coldest part of the molecular cloud \citep[e.g.,][]{Giannetti2014}. In this case we could not recover the high density part, whereas in the western subregion the feedback effects from the UC\ion{H}{ii} have released the $^{13}$CO molecules from the dust grains. As the line is not excited at very low density, altogether $^{13}$CO only traces the intermediate density gas between the dense star forming cores, and so these do not show up in the N-PDF (see e.g., \citealt{Ossenkopf2002} for a similar effect in the $\Delta$-variance.)

We estimated the $^{13}$CO depletion factor by comparing the molecular cloud column density we derived from $^{13}$CO with the H$_2$ column density derived from the ATLASGAL dust continuum. Four ATLASGAL dense clumps (AGAL036.406+00.021, AGAL036.666-00.114, AGAL036.826-00.039, and AGAL036.839-00.022) that have velocities within the respective $^{13}$CO velocity range are located in the eastern subregion \citep{Urquhart2018}. We smoothed the ATLASGAL image into the same pixel size and angular resolution calculated the column density of molecular hydrogen following Eq.~15 in \citet{Giannetti2014}. The same dust opacity $\kappa=1.8$~cm$^{2}$~g$^{-1}$ as that used by \citet{Giannetti2014} is employed, and we assume the dust temperature from Herschel Hi-GAL results \citep{Marsh2017}. The $^{13}$CO depletion factors we derived are between 2 and 4, with a mean value of 3, which agrees with the numbers \citet{Giannetti2014} derived for infrared dark sources. Therefore, the high column density part of the N-PDF in the eastern subregion could be underestimated due to $^{13}$CO depletion.  

\subsection{Evolutionary stages}
\label{sect_evolutionary_stages}

As mentioned in Sect.\,\ref{HISA_column_density}, we observe a significant difference in the distribution of the H$_2$ column density for the eastern and western subregion of the filament. The eastern subregion shows a more diffuse column density distribution, whereas the western subregion reveals several high column density peaks. Hence, the western subregion reveals a power-law in the N-PDFs shown in Fig.\,\ref{fig_column_density_pdf_left_and_right} and the eastern subregion shows a log-normal shaped PDF. Furthermore, we see a luminous UC\ion{H}{ii} region within the western subregion, whereas the eastern subregion does not harbor significant continuum emission. The ATLASGAL survey \citep{Schuller2009} shows strong extended submillimeter emission with a group of star forming clumps in the western subregion, whereas only a few unresolved continuum clumps are found with velocities within the respective $^{13}$CO velocity range in the eastern subregion \citep{Ragan2014, Urquhart2018}. All these different tracers are indicative of active high mass star formation with strong feedback effects on the western side of the filament, while the eastern side shows no effects of feedback. 


The kinematics of the HISA do not exhibit significant differences between the eastern and western subregion. However, comparing the HISA and $^{13}$CO kinematics shows an interesting difference. The eastern subregion shows similar peak velocities for the \ion{H}{i} and CO, whereas the western subregion reveals a difference of $\sim$4\,km\,s$^{-1}$. Using the newly developed histogram of oriented gradients (HOG) tool, \citet{Soler2018} confirmed morphological correlation between the \ion{H}{i} and $^{13}$CO emission in velocity channels separated by 3 to 4~km~s$^{-1}$ towards the western subregion. The \ion{H}{i} at $\sim$54~km~s$^{-1}$ is spatially correlated with $^{13}$CO emission at $\sim$57.5~km~s$^{-1}$ \citep[Fig.~15 in ][]{Soler2018}. They further demonstrated by applying the HOG analysis on synthetic observations from MHD simulations that this velocity offset between $^{13}$CO and \ion{H}{i} could arise from more general molecular cloud formation conditions. Another explanation for this velocity offset could be the feedback from the expanding UC\ion{H}{ii} region G34.256+0.146. A significant amount of the $^{13}$CO gas at $\sim54$~km~s$^{-1}$ has been driven away by the radiation from the forming high-mass star.

Studying the cold \ion{H}{i} column density, we found higher column density peaks on the eastern side in comparison to a more diffuse \ion{H}{i} column density structure on the western side. The maximum spin temperature also shows smaller values on the eastern side. This might be an indication of a young, colder and more dense \ion{H}{i} cloud on the eastern side in comparison to a more evolved cloud on the western side. It is possible that the dense \ion{H}{i} cloud on the eastern side of the filament is about to become a dense molecular cloud, forming high density peaks and subsequently form stars. However, further observations or simulations are needed to support this hypothesis.


\subsection{Uncertainties for the determined HISA properties}
\label{sect_Uncertainties_in_the_HISA_description}
Several factors introduce uncertainties to the determined properties of the HISA features. In the following we will discuss three contributions: the ratio of foreground to background emission -- factor $p$, different methods to determine the background emission and the assumption of the spin temperature $T_{\rm HISA}$.

\begin{figure}
\centering
 \includegraphics[width=0.4\textwidth]{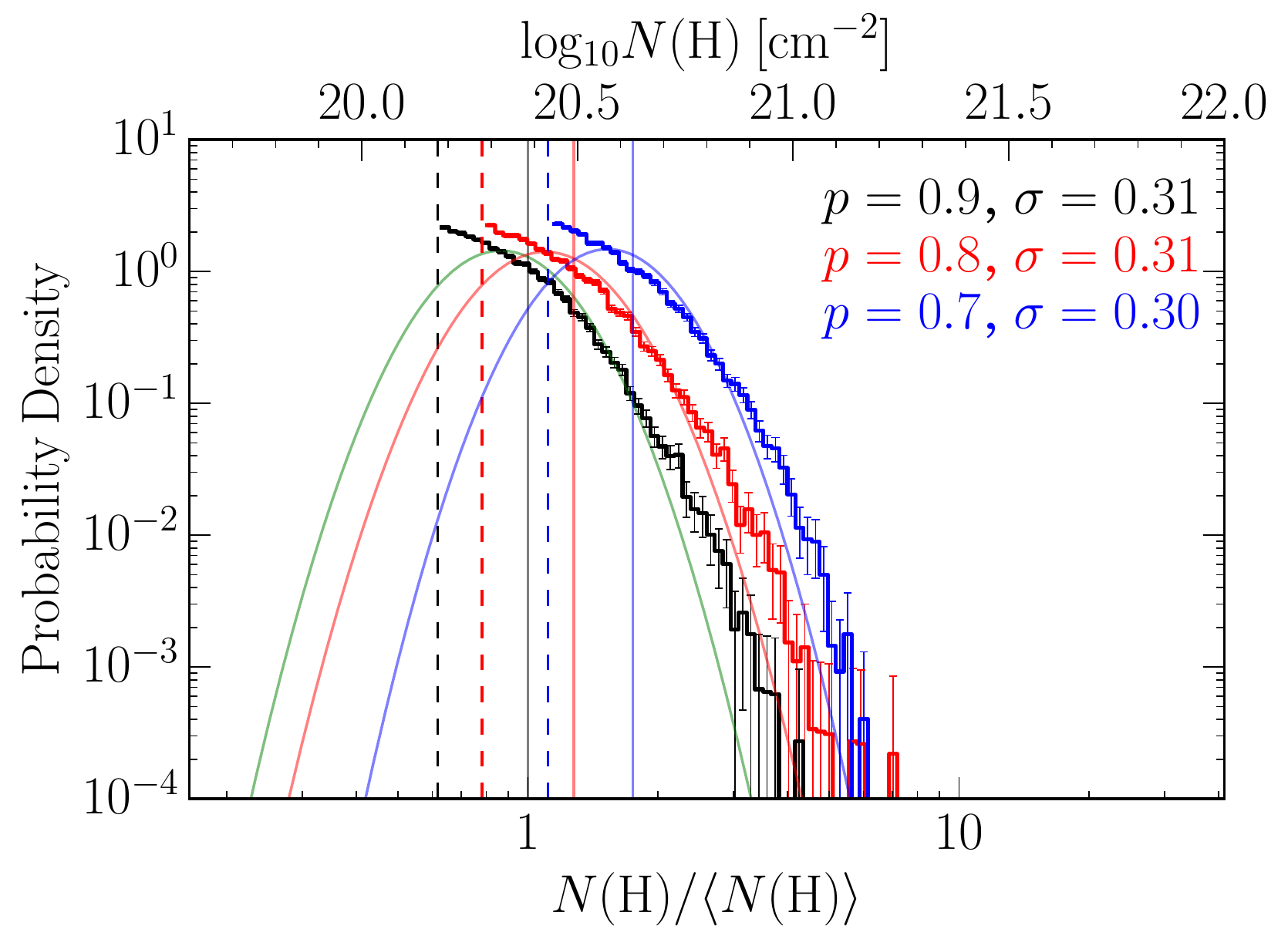}\\
 \includegraphics[width=0.4\textwidth]{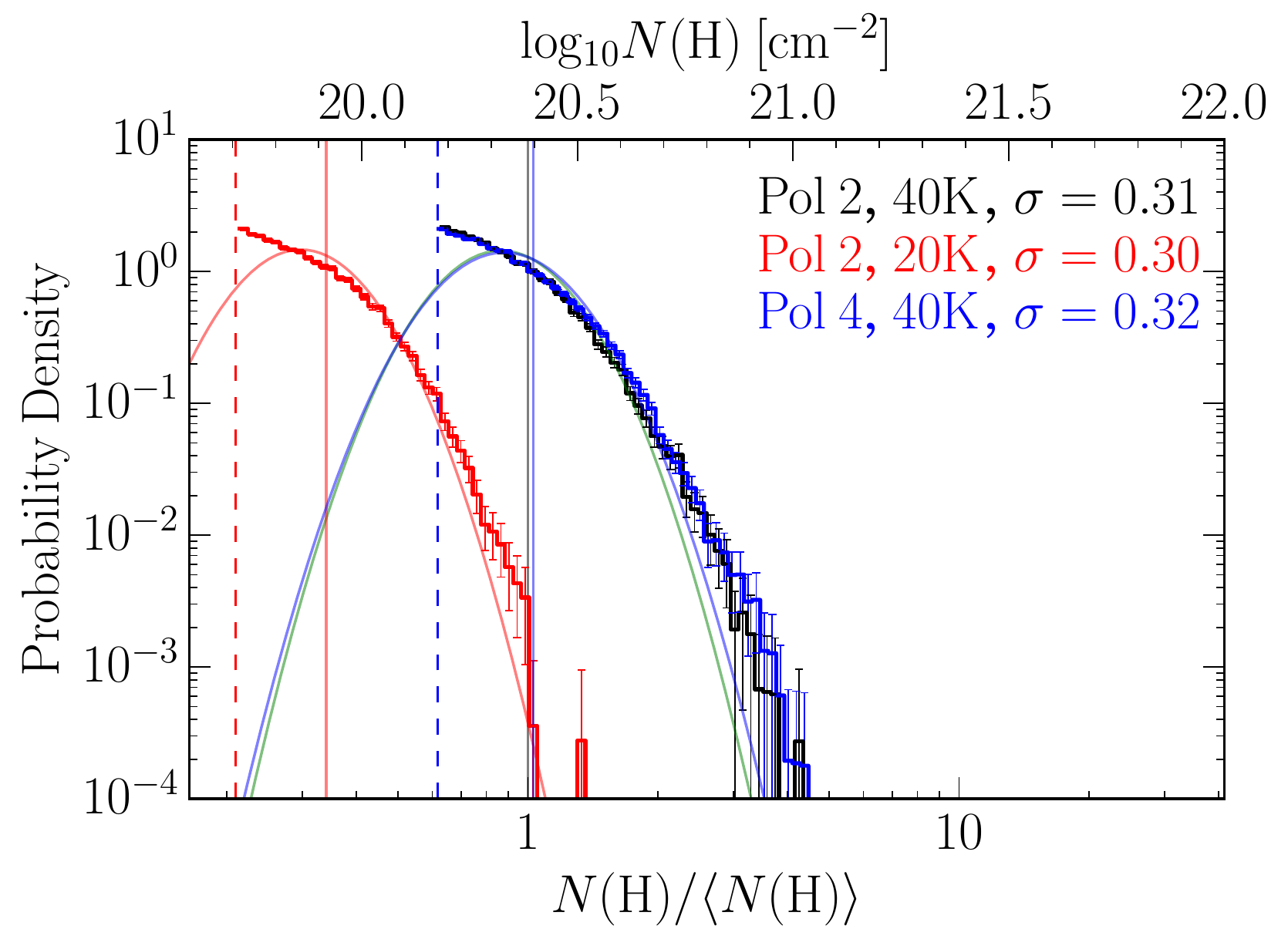}
\caption{Top-panel: N-PDFs for the atomic hydrogen for different $p$. The PDF is measured over the entire filament indicated with a red polygon in Fig.\,\ref{fig_column_density}, assuming a constant spin temperature of $T_{\rm{HISA}}=40$\,K. Bottom-panel: N-PDFs for the atomic hydrogen for different spin temperature and different polynomial functions fit to get $T_{\rm off}$. The dashed vertical lines in each panel indicate the column density threshold, and the solid vertical lines indicate the mean column densities. The bottom x-axes in both panels are the relative column densities for the N-PDF of ``Pol 2, $p$=0.9, 40K'', i.e., the black histogram in both panels. Other N-PDFs are shifted to the correct absolute column density for comparison. }
  \label{fig_column_density_pdf_compare}
\end{figure}

Fig.~\ref{fig_Ts_vs_tau} also shows that for a fixed spin temperature $T_{\rm HISA}$, the larger the value of $p$ is, the lower the optical depth $\tau$ is, hence the lower the column density is. Depending on the $p$ value we choose (between 0.7 and 0.9), the column density can change by at most a factor of $\sim2$. This is shown in Fig.\,\ref{fig_column_density_pdf_compare} for N-PDFs of the entire filament assuming three different values for $p$. The column density structure stays almost constant, but the actual values are shifted for different $p$ values.

As discussed in Sect.\,\ref{sect_background_estimate_T_off}, the chosen method for the background estimate can influence the absorption depth of the HISA feature and thus the column density. We showed in Sect.\,\ref{sect_background_estimate_T_off} that the best method is a polynomial fit to the \ion{H}{i} spectra and interpolate for the HISA feature. The difference for a second or fourth order polynomial is negligible for most regions. As we can see in the bottom panel of Fig.~\ref{fig_column_density_pdf_compare}, the column density structures of the second and fourth order polynomial are almost identical, the mean column density of the fourth order polynomial fit is $\sim3\%$ higher than the one of second order. Thus we chose to use a second order polynomial fit for $T_{\rm off}$.

Another important factor is the assumption of a constant spin temperature for the cloud. This is obviously a poor assumption, but using additional measurements we can constrain the range of the spin temperature. The most important one is the upper limit for the spin temperature introduced in Sect.\,\ref{sect_maximum_spin_temperature}. Using this information, we can constrain the spin temperature to values of $T_{\rm{HISA}}<70$\,K for the majority of the HISA features. For the CNM mass estimation given in Sect.\,\ref{sect_mass_estimate}, we assumed a spin temperature of $T_{\rm{HISA}}=20$ and 40\,K. As seen in Fig.\,\ref{fig_column_density_pdf_compare}, the N-PDF does not change significantly, but higher spin temperatures result in larger column densities and masses. In Sect.\,\ref{sect_mass_estimate}, we showed that the mass is about a factor of three larger for $T_{\rm{HISA}}=40$\,K with respect to that assuming  $T_{\rm{HISA}}=20$\,K.

Furthermore, we performed our HISA extraction method on model images (see Appendix~\ref{app_pdf_width}) to estimate the uncertainty brought in by the $T_{\rm off}$ fitting method to the width of the N-PDF. The test results show that the ``instrument broadening'' introduced by the fitting method to the N-PDF is not very large and the log-normal width of the CNM we derive in the paper is robust.

In summary, it is difficult to quantify exactly the uncertainty of the CNM column density and mass. Considering all assumptions, the CNM mass has an uncertainty of a factor of 2-3 (dominated by the uncertainty of $T_{\rm S}$), which is similar to the H$_2$ mass uncertainty based on $^{13}$CO. However, we showed that the shape of the column density PDF is robust.

\begin{table}[htbp]
\caption{Results of the fits to the N-PDFs of CNM for different parameters (Fig.~\ref{fig_column_density_pdf_compare}).}
\centering
\begin{threeparttable}
\label{table_npdf_cnm}
\begin{tabular}{l c c c c c} 
\hline\hline
Pol. & p&$T_{\rm HISA}$&$N$ Threshold&$\langle N$(H)$\rangle$ & width\\
           & & [K]& log$_{10}{(N{\rm(H)})}$  & log$_{10}{(N{\rm(H)})}$ &$\sigma$ \\
\hline\vspace{-0.35cm}\\
Pol 2 &$0.9$ & 40 & 20.18 &20.39& 0.31\\
Pol 2 &$0.8$ & 40 & 20.28 &20.49& 0.31\\
Pol 2 &$0.7$ & 40 & 20.43 &20.63& 0.30\\
Pol 2 &$0.9$ & 20 & 19.71 &19.92& 0.30\\
Pol 4 &$0.9$ & 40 & 20.18 &20.40& 0.32\\
\hline   
\end{tabular} 
\end{threeparttable}
\end{table}


\section{Conclusions}
We studied atomic and molecular gas components of the giant molecular filament GMF38a. The molecular component is traced via observations of $^{13}$CO, whereas the cold atomic gas is observed via HISA features and \ion{H}{i} 21~cm emission. The main results can be summarized as:

\begin{enumerate}
\item We extracted HISA by estimating the background emission with different methods. For the observed giant filament, a polynomial fit of second order to the neighboring channels of each HISA is the most reliable method from the different options we tested to estimate the background emission. 
\item The HISA features and the $^{13}$CO emission are spatially correlated. While in the eastern subregion they correlate well, the peak velocities of the two tracers show an offset of $\sim$4\,km\,s$^{-1}$ for the western subregion of the filament.
\item Although the linewidth ratio between HISA and $^{13}$CO is around unity, the Mach number estimation shows that the $^{13}$CO emission is dominated by supersonic turbulent motions, whereas a large fraction of the CNM is at subsonic or transonic velocities.
\item Assuming a spin temperature of $T_{\rm{HISA}} = 40$\,K for the \ion{H}{i}, we determined the column densities of the cold dense \ion{H}{i} and compared them to the H$_2$ column density distribution derived from the $^{13}$CO emission. The column density peaks do not coincide and the \ion{H}{i} column density shows in general a diffuse structure. The H$_2$ column density reveals prominent peaks in the western subregion of the filament whereas the eastern subregion appears more diffuse. The CNM-to-H$_2$ column density ratio varies between 0.5 to 25\% with a median value of $\sim9$\%. The outer layer of the filament exhibits a higher ratio than that toward the cloud centers. The surface density of atomic hydrogen peaks at $\sim 14-23~M_\odot$~pc$^{-2}$ (corresponding to $\sim1.8-2.9\times10^{21}$~cm$^{-2}$). Furthermore, the mass traced by HISA is only $\sim3$\% of the molecular mass, and $\sim1.6$\% of the mass traced by atomic \ion{H}{i} emission.
\item Studying the N-PDFs, we are able to provide constraints on the physical processes within the cloud. The location of the HISA N-PDF is strongly dependent on the assumed parameters, but the width is not. The N-PDFs of CNM, \ion{H}{i} emission, and H$_2$ can be described by a log-normal function, which indicates turbulent motions as the main physical driver. Only the H$_2$ column density of the western subregion within the filament is characterized by a high column density power-law structure, consistent with the observed star formation activity. Adding the column density maps of all three tracers (CNM, \ion{H}{i} emission, and H$_2$) up, we generated a column density map of ``all gas''.
\item We hypothesize that the eastern and western side of the filament represent different evolutionary stages. The eastern side represents an earlier stage, which is currently forming a dense molecular cloud out of the atomic reservoir. As we do not observe high molecular column density peaks, the \ion{H}{i} shows low spin temperatures and high column densities. In contrast, the western side of the filament shows high H$_2$ column density peaks, signs of active star formation, such as UC\ion{H}{ii} regions, and in general a warmer and less dense atomic counterpart. These differences provide interesting constraints for theoretical models and simulations of the formation of molecular clouds. 
\end{enumerate}

\begin{acknowledgements}
The National Radio Astronomy Observatory is a facility of the National Science Foundation operated under cooperative agreement by Associated Universities, Inc. Y.W., H.B., S.B., and J.D.S. acknowledge support from the European Research Council under the Horizon 2020 Framework Program via the ERC Consolidator Grant CSF-648505, and R.S.K. via the ERC AdvancedGrant 339177 (STARLIGHT). H.B., S.C.O.G., and M.R. acknowledge support from the Deutsche Forschungsgemeinschaft in the Collaborative Research Center (SFB 881) ``The Milky Way System'' (subproject B1, B2, B8). R.S.K. and S.C.O.G also acknowledge support from the DFG via Germany’s Excellence Strategy EXC-2181/1 - 390900948 (the Heidelberg STRUCTURES Cluster of Excellence). This work was carried out in part at the Jet Propulsion Laboratory which is operated for NASA by the California Institute of Technology. N.S. acknowledges support by the french ANR and the german DFG through the project ``GENESIS'' (ANR-16-CE92-0035-01/DFG1591/2-1). F.B. acknowledges funding from the European Union’s Horizon 2020 research and innovation program (grant agreement No 726384). This research made use of Astropy and affiliated packages, a community-developed core Python package for Astronomy \citep{astropy2018}, Python package {\it SciPy}\footnote{\url{https://www.scipy.org/}}, APLpy, an open-source plotting package for Python \citep{robitaille2012}, and software TOPCAT \citep{taylor2005}. The authors thank the anonymous referee for the constructive comments that improve the paper.

\end{acknowledgements}

\bibliographystyle{aa}
\bibliography{references.bib}

\begin{appendix}

\section{All gas}
\begin{figure*}[htpb]
\centering
 \includegraphics[width=0.9\textwidth]{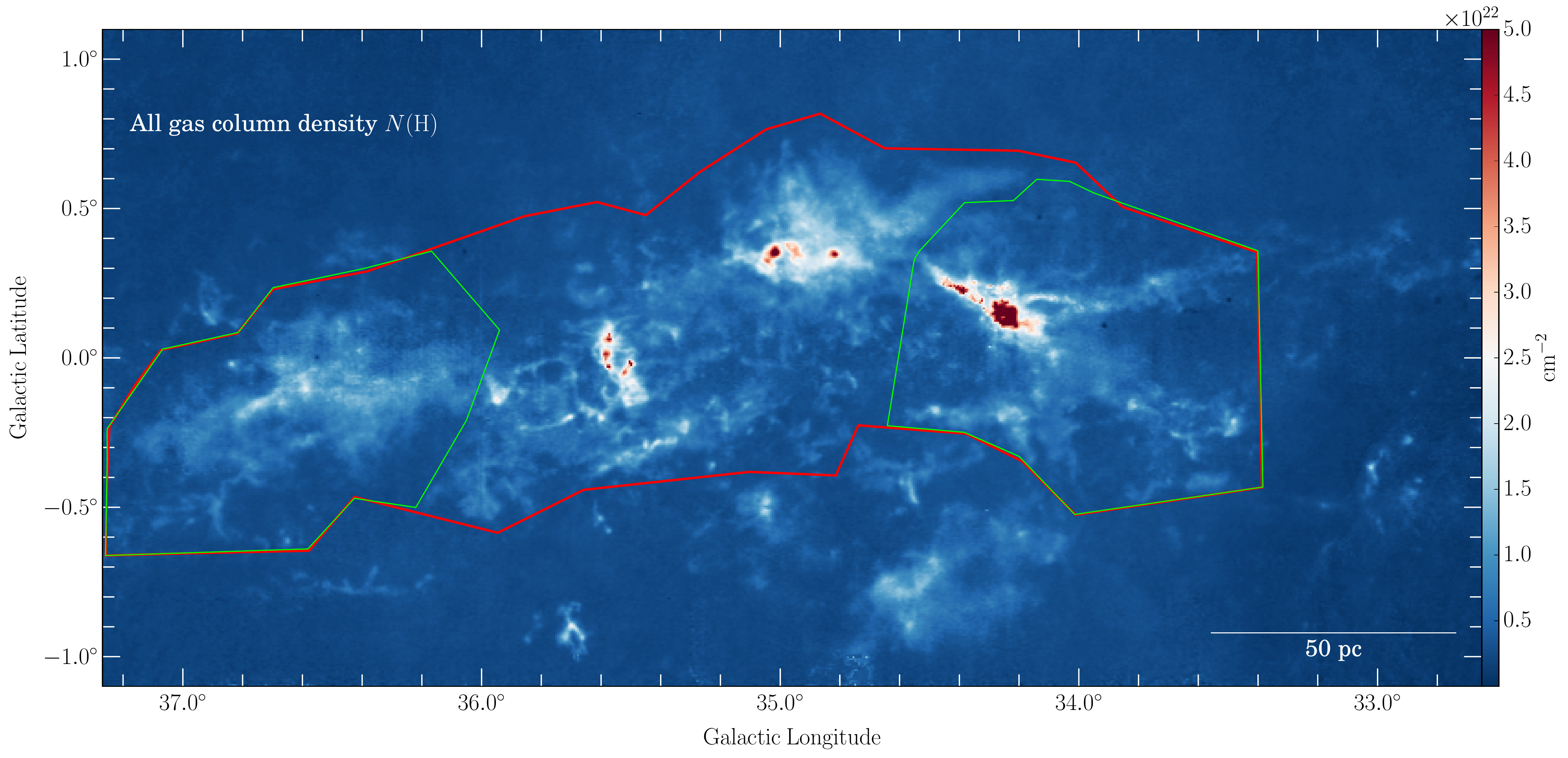}
\caption{The column density map of the gas combined molecular component (from $^{13}$CO), CNM (from HISA) and WNM+CNM (from \ion{H}{i} emission) (see Sect.\,\ref{sect_H2_column_density_estimate}). The red and green dashed polygons mark the region for the mass estimates and the column density PDF measurements (only molecular component and CNM) shown in Figs.\,\ref{fig_column_density_pdf_overview} and \ref{fig_column_density_pdf_left_and_right}, respectively.}
  \label{fig_column_density_all_gas}
\end{figure*}

\section{CNM N-PDF width test}
\label{app_pdf_width}
To test how much our HISA identification and fitting method broadens the N-PDF of the CNM, we tested our HISA extraction method on artificial \ion{H}{i} emission maps with absorption features. 

We made a model \ion{H}{i} map with a relative uniform peak $T_{\rm B}\sim85$~K with a noise of $\sim4$~K, which is similar to the real data we have. The \ion{H}{i} spectra in the model map have a random linewidth varying between $\sim$30 to 45~km~s$^{-1}$, and a random peak velocity between 47.5 to 52.5~km~s$^{-1}$. We also made a model continuum map with a relative uniform $T_{\rm B}\sim13$~K with a noise of 0.7~K, and both values are similar to the diffuse continuum emission flux and the noise level of the real data. Both model images have the same pixel size (10\arcsec) and beam size (40\arcsec) as the real data.

\begin{figure}[htpb]
\centering
 \includegraphics[width=0.4\textwidth]{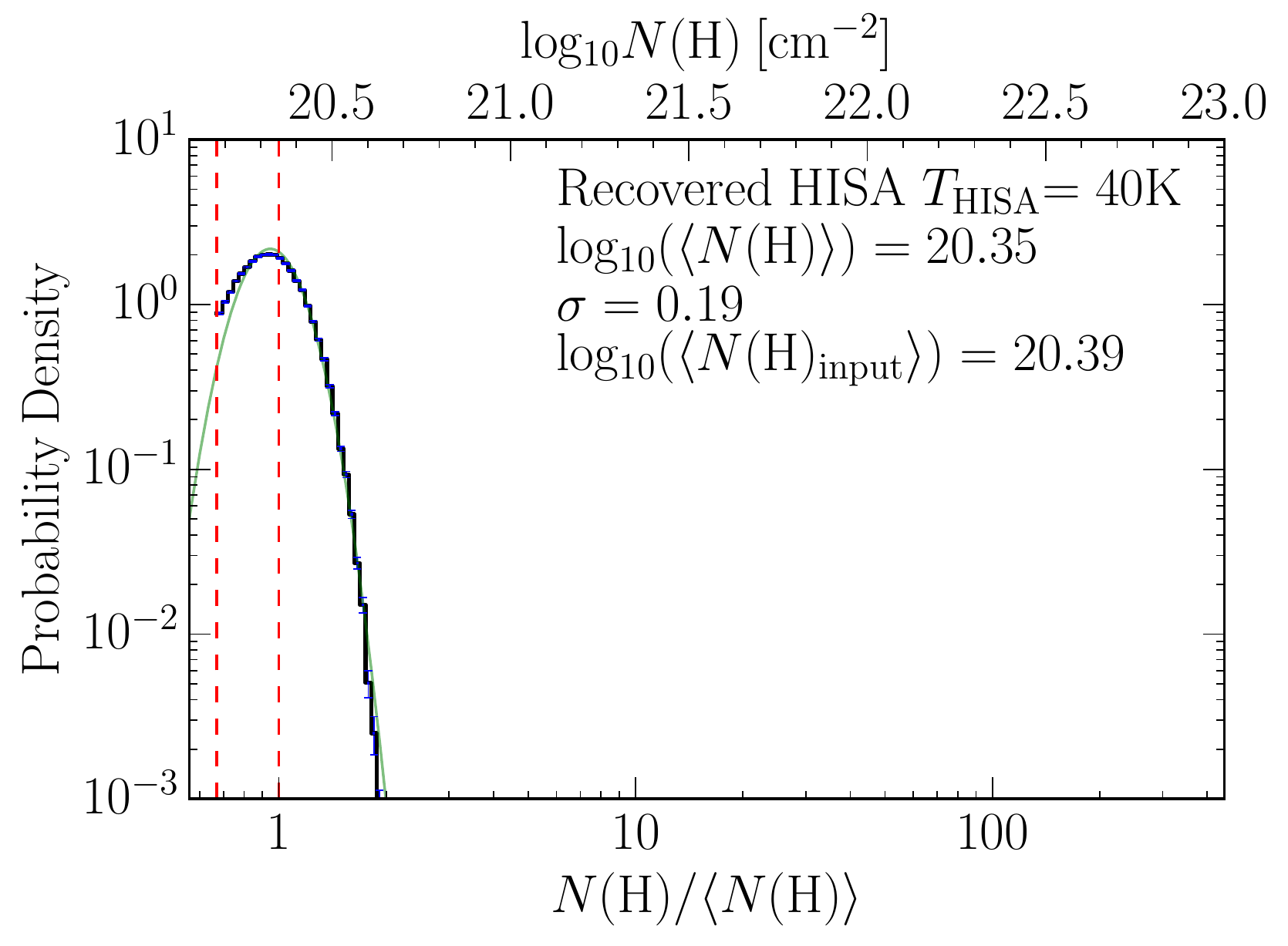}
 \includegraphics[width=0.4\textwidth]{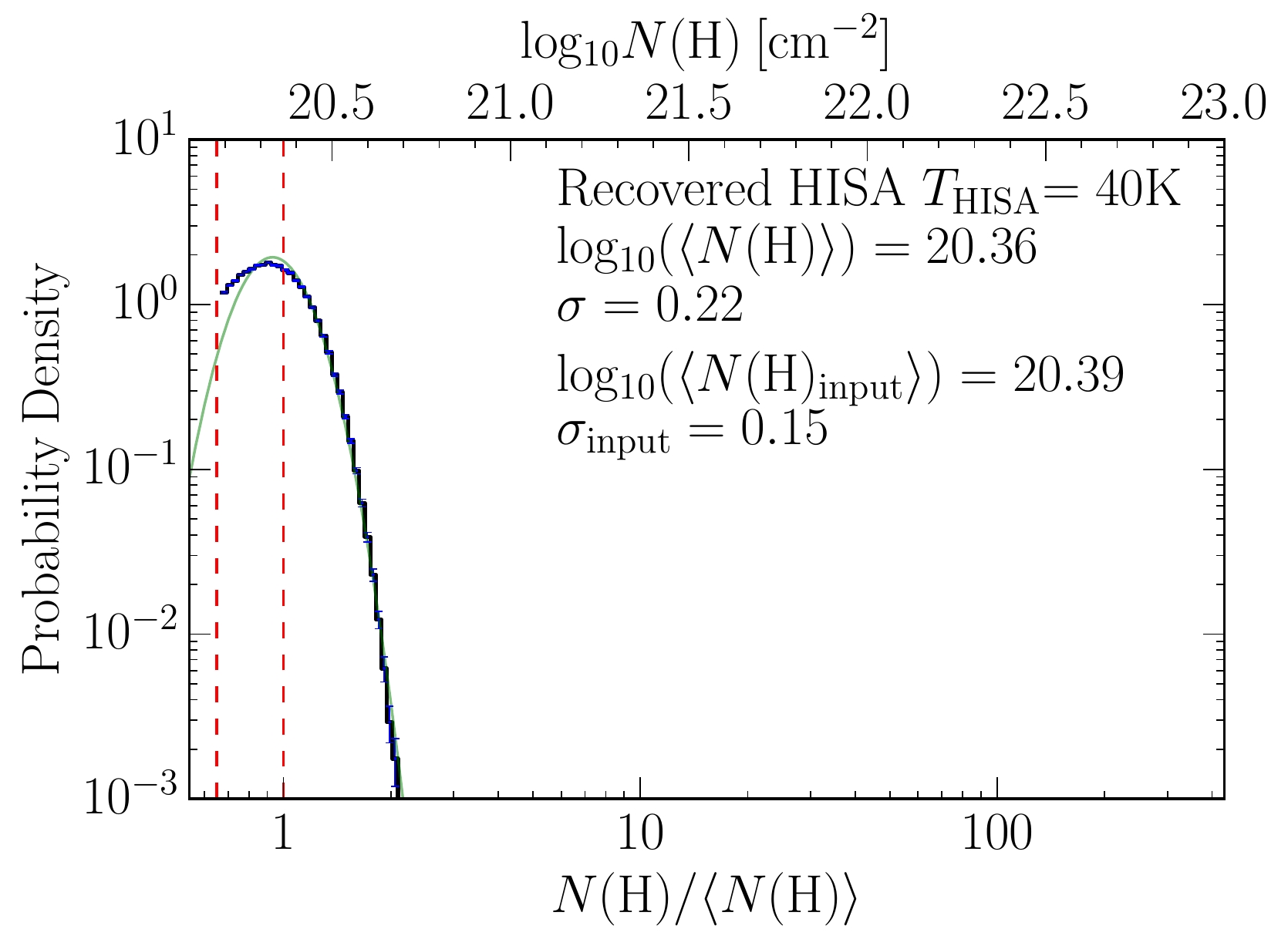}
\caption{The N-PDFs of the CNM recovered from the tests.} 
  \label{fig_model_pdf}
\end{figure}

For the first test, we generate artificial absorption features from a single column density value with $T_{\rm HISA}=$40~K and $p=0.9$, and added them into the model \ion{H}{i} image. This column density value equals to the mean CNM column density of the GMF38a in Table~\ref{table_npdf} (log$_{10}{(N{\rm(H)})}$ = 20.39~cm$^{-2}$). The artificial absorption features peak at 50~km$^{-1}$ and have a linewidth of 5~km~s$^{-1}$. Following the method described in Sect.~2 and 3, we extracted the HISA spectra, estimated the column density and construct the N-PDF shown in the top panel of Fig.~\ref{fig_model_pdf}. The absorption features we put in the model image all have the same column density, so the modeled N-PDF is a delta function. Due to noise and the uncertainty we brought in through our HISA extraction method, the N-PDF we derived has a width of 0.19, and about 10\% lower mean column density than the input one. 

For the second test, instead of a single column density value, we generate artificial absorption features from a log-normal distribution column density with $T_{\rm HISA}=$40~K and $p=0.9$, and added them into the model \ion{H}{i} image. The input log-normal distribution has a width of 0.15, and a peak at the mean CNM column density of the GMF38a in Table~\ref{table_npdf} (log$_{10}{(N{\rm(H)})}$ = 20.39~cm$^{-2}$). Similarly, we ran our procedure on the model image and the N-PDF is shown in the bottom panel of Fig.~\ref{fig_model_pdf}. The input width of 0.15 is broadened to 0.22, and the mean column density is also about 10\% lower than the input one.

Both tests demonstrate that the ``instrument broadening'' contribution to the N-PDF of our method is not very large and the log-normal width of the CNM we derive in the paper is robust.

\end{appendix}

\end{document}